\documentclass[aps,pre,twocolumn,superscriptaddress,epsf]{revtex4-1}
\usepackage{latexsym}
\usepackage{amssymb,amsmath,amsbsy}
\usepackage{graphicx,color}
\usepackage{bm}
\usepackage{amssymb,amsmath,amsbsy}
\usepackage{graphicx,color}
\usepackage{enumerate}
\usepackage[english]{babel}                                                        
\usepackage{euscript}
\usepackage{upgreek}
\usepackage{textcomp}
\usepackage{color}

\allowdisplaybreaks

\def\mv{{\bm v}}

\def\mp{{\bm p}}
\def\mq{{\bm q}}
\def\mpar{{\bm \partial}}
\def\mf{{\bm f}}

\begin{document}


\title {Superfluid Phase Transition with Activated Velocity Fluctuations: Renormalization Group  Approach}

\author{Michal Dan\v{c}o} 
 \affiliation{Institute of Experimental Physics, SAS, Ko\v{s}ice, Slovakia}
 \affiliation{Joint Insitute for Nuclear Research, Dubna,  Russia}
\author{Michal Hnati\v{c}}
\affiliation{Institute of Experimental Physics, SAS, Ko\v{s}ice, Slovakia}
\affiliation{Joint Insitute for Nuclear Research, Dubna,  Russia}
\affiliation{Faculty of Sciences, P.J. \v{S}afarik University, Ko\v{s}ice, Slovakia}
\author{Marina V. Komarova}
\affiliation{Department of Theoretical Physics, St. Petersburg University,\\Ulyanovskaya 1, St. Petersburg, Petrodvorets, 198504 Russia}
\author{{Tom\'a\v{s} Lu\v{c}ivjansk\'y}}
\affiliation{Faculty of Sciences, P.J. \v{S}afarik University, Ko\v{s}ice, Slovakia}
\affiliation{Fakult\"at f\"ur Physik, Universit\"at Duisburg-Essen, D-47048 Duisburg, Germany} 
\author{Mikhail Yu. Nalimov}
\affiliation{Department of Theoretical Physics, St. Petersburg University,\\Ulyanovskaya 1, St. Petersburg, Petrodvorets, 198504 Russia}

\date{\today}

 \begin{abstract}
 A quantum field model that incorporates Bose-condensed systems near
 their phase transition into a superfluid phase and velocity fluctuations is proposed. 
 The stochastic Navier-Stokes equation is used for a generation of the velocity fluctuations. 
 As such this model generalizes model F of critical dynamics. 
 The field-theoretic action is derived using {the} 
 Martin-Siggia-Rose formalism and path integral approach.
 The regime of equilibrium fluctuations is analyzed within perturbative renormalization
group method.  The double $(\epsilon,\delta)$-expansion scheme
is employed, where $\epsilon$ is a deviation from space dimension $4$ and $\delta$
describes scaling of velocity fluctuations. The renormalization procedure is performed
to the leading order. 
The main {corollary} gained from 
 the  analysis  of the thermal equilibrium regime suggests that one-loop calculations of
 the presented models are not sufficient to make a definite conclusion about the stability of fixed points. 
We also show that 
critical exponents are drastically changed as a result of the turbulent background
and critical fluctuations are in fact destroyed by the developed turbulence fluctuations. 
{The} scaling exponent of effective 
viscosity is calculated and agrees with expected value $4/3$. 
\end{abstract}

\pacs{}

\maketitle

%
%
\section{Introduction}
 Non-equilibrium physics \cite{Tauber2014,HHL08} constitutes an interesting research topic 
 to which a lot of effort has been devoted in last decades. In general
 such problems are difficult to solve exactly. However, a great simplification is
 possible near continuous phase transitions where
  new symmetry related to scale invariance appears. An immediate hallmark of it
  is divergence of the correlation length, which results into an importance
  of fluctuations on all length scales. The system then effectively forgets
  about microscopic details and can be described by a few coarse grained quantities.
 
  The liquid-vapour critical point, $\lambda$-transition 
  in superfluid helium  $\mbox{ }^4$He and transition between ferro- and paramagnetic phase near Curie temperature in 
  ferromagnetic materials belong to famous examples of continuous phase transitions. From
  an experimental point of view \cite{Folk06}
  a special role is devoted to the phase transition of $\mbox{ }^4$He where one can
  approach criticality closer than in any other system. However, special
 techniques have to be applied in order to overcome slow dynamic transitions.
  Interests {in} the theoretical investigation of Bose-con\-den\-sa\-tes in
  {the} superfluid 
  state {retrieve attention} after recent experimental 
  achievements in con\-den\-sa\-tion of diluted inert gases \cite{gas}. Superfluidity appears 
  at the phase transition lambda point,  where  the viscosity of the fluid vanishes \cite{Barenghi08,Altland}. 
  However, the critical dimension of {the} viscosity coefficient, i.e. the
  law defining  its behavior in the limit $\nu\rightarrow 0$, is not determined
  yet. This drawback is explained  by the fact that traditionally the critical 
  dynamics in the vicinity of the lambda point is described {by} model E or F in
   the standard terminology \cite{HH77}. In the  framework of {a} 
  traditional  construction of aforementioned models 
  the velocity, whose dynamics is described by the Navier-Stokes equation, is infrared 
  (IR) irrelevant. As a result, the vis\-co\-si\-ty  drops out and clearly cannot be analyzed.  
 
  From classical theory of fluids \cite{Frisch,Landau_fluid} it is well-known that vanishing viscosity leads
  to the phenomenon of turbulence. A genuine property of turbulent flows with continuous
  phase transition is scale invariance. In an inertial interval power laws 
  are generically observed and independent of viscosity (second Kolmogorov hypothesis). These findings are
  corroborated in the celebrated Kolmogorov works \cite{Kolmogorov1,Kolmogorov2}. 

  A phenomenon of turbulence has been discovered and lately analyzed also in other than
  its original context, among others in
 high energy physics \cite{AM06,MSW07,Khach08,Scheffler09,BSS09,BS11,Schlichting12,BSSV14}, inflation cosmology 
  \cite{MT04,BTW06,VPS10}, ultracold gas \cite{NSSG12} or quantum turbulence (QT)
 \cite{Vinen02,Vinen06,KT07,SS12,BSS14}. Due to a progress in experimental methods
  the latter was studied also directly
  \cite{Raman99,WG08,Bradley08,Salort}. 
  For the  turbulence in superfluids the quantum effects are of utmost importance. 
  From a theoretical point of view the zero temperature (ground state) 
 of bosonic superfluid 
  is well described by the non-linear
  Schr\"odinger equation, also none as Gross-Pitaevskii equation \cite{Gross61,Pitaevskii61}. Different
 theoretical techniques  \cite{Nemirovskii13} can be employed in order to study QT. 
 Various aspects in connection with turbulence were recently analyzed \cite{KT07,NSG11,SNG12,MGP14}.
  
 A common feature of all these studies is a concentration on the quantum state
 of the superfluid state and thus operate well below the critical temperature (to the left
 from $\lambda$-line). 
 We would like to point out that our aim is to study behavior of phase transition in liquid helium above
 critical temperature, where all quantum effects can be neglected. 
 According to the classical work on critical dynamics 
 \cite{HH77} to a given static universality class, different dynamic classes can be
 assigned.  
 At the present time, there is no general {consensus}
  which dynamic model (E or F)  is
  {genuine} from the point of view of experimentally measurable  quantities. 
  Both models E
  and F are developed from  model C \cite{HH77,Vasiliev} by adding
  new interaction terms. Model 
  F reduces to model E as an appropriate coupling constant ($g_{2}$ in our notation)
  equals zero. In {the} corresponding static model, one of the $\upomega$ {indices} 
  coincides directly with 
  the famous, experimentally measurable  index $\alpha$ \cite{Vasiliev}. The index $\alpha$
  was calculated in the framework of {the}
  renormalization group approach (RG) using resummation
  procedure \cite{Kleinert} up to the four-loop perturbation precision and was measured in the famous 
  Shuttle experiment \cite{shuttle}. {The} present-day
  value  $\alpha =-0.0127$ is {generally accepted}.
  The negativity of the index $\alpha$ ensures $g_{2}^*=0$ for the stable fixed point. 
  That means that the stability of model E can be considered as a particular realization of model F.

 
 In vicinity of a critical point the correlation length diverges. Due to the
 small value of viscosity any small velocity fluctuation can be considerably
 enhanced and thus large Reynolds number is to be expected. 
According to Kolmogorov hypothesis \cite{Kolmogorov1,Kolmogorov2} there is an inertial interval in which
transfer of energy from large to small scales takes place. In this interval
homogeneous isotropic turbulence is realized.
 We would like to make an important note with respect to
 a turbulent regime. 
We assume that
there is large scale behavior of the fluid near the outer boundary of the system. One can
imagine that some large vortex of the system size is created, i.e., that energy is 
pumped into a system at $L\rightarrow\infty$ scales. The injected energy  is
then transported via non-linearities in Navier Stokes equation
from the outer to smallest scales. 
This transport takes place in aforementioned inertial interval where scaling behavior
is observed \cite{Frisch}.
 In what follows we employ the perturbative RG technique in order to gain
 information about these velocity fluctuations on the critical behavior. 
 It is possible to proceed along different ways. One that
 is more difficult is based on the analysis of composite operators \cite{Vasiliev}.
 As was mentioned in \cite{Krasnov} composite operators corresponding to the velocity field
 are very complicated objects
  already in model E. 
 Also IR irrelevance for model F has been demonstrated
 only for small values of $\epsilon$. Multiloop calculations of composite
 operators with subsequent resummation is even more complicated task
 than the standard calculation of beta functions and critical exponents.
 After initial efforts in the 1970-80s in the theory of critical phenomena
 we are not aware of any other work whose aim is similar. It is very probable
 that this remains so also in the near future.
 We address this question from a different point of view.
 Instead of using only one expansion parameter (deviation from
 upper critical dimension) we introduce an additional one
 related to the scaling of a fluctuating velocity field. Our approach is
 motivated by other works
  \cite{Nalimov,AHKV05,Honkonen1}
 in which this technique has led to new and interesting results.
  After successful
 renormalization we set expansion parameters to their physical values, which
 is a common procedure in perturbative RG approach \cite{Zinn,Vasiliev}. For a complete
 solution of our problem it is necessary to perform multiloop calculation and
 subsequent resummation of diagrams using RG equations. 
 Though feasible and simpler than computation of composite operators, a multiloop
 calculation is well beyond a scope of this work.
 At this place it is {advantageous} to make an important remark. There exist two RG fixed points in the 
  dynamic model~E,  which are candidates to the {possible IR stable regimes}
  (see, e.g., \cite{Vasiliev}).
  On one hand,  two-loop calculations \cite{Dominicis}  do not lead to the decision how 
  to choose the true fixed point  because of the lack of accuracy in the $\upomega$ calculation.
   On the other hand, the value of the $\upomega$ index depends on the chosen dynamic model.   
  For the same fixed point the $\upomega$ value obtained in the framework of model F can 
  differ from the analogous $\upomega$ in model E. Moreover, an inclusion of hydrodynamic
  {fluctuations} can
  enhance this difference. 
   
  Model E with {activated} hydrodynamic modes
  was proposed and investigated by the RG method in 
  \cite{Danco}. Particularly, hydrodynamic modes were shown to give significant 
  contributions to the $\upomega$ index, which is crucial for a general stability analysis.
   Therefore, an investigation of the most general model F extended by velocity
  fluctuations is highly actual and desired. 
By the word general we would like to stress that model F contains
all possible relevant terms, in which all IR irrelevant terms have been dropped.

 This work is organized as follows. In Section \ref{Sec:intro}, we overview
 model F in the framework
 of the microscopic description. In Section \ref{Sec:action},  the dynamic equations for the most general case
 are derived. The stochastic model given by these equations, stochastic Navier-Stokes
 equation, and suitable asymptotic and retar\-da\-tion conditions are reformulated  as 
 an effective field-theoretic model with Martin-Siggia-Rose action \cite{MSR} and subsequently analyzed. 
 The ultraviolet (UV) renormalization of the model and the elaborated algorithm for the
 calculation of the renormalization constants {are} described in Section
 \ref{Se:RG}. The fixed points of
 the renormalization group approach (RG) are calculated and classified together with 
 their stability regions and possible scaling regimes in Section \ref{Sec:fixed}. The conclusions  and 
  results of the one-loop calculations of Feynman graphs are presented in
 Sections \ref{Sec:conclusion} and \ref{Sec:appendix}, respectively. 
%
%
{\section{Microscopic background of the standard F model} \label{Sec:intro}}
 The microscopic background of critical dynamics for superfluid Bose-condensed 
 systems was considered {previously} in \cite{Komarova}.
 The large scale effective model and corresponding stochastic equations were described 
 directly in the framework of time dependent Green functions at finite temperature. 
 {From a microscopic point of view it was the case of model F that needs 
  activation} of hydrodynamic
 modes. In this section, we present the basic arguments in order {to account for them}.
 Besides, it is useful to give a physical meaning of the fields and parameters of the model.
 
 Let us start with the action  {for a} quantum-field model \cite{Abrikosov}
 \begin{equation}
   \label{1}
  S=\uppsi ^+\left(\partial _\tau-\frac {\Delta }{2m_0}-\mu\right)\uppsi +\frac
  {\uplambda}2(\uppsi ^+\uppsi)^2.
\end{equation}
Here $\uppsi^+({\bm x},\tau)-$,  $\uppsi ({\bm x},\tau)-$fields appear as a path integral
representation of quantum-field operators for Bose-particles,
${\bm x}$ is a $d$-dimensional coordinate, $\tau$ is a 
complex parameter whose real and imaginary parts are constructed in the time variable
and the temperature.
The symbol $\Delta$ is the coordinate Laplace operator 
$\partial^2 {\equiv \mpar\cdot\mpar }$, $m_0$ is 
a particles mass, and $\mu$ is a chemical potential. The local "density-density" 
form of the interaction, whose intensity is given by the coupling constant $\lambda$,  is {a} usual but not essential approximation. All 
necessary integrations in ${\bm x}$ and $\tau$ are implied. We will not discuss 
here the integration contour in the $\tau$ plane \cite{Keldysh} because it is not essential for our analysis. 

Let us introduce {a cutoff parameter} $\Lambda $
dividing the full momentum space $\bm{p}\in \mathbb{R}^d$ into small and
large momentum regions, $|{\bm p}|<\Lambda$ and $|{\bm p}|\ge\Lambda$, respectively. 
By analogy with the Brownian motion, the large scale phenomena are macroscopic in nature and 
form a mean field, just as small scale phenomena {are responsible for}
stochasticity. So one divides  
the initial field variables into the hard and soft components (the momentum representation is assumed)
\begin{align*}
 &\uppsi(\bm{p},\tau) =\upphi +\upxi, \\
 &\upphi=\uppsi(\bm{p},\tau)\theta(\Lambda-|\bm{p}|),\quad 
 \upxi=\uppsi(\bm{p},\tau)\theta(|\bm{p}|-\Lambda)
\end{align*}
and analogous relations for $\uppsi^+$ field.

The {average value of the field} is considered to be an order parameter for the
given dynamic model of critical phenomena \cite{HH77, Vasiliev}. Let us introduce 
the notation $\langle...\rangle$ to describe averaging with respect to the
soft fields with $\exp (-S)$ 
 {serving as a} distribution function. The averaging with respect to the hard fields
 $\upxi$, $\upxi ^+$ is postponed, the 
related terms will be responsible for the random force in the stochastic equations.
 {The} dynamic equation for the soft
 mean fields $\upphi, \upphi ^+$ can be written with the help of the Schwinger-Dyson equation
\begin{align}
  \label{2}
  \left(\partial _\tau+\frac {\Delta}{2m_0}+\mu\right)\langle\upphi^+\rangle 
  &=\uplambda \langle\upphi^+ \upphi \upphi^+ 
  +2\uplambda \langle\upphi^+\rangle\upxi\upxi^+ \nonumber\\
  &+\uplambda 
  \langle\upphi \rangle\upxi^+\upxi^+ +\uplambda \langle\upxi ^+\upxi \upxi ^+\rangle.
\end{align}
An analogous complex conjugated equation with simultaneous change of
the sign at the $\partial_\tau$-term is also satisfied.

We observe that {the term} $\langle\upxi ^+\upxi\upxi ^+\rangle$ 
{takes a form of an} 
additive random force for the soft fields. 
Such systems are successfully described by the functional Legendre
transformation \cite{Vasiljev1}.
Let us introduce $\upalpha, \upalpha ^+$ fields by
$\upalpha \equiv \langle\upphi \rangle$, $\upalpha ^+ \equiv \langle\upphi ^+\rangle$
and let $\Gamma(\upalpha,\upalpha^+)$  be a functional  which is obtained by the
standard Legendre 
transformation of the  generating functional for connected graphs.
In terms of the fields $\upalpha, \upalpha ^+$ Eq. (\ref{2}) takes the form
 \begin{multline}
  (\partial_\tau+{\Delta}/{2m_0}+\mu)\upalpha^+ =\uplambda\upalpha\upalpha^+\upalpha^+ 
  +\uplambda\upxi \upxi^+\upxi ^+    \\
  +\uplambda(2\upalpha^+\upxi\upxi^+ +\upalpha\xi^+\upxi^+
  +2{\cal K}_{1}\upalpha^++ {\cal K}_{2}\upalpha+{\cal L}),
  \label{3}
\end{multline}
{where the following Feynman diagrams appear}
\begin{equation}
  \label{KL}
  {}^{\displaystyle
  {\cal K}_{1}=}\includegraphics[width=6mm]{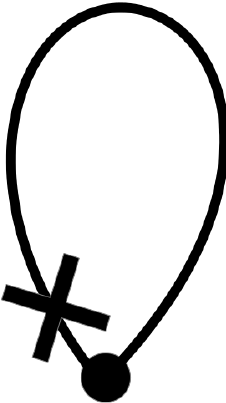}
  {}^{\displaystyle,\quad
  {\cal K}_{2}=}\includegraphics[width=6mm]{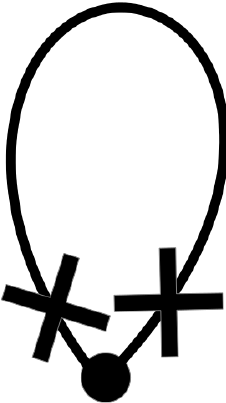} 
  {}^{\displaystyle , \quad \EuScript{L}=}
  \includegraphics[width=14mm]{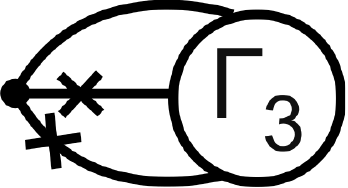}.
\end{equation}

In a sense, Eq.~(\ref{3}) is an extended Gross--Pitaevskii equation with 
a random microscopic force $\uplambda\upxi \upxi^+\upxi ^+ $ and additional
loop terms.
Graphs in Eq.~(\ref{KL}) are constructed from full vertices denoted as 
$\Gamma_3\equiv \delta ^3 \Gamma /(\delta \upalpha)^3$ with 
all possible cross symbols arrangements at the variating $\alpha $ fields. 
The propagator lines are connected graphs of the correlators 
$\langle\upphi\upphi\rangle$, $\langle\upphi\upphi^+\rangle$,
$\langle\upphi^+\upphi\rangle$, $\langle\upphi^+\upphi^+\rangle$ 
with the cross {marked} $\upphi^+$ fields. The {matrix of} propagators is
determined by $(-\Gamma _2)^{-1}$, where $(-\Gamma_2)$ is hermitian matrix with the elements
\begin{align}
  \label{4}
  \nonumber
  -\frac {\delta ^2\Gamma}{\delta \upalpha ^{2}} & = \uplambda
  \upalpha ^{+2} +\uplambda \xi ^{+2}+\mbox {loop terms},\\
  \nonumber
  -\frac {\delta ^2\Gamma}{\delta \upalpha ^+ \delta \upalpha
  }
  & = -\partial _\tau -\frac {\Delta}{2m_0}-\mu +2\uplambda m+\mbox
  {loop terms},
\end{align}
where $ m\equiv \langle\uppsi ^+\uppsi\rangle =\upalpha^+\upalpha +\xi ^+\upxi+{\cal K}_1$.
The {explicit} expressions for the loop contributions can be
found in \cite{Komarova}.

The field $m$ corresponds to a linear combination of internal energy and
density of models E and F \cite{HH77,Folk06}.
Its dynamic equation has obviously the following form:
\begin{equation*}
  \partial _\tau m =\langle\uppsi^+\partial_\tau\uppsi+\uppsi\partial_\tau\uppsi^+\rangle=
  \biggl\langle\uppsi  \frac {\Delta}{2m_0}\uppsi^+\biggl\rangle -
  \biggl\langle \uppsi ^+ 
  \frac {\Delta}{2m_0}\uppsi\biggl\rangle
\end{equation*}
that can be rewritten in terms of $\upalpha ^+$, $\upalpha$ variables as follows:
\begin{align}
\nonumber
  2m_0\partial _\tau m &=\upalpha ^+{\Delta}\upalpha -\upalpha{\Delta}\upalpha ^+ + 
  \includegraphics[width=0.6cm]{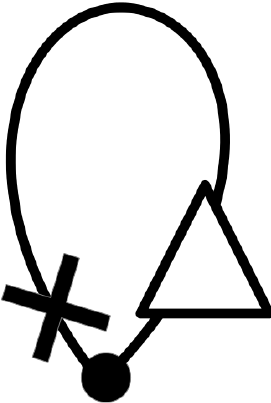}-\includegraphics[width=0.75cm]{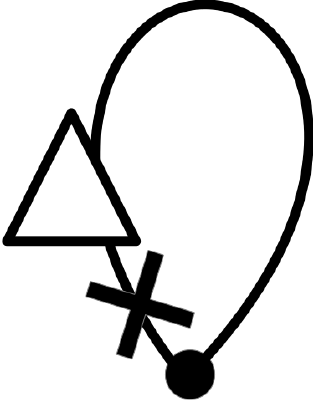}+\\
  &+(\upxi ^+\Delta \upxi-\upxi \Delta \upxi ^+).
  \label{5}
\end{align}
The $\Delta$ symbol in the loop contributions denotes the Laplace operator.
%
%
An analogous equation is fullfilled by the local energy density, i.e., by the
quantity
$\partial_i\psi^+\partial_i\psi$. Moreover due to presence of two derivatives
one can argue that it is in fact less relevant from the RG point than the field $m$.
%
%

Equations (\ref{3}) and (\ref{5}) can be rigorously reduced to the
usual stochastic equations of  model F \cite{Komarova}. To this end, one
needs to consider the loop contributions in a perturbative fashion and expand all
obtained diagrams in fields, external momenta and frequencies as in {the} usual theory of critical 
phenomena. In general, the probability distribution of random forces is not Gaussian.
 Nevertheless, in the
critical region it can be reduced to the white noise. 

In order to {comply} with the standard notation of model F, we
rename the $\upalpha$, $\upalpha^+$
fields as $\psi$, $\psi^+$.\\

{\section{The action and dynamics of model F with hydrodynamic modes activated}
\label{Sec:action} }
A standard way for constructing models of critical dynamics is based on 
the  Poisson bracket construction \cite{HH77,Folk06} using the correspondence
principle. This method can be
used to derive equation of motion for macroscopic observables that have their 
microscopic counterparts. When this is not the case (e.g. entropy) one
must proceed in a different fashion and employ symmetry operations and
related group generators to derive the Poisson brackets. In this work we rely on the latter
approach, whose details are discussed in \cite{Vasiliev}.
 
 In {the}  terminology proposed in \cite{HH77} model F of
critical dynamics
is described 
by the order pa\-ra\-me\-ter of  conjugated fields $\psi({\bm{x}},t)$, $\psi^{+}({\bm{x}},t) $ that 
are averages of the Bose-particle field  operators,  an external magnetic field $h_0({\bm{x}},t)$,
and a field $m({\bm{x}},t)$ 
connected with {temperature fluctuations in} the system. The 
dynamics of all these fields is given by the Langevin equations 
\begin{align}
  \label{eq}
  \nonumber
  \partial_t\psi^{\phantom{+}} & = f_{\psi}^{\phantom{+}}+
  \lambda_0(1+ib_0)\frac{\delta S_F}{\delta \psi^+}+ i\lambda_0 g_{03}
  \psi^{\phantom{+}}\frac{\delta S_F}{\delta m},\\
  \nonumber
  \partial_t\psi^+ & = f_{\psi}^++\lambda_0(1-ib_0)\frac{\delta S_F}{\delta 
  \psi^{\phantom{+}}}-i\lambda_0 g_{03}\psi^+\frac{\delta S_F}{\delta m},\\
  \nonumber
  \partial _tm & = f_m-\lambda_0 u_0\partial ^2 \left(\frac{\delta S_F}{\delta m}\right)
   +i\lambda_0 g_{03} \nonumber\\
   &\times \left(\psi^+\frac{\delta S_F}{\delta \psi^+}-
  \psi\frac{\delta S_F}{\delta \psi}\right).
\end{align}
The static action $S_F$ is defined as
\begin{align*}
  S_F &= \int \mathrm{d}^d {\bm{x}} \int \mathrm{d}t
  \Big(\psi^+\partial^2\psi-\frac{1}{2}m^2+mh_0,\nonumber \\
  &-\frac{1}{4}g_{01}(\psi^+\psi)^2+g_{02}\psi^+\psi m \Big).
\end{align*}
The random forces  $f_{\psi},   f_{m}$ are assumed to be Gaussian random
variables with zero means and correlators $D_{\psi},  D_{m}$ with {the}
white-noise correlations in time. 
Their time-momentum $(t, p\equiv |\mp|)$ representation then reads
\begin{equation}
D_{\psi}(p, t, t')=\lambda_0 \delta(t-t'),\quad
D_{m}(p, t,t')=\lambda_0 u_0 p^2 \delta(t-t').
\label{mnoise}
\end{equation}

The constants $g_{01}, g_{02}$ and $g_{03}$ define the intensity of 
(self)interactions of the order parameter and $m$ field; the parameters
$\lambda_0$ and $u_0$  relate to {the} diffusion coefficient, $b_0$ is an intermode coupling.
All these parameters are marked {with the subscript ``0''} to distinguish them from their 
renormalized counter-partners below.

At  $g_{02}=b_0=0$, the set of {Eqs.} (\ref{eq}) and (\ref{mnoise}) is transformed into
the equations for  model E.  
As stated  by {De}~Dominicis \cite{Dominicis} and A.~N.~Vasil'ev 
\cite{Vasiliev} it probably represents {the} IR stable limit of the initial model F. 

At the transition to the superfluid hydrodynamics, i.e. at the limit  $\nu_0 \to 0$, the 
Reynolds number increases
$\mathrm{Re}\equiv L V/\nu_0\to\infty$, here $\nu_0$ denotes a coefficient of molecular 
viscosity, $L$ is an outer length of turbulence, $V$ is a characteristic (mean) 
velocity. Then, one {necessarily} meets  with {a} phenomenon 
of developed turbulence \cite{Monin,Frisch}. 
Unfortunately, in the above models E and F velocity field and viscosity contributions 
{are not taken into account} because of their IR irrelevance. The corresponding dynamic 
effects are not investigated in the vicinity of phase transition point yet. We will 
return to {the} IR irrelevance discussion with the canonical dimension analysis below. 
 
The stochastic dynamic model with the hydrodynamic modes activated in the vicinity
of the lambda point was proposed in \cite{Krasnov}. Random velocity 
fluctuations around a mean velocity $\bm{V}$ are represented by {the} velocity field
$\mv ({\bm{x}},t)$ which is now taken into consideration and assumed to be
transversal, $\mathrm{div} \mv= 0$. 
%
%
In this paper we restrict our attention to the case of incompressible fluid mainly because
under usual circumstances the fluid velocity is much smaller than the
sound velocity. In such case \cite{Landau_fluid} the fluid is virtually incompressible. Though
it is feasible to include also a longitudinal velocity part 
\cite{ANU97,AHN99} into a model, 
 such problem is much more demanding from the computational point of view. Moreover
 it also brings about other physical effects as sound in turbulent media, shock waves etc. 
 Already for critical dynamics near the liquid-gas transition the elimination of sound
 modes is not completely trivial \cite{AV98}.
 These and related issues are left for future research.
%
%

The equations were derived in accordance with {the} equilibrium static limit 
\cite{Vasiliev} and Galilean invariance and can be written in a compact notation  
\begin{align}
  \nabla_t \varphi_a &= \eta_a+(\alpha_{ab} + \beta_{ab}) \frac{\delta {S}^{st}}{\delta \varphi_b}, 
  \quad \nabla_t \equiv \partial_t+\mv\cdot\mpar,
  \label{eqq}
  \\
  \nonumber
  {S}^{st}& = S_F- \frac{1}{2}\int \mathrm{d}^d{\bm{x}} \int\mathrm{d}t \,\mv^2,
\end{align}
with the set of fields $\varphi_a\in \{\psi, \psi^+, m, \mv\}$ and the set of random forces
$\eta_a\in\{f_{\psi}, f_{\psi^+}, f_m, \mf_v\}$.
The tensor $\alpha$ is a symmetric matrix of Onsager coefficients
and tensor $\beta$ represents an anti-symmetrical matrix of streaming coefficients. According to the definition
(see Section~5.9 in \cite{Vasiliev}),  their
real-space representation reads :
\begin{align}
\nonumber
\alpha_{ab} &= \left( \begin{array}{cccc}
0 & \lambda_0 & 0 & 0 \\
\lambda_0 & 0 & 0 & 0\\
0 & 0 & -\lambda_0 u_0 \partial^2 & 0 \\
0 & 0 & 0 & -\nu_0 \partial^2  \end{array} \right),\\
\label{beta1}
\beta_{ab} &= \left( \begin{array}{cccc}
0 & i \lambda_0 b_0 & i \lambda_0 g_{03} \psi & \psi \partial \\
-i \lambda_0 b_0 & 0 & -i \lambda_0 g_{03} \psi^+ & \psi^+ \partial \\
-i \lambda_0 g_{03} \psi & i \lambda_0 g_{03} \psi^+ & 0 & m \partial \\
- \psi \partial & - \psi^+ \partial & - m\partial & 0  \end{array} \right).
\end{align}
In fact, this is a generalization of the standard Langevin equation (\ref{eq}) due 
to replacing of the partial derivative $\partial_t$ by the Lagrangian
derivative $\nabla_t$. Indeed, 
the terms $-\partial _i(v_i\psi)$, $\partial _i(v_i m)$, $\partial_i(v_i\mv)$ 
are essential to {account for}
Galilean invariance, but they have to enter only as inter-mode coupling contributions  
in $\nabla_t\psi$ and $\nabla_t m$ equations. That is exhibited in the last column 
of the $\beta_{ab}$ matrix (\ref{beta1}). Then, due to  the
antisymmetry condition on $\beta_{ab}$, the terms in
{equation  for} $\nabla_t \mv$ require a {form} corresponding
to the last line in $\beta_{ab}$ \cite{Krasnov, Danco}.

Finally, the dynamic equations (\ref{eqq}) have the following form: for $\psi$ field
\begin{align}
  \label{ps1}
  \partial_t \psi  + { \partial_i (v_i} \psi) &= f_{\psi}+ \lambda_0(1+ib_0)[\partial^2 \psi \nonumber \\  
  &- g_{01}(\psi^+\psi)\psi/3 +  g_{02}m\psi ] \nonumber\\
  &+ i\lambda_0 g_{03} \psi [g_{02} \psi^+ \psi - m + h_0] ,
\end{align}
 for $m$ field
\begin{align}
  \partial_t m + {\partial_i (v_i} m) & =   f_m- \lambda_0 u_0 \partial^2 [g_{02} \psi^+ \psi -
  m + h_0] \nonumber \\
  &+ i \lambda_0 g_{03}[\psi^+ \partial^2 \psi - \psi \partial^2 \psi^+] , 
  \label{mm}
\end{align}
 for {the velocity field} $\mv$ 
\begin{align}
  \partial_t \mv + {\partial_i (v_i} \mv) & = \mf_{v}+\nu_0 \partial^2 \mv \nonumber \\
  &-c\psi^+ \mpar [\partial^2 \psi - g_{01}(\psi^+\psi)\psi/3 + g_{02}m\psi] \nonumber \\
  &- c\psi \mpar [\partial^2 \psi^+ - g_{01}(\psi^+\psi)\psi^+/3 + g_{02}m\psi^+] \nonumber \\
  &- c m\mpar[g_{02} \psi^+ \psi - m + h_0]  
  \label{eq0} 
\end{align}
and equation for $\psi^+$ field is given by complex conjugation of Eq. \ref{ps1}.
The model is extended here by the parameter $c$ in the last equation. It turns 
out to be convenient in the IR analysis below. The substitution $c=1$ corresponds 
to the original stochastic problem (\ref{eqq}).

The noise correlator of the force $\mf_{v}$ can be expressed in the form
\begin{equation}
D_{v}(p, t, t')=g_{04}\nu_0 ^3p^{\epsilon -\delta} \delta(t-t')
\label{vnoise}
\end{equation}
in the space with dimension  $d=4-\epsilon$. The additional exponent 
$\delta$ {allows for} a deviation from {the}
Kolmogorov turbulent regime \cite{Frisch,pismak}.

In fact, there are two physically pos\-sib\-le and interested regimes. 
The first one is the regime with hydrodynamic fluctuations near thermodynamic equi\-lib\-rium 
that corresponds to the values $\epsilon =1$, $\delta =-1$, $g_{04}=1/\nu_0^2$.
The second one is the Kolmogorov turbulent regime with $\epsilon =1$, $\delta = 4$.
In this case the noise  (\ref{vnoise}) imitates the energy injection to the system from 
{a}
range of the largest eddies \cite{Reynolds, Kolmogorov1,Kolmogorov2}, 
the constant $g_{04}$ {can be interpreted as an} energy dissipation rate per unit
mass (see, e.g. \cite{pismak}) and can be measured experimentally. The
most advanced approach to the study of developed turbulence is {an} 
investigation of its universal characteristics in the inertial interval, that
is an intermediate interval of wave numbers  (see, e.g. \cite{Monin}). 

Let us discuss the IR irrelevance of hydrodynamic modes (see \cite{HH77,Dominicis}). 
{The} stochastic problem described by the set of equations (\ref{eqq}) 
with $\eta$-force noise (\ref{mnoise}), (\ref{vnoise})
$$\left\langle \eta_a\eta_b\right\rangle=D_{ab}
,\qquad 
D_{ab}=\left( \begin{array}{cccc}
0             & \lambda_0  & 0                      & 0 \\
\lambda_0     & 0          & 0                      & 0 \\
0             & 0          & -\lambda u_0\partial^2 & 0 \\
0             & 0          & 0                      & D_v
\end{array} \right)$$
can be transformed into the field theoretic model by 
{the} means of the Martin-Siggia-Rose (MSR) mechanism \cite{MSR} with the Dominicis-Janssen action
\begin{equation}
\label{DJ}
S=\varphi'_a D_{ab}\varphi'_b +\varphi'_a\left(-\nabla_t\varphi_a+(\alpha_{ab}+\beta_{ab})\frac{\delta {S}^{st}}{\delta \varphi_b}
\right) \, ,
\end{equation}
where each $\varphi_a$ field gets a {complementary} 
field $\varphi'_a \in\{{\psi ^+}', \psi', m', \mv'\}$ and
the proper equation of the system (\ref{eqq}) stands in large brackets. 
Auxiliary $\varphi'_a$ fields  in formula (\ref{DJ}) appear in the MSR transformation procedure as
in the usual case of stochastic dynamic models.
These new fields are interpreted as response field variables \cite{Jan76,Dom76}. 
Thus, the constructed action is Gaussian with respect to $\varphi'_a$ fields, the first 
term in (\ref{DJ}) describes the contributions of all stochastic noises. All necessary 
integrations over chosen variables
(space and time, wave vectors and frequency or combined cases) and sum over vector 
indices are implicitly assumed.

The standard calculation \cite{Vasiliev,Zinn} of canonical dimensions for all fields and pa\-ra\-me\-ters
of the action $S$
is the most straightforward way to analyze IR irrelevance. The canonical dimensions of 
the model are presented in Table~\ref{QQQ}.
The momentum dimension $d^p$ and the frequency one $d^{\omega}$ can be determined 
in\-de\-pen\-dent\-ly. The total dimension $d=d^p + 2 d^{\omega}$ is determined due to 
the dispersion relation $i\omega \sim p^2$ between frequency $\omega$ and {the} 
momentum vector $\mp$ .

\begin{table}[h]
\begin{center}
\begin{tabular}{||c||c|c|c|c|c|c|c||}
  \hline \hspace{2.3mm}
	&&&&&&\\[-3mm]
$F$ & $p$, $1/x$ & $\omega$, $1/t$ & $\psi, \psi^{+}$ & $\psi',\psi^{+'}$ & $m, m'$ & $v$ & $v'$
  \\[2mm] \hline \hline 
	&&&&&&&\\[-3mm]
$d_F^p$ &1&0& $\frac{d}{2} - 1$ & $\frac{d}{2} + 1$ & $\frac{d}{2}$ & $-1$ & $d+1$ 
  \\ [2mm]\hline 
	&&&&&&&\\[-3mm]
$d_F^\omega$&0 &1& $0$ & $0$ & $0$ & $1$ & $-1$ 
	\\[2mm]  \hline
	&&&&&&&\\[-3mm]
$d_F$ & 1&2&$\frac{d}{2} - 1$ & $\frac{d}{2} + 1$ & $\frac{d}{2}$ & $1$ & $d-1$ 
	\\[2mm]  \hline   \hline
	&&&&&&&\\[-3mm]
$F$ &$\lambda_0, \nu_0$ & $u_0, b_0$ & $g_{01}$ & $g_{02}, g_{03}$ & $g_{04}$&$c$&
  \\[2mm] \hline \hline 
	&&&&&&&\\[-3mm]
$d_F^p$& $-2$ & $0$ & $4-d$ & $\frac{4-d}{2}$ & $\delta$&$-(2+d)$ &
  \\ [2mm]\hline 
	&&&&&&&\\[-3mm]
$d_F^\omega$&$1$& $0$ & $0$ & $0$ & $0$&$2$ &
	\\[2mm]  \hline
	&&&&&&&\\[-3mm]
$d_F$ &$0$& $0$ & $4-d$ & $\frac{4-d}{2}$ & $\delta$ &$2-d$ &
	\\[2mm]  \hline
\end{tabular}
\caption{Canonical dimensions of the fields and parameters for
model F~with hydrodynamic modes activated.}
\label{QQQ}
\end{center}
\end{table}

The first term of action (\ref{DJ}) at $\varphi'_a=\mv'$ corresponds
to the $\mv'D_v\mv'$ contribution
and represents  the influence of random velocity fluctuations on the critical
behavior of the system. It is proportional to the constant $g_{04}$ that
stands in the noise $D_v$ (\ref{vnoise}) and mimics stochasticity of the velocity field.
In the case of the thermal equilibrium regime, as $\delta=-1$, the canonical 
dimension of $g_{04}$ is equal to $-1$ (see Table~\ref{QQQ}).
Then the action term discussed is IR irrelevant. In other words, the random 
velocity fluctuations do not {affect} large-scale (infrared)  asymptotics 
of all physically relevant and experimentally measurable quantities.
Moreover, for all $d>2$ the ca\-no\-ni\-cal dimension of the parameter $c$ is negative.
Therefore, all corresponding terms of the action are irrelevant and again can be omitted. 
{As opposed to the situation in Eq.} (\ref{eq0}), the IR relevant 
dynamics of the $\mv$ field corresponds
{in this case} to the Navier-Stokes equation
\begin{align}
\nonumber
\partial_t \mv + \partial_i (v_i \mv)  = \nu_0 \partial^2 \mv .
\end{align}
As a result, the velocity field $\mv$ is not stochastic {anymore},  its
role in action (\ref{DJ}) is reduced to the role of an external field. It does 
not affect the critical behavior of the model in full accordance with \cite{HH77, Dominicis}.
 
To keep in {play} the stochastic hydrodynamic fluctuations, we 
propose a scenario similar to
the developed turbulence \cite{Nalimov,Hnatic1,Ronis,Jurcisin} and
chemical reaction kinetics \cite{Honkonen1}.
The {remedy} proposed in \cite{Nalimov,Ronis} is a construction of a double
expansion in two {small expansion} parameters $\epsilon$ and $\delta$. 
{In other words it constitutes}
 a generalization of {the} famous Wilson $\epsilon-$expansion \cite{Kogut,Vasiliev,Zinn}. 
{For the logarithmic theory} $\epsilon=\delta=0$, $d=4$, all terms of action (\ref{DJ})
proportional to $c$ can be omitted as the related canonical dimensions are negative. 
Hence the IR relevant dynamics of the $\mv$ field is the following:
\begin{align}
\partial_t \mv + \partial_i (v_i \mv)  = \nu_0 \partial^2 \mv +f_v,
\label{eq1}
\end{align}
the dynamic equations for the basic fields $\psi$, $\psi^+$, $m$ (\ref{ps1}), (\ref{mm}) 
are not changed in {the} IR limit.
The  {remaining} coupling constants are dimensionless simultaneously at the starting point;
the model is logarithmic.

In the following steps one calculates perturbative expansion in $\epsilon$ and $\delta$ powers. 

Note that the viscosity as well as the stochasticity enter the equations
{in a nontrivial fashion}.
So the scaling behavior of $\nu_0$ can be analyzed {for both} thermal equilibrium regime 
and Kolmogorov {regime, respectively}. Let us stress that it is not a
turbulence of superfluids but a developed 
turbulence of random medium with the fields $\psi$, $\psi^+$ and $m$ emerging in the
vicinity of phase transition point as a passive admixture.
{\section {Renormalization and renormalization constants}
\label{Se:RG}}
{Let us refer to model F with activated} hydrodynamic modes as model F${}_{h}$.
{Its field-theoretic action reads}
\begin{align}
\label{SF}
\nonumber
S &=2\lambda_0  {\psi ^+}'\psi ' -\lambda_0 u_0 m'\partial ^2m' +\mv' D_v\mv' \\
\nonumber
&+{\psi ^+}'\{-\partial_t\psi -\partial_i(v_i\psi )+\\
\nonumber
&+\lambda_0 (1+ib_0)[\partial^2\psi - {g_{01}}(\psi ^+\psi )\psi /3  +g_{02}m\psi ]\\
\nonumber
&+ i\lambda_0  \psi [g_{07}\psi ^+\psi -g_{03}m+g_{03}h_0] \}\\ 
\nonumber
& +\psi ' \{-\partial_t \psi ^+ -\partial_i(v_i\psi ^+ ) \\
\nonumber
&+\lambda_0 (1-ib_0)[ \partial ^2\psi^+ - g_{01}(\psi ^+\psi)\psi^+/3+g_{02}m\psi ^+]\\
\nonumber
&- i\lambda_0  \psi ^+[g_{07}\psi ^+\psi -g_{03}m+g_{03}h_0]  \}\\
\nonumber
&+m'\{-\partial _t m-\partial_i(v_i m) -\lambda_0 u_0\partial ^2[-m+g_{06}\psi ^+\psi\\
\nonumber
&+h_0]+i\lambda g_{05}[\psi^+\partial^2\psi-\psi\partial^2\psi^+]\} \\
&+\mv'\{-\partial _t\mv+\nu_0 \Delta \mv-\partial_i(v_i\mv)\},
\end{align}
{where the following relations between charges (coupling constants) are fulfilled}:
\begin{align}
\label{dog}
 g_{05}&=g_{03},\quad g_{06}=g_{02},\quad g_{07}=g_{02}g_{03}. 
\end{align}
In the framework of {the double} $(\epsilon,\delta)$
expansion the logarithmic theory {does not have a static limit which} changes 
the renormalization scheme.We stress  that the introduction of the new coupling constants 
$g_{05}$, $g_{06}$, $g_{07}$ restores the multiplicative renormalizability of the model.

The {following} notes are essential for {the} renormalization procedure.

The fields $\psi$, $\psi ^+$, $m $ can be considered as passive scalars for the $\mv$ field because graphs with external $\mv$, $\mv'$ lines do not include internal lines of other fields. 

Renormalization constants    of the terms $\varphi_a' {\partial_i(v_i }\varphi_a^+)$ 
and  $\varphi_a'\partial _t \varphi_a^+ $,  for which the generic notation $Z$ is used below,  are the same for
$\varphi_a=\psi,\psi^+,m$ due to the Galilean invariance. There are no 
counterterms of the corresponding terms at
$\varphi_a=\mv$, as in the usual developed turbulence theory.
The nonlocal counterterms of $\mv'D_v\mv'$ type are absent.
The renormalization constants of the $\psi , \psi ^+$ fields can be chosen
 {to be} real due to the symmetry
of the action with respect to a transformation $\{\psi , \psi '\}\to \mathrm{e}^{ir}\{\psi , \psi '\}$,
$\{\psi ^+ , \psi '^+\}\to \mathrm{e}^{-ir}\{\psi ^+, \psi '^+\}$, similarly to 
 the static theory \cite{Vasiliev}.

The field theoretic action of the theory (\ref{SF}) obviously has to be real. Introduction of the independent 
coupling constants $g_{03}$, $g_{06}$, $g_{07}$ allows us to make a model 
multiplicatively renormalizable. All renormalization constants for parameters and fields of the mode are real except $Z_{\psi '}, Z_{\psi '^+}$.
{It is convenient to express} the constant $\nu_0$ as $\nu_0=u_{01}\lambda_0$ 
{via a} new
dimensionless coupling constant $u_{01}$.
The relations between bare and renormalized constants then read
\begin{align}
  & \lambda_0 = \lambda Z_\lambda, & &\nu_0 = \nu Z_\nu, \nonumber \\
  &  u_0 = u Z_u, & &b_0 = b Z_b, \nonumber \\
  & g_{01} = g_1 \mu^\epsilon Z_{g_1}, & &g_{02} = g_2 \mu^{\epsilon/2} Z_{g_2}, \nonumber \\
  & g_{03} = g_3 \mu^{\epsilon/2} Z_{g_3}, & &g_{04} = g_4 \mu^\delta Z_{g_{4}}.
  \label{eq:ren_param}
\end{align}
Renormalization of the fields is achieved through the replacements
\begin{align}
   & \psi \rightarrow \psi Z_\psi, & &\psi_+ \rightarrow \psi_+ Z_{\psi^+},\nonumber \\
   & \psi' \rightarrow \psi' Z_{\psi'}, & &{\psi^+}' \rightarrow {\psi^+}' Z_{{\psi^+}'},\nonumber \\
   & m \rightarrow m Z_m, & &m' \rightarrow m' Z_{m'}, \nonumber \\
   & v \rightarrow v Z_v, & &v' \rightarrow v' Z_{v'}.
   \label{eq:ren_fields}
\end{align}
The renormalization constants $Z$ are calculated in {the} MS scheme \cite{Zinn}. In this scheme 
they can be represented in the form $Z=1+[Z]$ where $[Z]$ denotes 
the pole part  in arbitrary linear combinations of the parameters $\epsilon$ and $\delta$.
In this notation the {computation} of the following counterterms:
\begin{align}
\nonumber
  &2\lambda [Z_1] {\psi ^+}'\psi ', &-& \lambda u [Z_2] m'\partial ^2m', &-&[Z_3] {\psi^+}'\partial_t\psi , \\
  \nonumber
  &+\lambda [Z_4]  {\psi^+}'\partial^2\psi ,&-&\lambda [Z_5] {\psi^+}'(\psi^+\psi)\psi/3, 
  &\lambda& [Z_6] {\psi^+}'m\psi , \\ 
  \nonumber 
  &-[Z_3]^*\psi ' \partial_t\psi^+ , &-&\lambda [Z_5]^*\psi '(\psi ^+\psi)\psi^+/3 , 
  &\lambda&  [Z_4]^*\psi '\partial^2\psi^+,  \\
  \nonumber 
  &\lambda [Z_6]^*\psi 'm\psi ^+ ,&\lambda& [Z_7] m' \partial ^2m,&\lambda&  m'\psi ^+[Z_8]\psi, \\
  &\mv'\nu [Z_9]\Delta v. &\mbox{}& &\mbox{}&
\label{SFR}
\end{align}
is necessary, where we have introduced counterterms $[Z_i]$  (index $i$ in the following
will always run from $0,1,\ldots,9$).
Further by comparing expressions (\ref{SF}) and (\ref{SFR}) the relations
between $[Z_i]$ and the renormalization constants
of the model  (\ref{eq:ren_param}), (\ref{eq:ren_fields}) can be derived in a straightforward fashion
\begin{align}
\nonumber
&Z_{\lambda}Z_{\psi +'}Z_{\psi '}=1+[Z_1],\quad Z_{\lambda}Z_uZ_{m'}^2=1+[Z_2],\\
&Z_{\psi +'}Z_{\psi }=1+[Z_3],\nonumber \\
&Z_{\lambda}Z_{\psi +'}Z_{\psi }(1+ibZ_b)=(1+ib)+[Z_4],\nonumber \\
\nonumber
&Z_{\lambda}Z_{\psi +'}Z_{\psi }((1+ibZ_b)g_1Z_{g_1}/3-i g_7 Z_{g_7})\\
\nonumber
&=(1+ib)g_1/3-ig_7+[Z_5]/3,\\
\nonumber
&Z_{\lambda}Z_{m}Z_{\psi +'}Z_{\psi }((1+ibZ_b)g_2Z_{g_2}-ig_3Z_{g_3})\\
\nonumber
&=(1+ib)g_2-ig_3+[Z_6],\\
\nonumber
&Z_{\lambda}Z_uZ_{m'}Z_m u=u+[Z_7],\quad Z_{m'}Z_m=1,\\
\nonumber
&Z_{\lambda}Z_{\psi +'}Z_{\psi }Z_{m'}(uZ_u(\mp+\mq)^2g_6Z_{g_6}-i({{q}}^2-{{p}}^2)g_5Z_{g_5})
\\
\nonumber
&=u(\mp+\mq)^2g_6-i({{q}}^2-{{p}}^2)g_5+[Z_8] ,\\
&Z_{\nu}=1+[Z_9],\quad Z_{\nu}=Z_{\lambda}Z_{u_1},\quad Z_{g_4}Z_{\nu}^3=1.
\label{Z}
\end{align}

In contrast to the renormalization constants for
parameters (\ref{eq:ren_param}) and fields (\ref{eq:ren_fields}) the counterterms $[Z_i]$ can contain both real and imaginary parts.
Moreover, the counterterm $[Z_8]$ depends in a nontrivial way on external momenta ${\mp}$  
and ${\mq}$ that are carried
by the fields $\psi(\mp)$ and $\psi^+(\mq)$, respectively. The {notation}
$p=|\mp|$, $q=|\mq|$  is assumed here.

The Feynman diagrammatic technique is based on the interaction vertices connected
by lines (propagators).
The propagators of the model have the following form:
\begin{align*}
  & \Delta_{mm}=\frac{2\lambda uk^2}{\omega^2+\lambda^2 u^2k^4},\quad
  \Delta_{m'm}=\frac{1}{i\omega+\lambda uk^2}=\Delta_{mm'}^*, \\
  & \Delta^{ij}_{vv}=\frac{g_4\nu^3 k^{\epsilon-\delta}P_{ij}^k}{\omega^2+\nu^2k^4},\qquad
  \Delta_{v'v}^{ij}=\frac{P_{ij}^k}{i\omega+\nu k^2}={\Delta_{vv'}^{ij}}^* , \\
  & \Delta_{\psi' \psi^+} =\frac{1}{i\omega+\lambda(1-ib)k^2} = {\Delta^*}_{\psi{\psi^+}'}, 
  \\
  & \Delta_{\psi^+\psi'} = \frac{1}{-i\omega+\lambda(1-ib) k^2} = \Delta_{{\psi^+}' \psi}^*,  
  \\
  & \Delta_{\psi\psi^+}=\frac{2\lambda}{(\omega-i\lambda k^2(1-ib))(\omega+i\lambda k^2(1+ib))} 
  =\Delta_{\psi^+\psi}^* .
\end{align*}
The interaction vertices \cite{Vasiliev} correspond to the vertex factors $V_{{\psi^+}' \psi v}$, $V_{{\psi^+}' \psi^+ \psi \psi}$,
$V_{{\psi^+}' m \psi}$,  $V_{m' m v}$, $V_{m' \psi^+ \psi}$, $V_{v' v v }$ 
plus their complex conjugates. Their explicit form can be easily obtained from the action (\ref{SF}).

A {prescription for the} multiplicative renormalization is now determined
by expressions (\ref{Z}). To the one-loop approximation  we  obtain the following relations:
\begin{align}
  \nonumber
  [Z_{\lambda}] &= \mbox{Re}([Z_4]-[Z_3](1+ib)),\\
  \nonumber
  [Z_b] &= [Z_4]-[Z_3](1+ib)-(1+ib)[Z_{\lambda}]/(ib),\\
  \nonumber
  [Z_{\psi'}] &= ([Z_1]-[Z_{\lambda}])/2-i \mbox{Im}([Z_3]),\\
  \nonumber
  [Z_{\psi}] &= [Z_3]-[Z_{\psi'}]^*,\\
  [Z_u] &= [Z_7]-[Z_{\lambda}],  \nonumber \\
  \nonumber
  [Z_{m'}] &= -[Z_m]=([Z_2]-[Z_{\lambda}]-[Z_u])/2,\\
  \nonumber
  g_1[Z_{g_1}] &=3\mbox{Re}{\biggl(}[Z_5]-([Z_{\lambda}]+2[Z_{\psi}]+[Z_3])\\
  \nonumber
  &\times((1+ib)g_1/3-ig_7)-ib[Z_b]g_1/3 {\biggl)},\\
  \nonumber
  g_7[Z_{g_7}] &=-\mbox{Im}{\biggl(}[Z_5]-([Z_{\lambda}]+2[Z_{\psi}]+[Z_3])\\
  \nonumber
  &\times \{(1+ib)g_1/3-ig_7)-ib[Z_b]g_1/3\\
  &-ib[Z_{g_1}]g_1/3\} {\biggl)},\nonumber \\
  \nonumber
  g_2[Z_{g_2}] &=\mbox{Re}\textbf{(}[Z_6]-((1+ib)g_2-ig_3)([Z_{\lambda}]+\\
  \nonumber
  &+[Z_m]+[Z_3])\textbf{)},  \nonumber \\
  [Z_{u_1}]&=[Z_{\nu}]-[Z_{\lambda}]  \nonumber \\
  \nonumber
  g_3[Z_{g_3}] &=-\mbox{Im}([Z_6]-((1+ib)g_2-ig_3)([Z_{\lambda}]+\\
  \nonumber
  &+[Z_m]+[Z_3])-ibg_2[Z_{g_2}]-ibg_2[Z_b])),\\
  \nonumber
  ug_6[Z_{g_6}] &=g_6u(2[Z_{\psi}]-[Z_{\lambda} ]-[Z_{m'}]-[Z_u])-\\
  \nonumber
  &-[Z_8]/4/q^2|_{p=q}, \\ 
  [Z_{g_4}] &=-3[Z_{\nu}]  \nonumber\\
  \nonumber
  g_5[Z_{g_5}] &=-g_5([Z_{\lambda}]+[Z_{m'}]+2[Z_{\psi}])- \nonumber\\
  &-i\partial _p[Z_8]|_{q=-p}/(2p),   \nonumber\\
  [Z_{\nu}] &=[Z_9].  
  \label{res}
\end{align} 
  The contributions of all graphs to the $Z_i$ 
constants are collected in Appendix. Graphs 1 and 2 contribute to
$Z_1$. Graphs 3 and 4 contribute to $Z_2$. Expression for $Z_3$ is a sum of 
contributions of  5-7 marked with symbol $\omega$. The contributions 
of the graphs \textnumero~5--7
to  $Z_4$ are marked by $p^2$; $Z_5$ is a sum of 
contributions of \textnumero~8--29; $Z_6$ corresponds
to \textnumero~30-34, $Z_7$ -- \textnumero~35-40; $Z_8$ -- \textnumero~41-47; $Z_9$ -- \textnumero~48, 49.
\smallskip

{\section{Fixed points}
\label{Sec:fixed} }
The anomalous dimensions $\gamma$ of {the} renormalization group 
equation ($\gamma$ functions henceforth) are defined as follows \cite{Vasiliev,Zinn}:
\begin{equation}
\gamma _{\alpha} =-\sum _i e_{g_i}\partial_{g_i}[Z_{\alpha}],\qquad i=1,2,\ldots7,
\end{equation}
where  $\alpha \in\{\psi, \psi', m, m', v', v',g_i, u, u_1, \lambda, \nu, b\}$.

It is appropriate to rescale the coupling constants 
\begin{align}
\nonumber
g_i/(8\pi ^2) &\to g_i \quad \mbox{{if} }i=1,4,7;\\
 g_i /\sqrt{8\pi^2} &\to g_i \quad \mbox{{if} }i=2,3,5,6.
\label{res_char}
\end{align}
Unfortunately, the one-loop approximation for renormalization constants as well
as for RG functions yields expressions that are too large to be published 
in this paper. Even the truncated system for model E${}_h$ 
(model F${}_h$ at $b=g_2=g_6=g_7=0$) published in \cite{Danco}
yields very cumbersome renormalization constants. The corresponding anomalous dimensions
are then 
\begin{align*}
  \gamma_{\lambda} & = \frac{3g_4 u_1^2}{8(1+u_1)} + \frac{g_3^2}{(1+u)^3 } + 
       \frac{g_3 g_5 u (2+u)}{(1+u)^3}, \\
  \gamma_u &= - \frac{g_3^2}{(1+u)^3} - \frac{g_3 g_5 (u^3 + u^2 - 3u -1)}{2u(1+u)^3}\\
    	   &+ \frac{3g_4 u_1^2 (1+u_1 - uu_1 - u^2)}{8u(1+u_1)(u+u_1)} ,\\
  \gamma_{g_3} & = - \frac{3g_4 u_1^2}{8(1+u_1)} - \frac{g_3^2}{(1+u)^3} + \frac{g_5^2}{4u} \\
  &- \frac{g_3 g_5 (1 + 3u + 11u^2 + 5u^3)}{4u(1+u)^3}, \\
  \gamma_{g_5} &= - \frac{3g_4 u_1^2(1+2u+2u_1)}{8(1+u_1)(u+u_1)}+\frac{g_3^2(2+9u+3u^2)}{2(1+u)^3}\\ 
       & - 
         \frac{g_3 g_5 (5u + 23u^2 + 9u^3 - 1)}{4u(1+u)^3} - \frac{g_5^2}{4u},\\
   \gamma_{g_1} &= - \frac{3g_4 u_1^2}{4(1+u_1)} - \frac{5 g_1}{3} - 
   \frac{6 g_3^2 g_5 (g_3 - g_5)}{u g_1 (1+u)} \\
   &+ \frac{2 g_3 (1+3u+u^2)(g_3 - g_5)}{(1+u)^3},\\
  \gamma_{u_1} &= - \frac{g_3^2}{(1+u)^3} - \frac{g_3 g_5 u (2+u)}{(1+u)^3} +
   \frac{g_4(1+u_1 - 3u_1^2)}{8 (1+u_1)}, \\
  \gamma_{m} &= \frac{g_3 g_5}{4u} - \frac{g_5^2}{4u}, 
  \qquad \qquad \gamma_{m'} = - \frac{g_3 g_5}{4u} + \frac{g_5^2}{4u} ,\\
  \gamma_{\psi} &= \gamma_{\psi^{+}} = \frac{3 g_4 u_1^2}{16 (1+u_1)} - 
  \frac{g_3 (g_3 -g_5)(2+4u+u^2)}{2(1+u)^3 } ,\\
  \gamma_{\psi'} &= \gamma_{\psi^{+'}} = - \frac{3 g_4 u_1^2}{16 (1+u_1)} +
  \frac{g_3(g_3-g_5)u(2+u)}{2(1+u)^3 } ,\\
  \gamma_{\nu} &= \frac{g_4}{8} ,\qquad \qquad \gamma_{g_4} = \frac{3g_4}{8}.
\end{align*}
A misprint in \cite{Danco} {is corrected} here.


As for RG -functions there are ten $\beta$-functions 
\begin{equation*}
\beta _{\kappa}=\kappa(-e_{\kappa} -\gamma _{\kappa}),\qquad \kappa \in\{g_i, u,u_1\},
\end{equation*}
where the canonical dimensions $e_{g_1}=e_{g_7}=\epsilon$, 
$e_{g_2}=e_{g_3}=e_{g_5}=e_{g_6}=\epsilon /2$,
$e_{g_4}=\delta$, $e_u=e_{u_1}=0$ correspond to the {dimensions} in Table~\ref{QQQ} and
formula (\ref{dog}).
The system of equations 
\begin{equation}
\label{fps}
\beta_\kappa=0,\qquad \kappa \in\{g_i, u,u_1\}
\end{equation}
has about $2^{10}$ different solutions; in principle each of them corresponds to a fixed point.
The stability of a point is determined by the set of eigenvalues $\upomega$ for the 
first derivative matrix $\Omega = \left\{\Omega_{ik} = \partial \beta_i / \partial g_k \right\}$, 
here
$\beta$ is the full set of $\beta_i$ functions and $g$ is the full  set of 
charges, $i,k\in \{g_i, u,u_1\}$. The IR asymptotic behavior is governed by the 
IR stable fixed points with a positive-{definite} $\Omega$ matrix.

It is important that the stability analysis yields different results as one takes into
consideration a different number of {the} perturbation order.
For example, in the standard model F  it was shown \cite{Dominicis, Vasiliev} that the
one-loop results did not lead to the correct IR fixed point.

The majority of the fixed points can be found only by the numerical calculations.
Some {fraction} of them can be immediately discarded, because they fall out of the region 
with admissible values for physical parameters. The calculation of {a} full 
solution for the system (\ref{fps}) has no sense because of 
 {the} stability problem discussed 
below. This is why
we have attempted to investigate the system specifically in the  different regimes, {rather 
than solving} it. 
Furthermore, we reduce the model with respect to Table~\ref{mod} in order to discuss
the relationship between different models of critical dynamics.

\begin{table}[h!]
\centering
\renewcommand{\arraystretch}{1.75}
\begin{tabular}{|c|c|c|}
  \hline
 Standard model F  & Model E${}_h$   & Standard model E   \\ \hline
 $g_4=0$ & $g_2=g_6=g_7=0$ & $g_2=g_4=g_6=g_7=0$ \\
$u_1=0$ & $b=0$ & $u_1=b=0$, $g_3=g_5$  \\
\hline
\end{tabular}
  \caption{Relationship between different models of critical dynamic.}
  \label{mod}
\end{table}
%
%
 Regarding searching for solutions of RG equations we would like to make the following
 comment. From the numerical point of view it is easier to look for IR stable 
 points, because solutions of flow equations (given by Gell-Mann-L\"ow equations)
 directly flow into the stable fixed points. However, it is practically impossible
 to determine unstable regimes in this way. Obviously, analytical
 solutions are of decisive importance.
%
%

{\subsection {Turbulent scaling regime, $\epsilon =1$, $\delta =4$}
\label{Subsec:turbulent}}
In this regime the numerical analysis reveals an IR stable fixed point 
\begin{align}
 & g_{4*}=10.(6), \quad u_*=1,\quad u_{1*}=0.7675919, \nonumber \\
 & b_*=g_{1*}=g_{2*}=g_{3*}=g_{5*}=g_{6*}=g_{7*}=0.
  \label{nontr}
\end{align}
The one loop approximation at this fixed point gives the following anomalous 
dimensions $\gamma_i$ and eigenvalues of the $\Omega$ matrix:
\begin{align}
\label{43}
  & \gamma_{\nu}=\gamma_{\lambda}=1.(3),\quad
  \gamma_{\psi}=-\gamma_{\psi'}=1.(3),\quad
  \gamma_{m}=\gamma_{m'}=0, \nonumber \\  
  & \upomega=\{2.087,\; 1.666,\; 0.833,\; 4,\; 2.921\}.
\end{align}

As was mentioned above, the many-loop calculations could change the stability of
the fixed points. The fixed points of model E${}_h$ turn out to be unstable in the context of model F${}_h$, but
this instability could appear at the one-loop
approximation only. Then let us include the fixed points of E${}_h$ model into 
consideration and overview them. 

The fixed points of model E${}_h$ were published in \cite{Danco}; the stable fixed 
points are listed in Tab.~\ref{tab:fp_stab}, the unstable ones in Tab.~\ref{tab:fp_unstab}.

\begin{table}[h!]
\centering
\renewcommand{\arraystretch}{1.75}
\begin{tabular}{|c|c|c|c|c|}
  \hline
  FP           & FP1     & FP2         &     FP3 & FP4 \\ \hline
  $g_1$        & $0$     & $0$         &    $\frac{3}{5}\varepsilon$  & $\frac{3}{5}\varepsilon$ \\ \hline
  $g_3$        & $0$     & $0$         &    $\varepsilon^{1/2}$   & $\varepsilon^{1/2}$ \\ \hline
  ${g_5}$ & $0$     & $0$         &    ${\varepsilon^{1/2}}$ 
   & ${\varepsilon^{1/2}}$  \\ \hline
  $g_4 $  & $0$     & $\frac{8\delta}{3}$ & $0$ & $\frac{8\delta}{3}$ \\ \hline
  $u$          & $0$     & $1$         & $1$  & $1$ \\ \hline  
  $u_1$        & $0$     & $\frac{1+\sqrt{13}}{6}$  & $0$ 
  & $0$  \\ \hline
\end{tabular}
  \caption{Stable fixed points for model E${}_h$.}
  \label{tab:fp_stab}
\end{table}

\begin{table}
\centering
\renewcommand{\arraystretch}{1.75}
\begin{tabular}{|c|c|c|c|c|c|}
  \hline
  FP           & FP5     & FP6         &     FP7 & FP8 & FP9\\ \hline
  $g_1$        & $0$     & $\frac{3\epsilon-2\delta}{5}$         &    $\frac{3\epsilon-2\delta}{5}$  & $0$ & $0$\\ \hline
  $g_3$        & $0$     & $0$         &    $0$   & $\varepsilon^{1/2}$ & $\varepsilon^{1/2}$\\ \hline
  ${g_5}$ & $\frac{\sqrt{2(-19+\sqrt{13})\delta + 18\varepsilon}}{3}$     & $\frac{\sqrt{2(-19+\sqrt{13})\delta + 18\varepsilon}}{3}$         &    $0$ 
   & ${\varepsilon^{1/2}}$ & ${\varepsilon^{1/2}}$ \\ \hline
  $g_4 $  & $\frac{8\delta}{3}$     & $\frac{8\delta}{3}$ & $\frac{8\delta}{3}$ & $0$ & $\frac{8\delta}{3}$\\ \hline
  $u$          & $1$     & $1$         & $1$  & $1$ & $1$\\ \hline  
  $u_1$        & $\frac{1+\sqrt{13}}{6}$     & $\frac{1+\sqrt{13}}{6}$  & $\frac{1+\sqrt{13}}{6}$ 
  & $0$ & $0$ \\ \hline
\end{tabular}
  \caption{Unstable fixed points for E${}_h$ model.}
  \label{tab:fp_unstab}
\end{table}

The detailed analysis of these points can be found in \cite{Danco} as well as the 
discussion about the IR stabilizing influence of the $g_1\psi^{+'} \psi^{+} \psi^2$ term and destabilizing contributions of the velocity
fluctuations. 

The charges $u$ and $u_1$ are not expansion parameters. 
In fact they denote so-called non-perturbative charges. 
As shown in previous works \cite{Dominicis,Ant00} from the RG perspective such parameters can 
 also acquire infinite values  in the fixed point and
there are no inconsistencies within perturbation theory. From a physical point of view
it is actually necessary to consider also a limiting case as $u\rightarrow\infty$ or 
$u_1\rightarrow\infty$ because
such regimes can be possible candidates for a stable point of model E.
{Therefore,} it seems reasonable to 
consider specific limits as their values tend to infinity. It yields additional 
fixed points. In the case {of} $u \to \infty$, the fixed points obtained are collected 
in Table~\ref{tab:fpu}. The case $u_1 \to \infty$ (case II) are presented in Table \ref{tab:fpu1}. The case when both charges $u$ and $u_1$ tend to infinity 
simultaneously can be found in Table \ref{tab:fpuu1}. 

\begin{table}
\centering
\renewcommand{\arraystretch}{1.75}
\begin{tabular}{|c|c|c|c|c|c|}
  \hline
  FP           & FP1$^{I}$     & FP2$^{I}$  &     FP3$^{I}$ & FP4$^{I}$ & FP5$^{I}$\\ \hline
  $g_1$        & $0$     & $\frac{3\epsilon}{5}$         &    $\frac{3\epsilon}{5}$  & $\frac{1}{5}(3\epsilon-2\delta)$ & $\frac{1}{5}(3\epsilon-2\delta)$\\ \hline
  $g_3^2/u$        & $0$     & $\frac{2\epsilon}{3}$         &    $\frac{2\epsilon}{3}$   & $0$ & $0$\\ \hline
  $g_5^2/u$        & $0$     & $\frac{2\epsilon}{3}$         &    $\frac{2\epsilon}{3}$ & $0$ & $2\epsilon - 2\delta$ \\ \hline
  $g_4$        & $0$     & $0$ & $\frac{8\delta}{3}$ & $\frac{8\delta}{3}$ & $\frac{8\delta}{3}$\\ \hline
  $1/u$          & $0$     & $0$         & $0$  & $0$ & $0$\\ \hline  
  $u_1$        & $0$     & $0$  & $0$ 
  & $\frac{1}{6}(1+\sqrt{13})$ & $\frac{1}{6}(1+\sqrt{13})$ \\ \hline
\end{tabular}
  \caption{Fixed points for model E${}_h$, $u \to \infty$}
  \label{tab:fpu}
\end{table}

\begin{table}
\centering
\renewcommand{\arraystretch}{1.75}
\begin{tabular}{|c|c|c|c|c|c|}
  \hline
  FP           & FP1$^{II}$     & FP2$^{II}$  &     FP3$^{II}$ & FP4$^{II}$ & FP5$^{II}$\\ \hline
  $g_1$        & $0$     & $0$         &    $\frac{3\epsilon}{5}$  & $\frac{3}{5}(\epsilon-2\delta)$ & $\frac{3}{5}(\epsilon-2\delta)$\\ \hline
  $g_3$        & $0$     & $0$         &    $0$   & $0$ & $0$\\ \hline
  $g_5$        & $0$     & $\sqrt{2\epsilon}$         &    $\sqrt{2\epsilon}$ & $0$ & $\sqrt{2(\epsilon-4\delta)}$ \\ \hline
  ${g_4/u}$        & $0$     & $0$ & $0$ & $\frac{8\delta}{3}$ & $\frac{8\delta}{3}$\\ \hline
  $u$          & $0$     & $1$         & $1$  & $1$ & $1$\\ \hline  
  $1/u_1$        & $0$     & $0$  & $0$ & $0$ & $0$ \\ \hline
\end{tabular}
  \caption{Fixed points for model E${}_h$, $u_1 \to \infty$}
  \label{tab:fpu1}
\end{table}

\begin{table}
\centering
\renewcommand{\arraystretch}{1.75}
\begin{tabular}{|c|c|c|c|c|c|}
  \hline
  FP           & FP1$^{III}$     & FP2$^{III}$  &     FP3$^{III}$ & FP4$^{III}$ & FP5$^{III}$\\ \hline
  $g_1$        & $0$     & $0$         &    $\frac{3\epsilon}{5}$  & $\frac{3}{5}(\epsilon-2\delta)$ & $\frac{3}{5}(\epsilon-2\delta)$\\ \hline
  $f_3$        & $0$     & $\frac{2 \epsilon}{3}$         &    $\frac{2 \epsilon}{3}$   & $0$ & $0$\\ \hline
  $f_5$        & $0$     & $\frac{2 \epsilon}{3}$         &    $\frac{2 \epsilon}{3}$ & $0$ & $2(\epsilon-3\delta)$ \\ \hline
  $f_4$        & $0$     & $0$ & $0$ & $\frac{8\delta}{3}$ & $\frac{8\delta}{3}$\\ \hline
  $w$          & $0$     & $0$         & $0$  & $0$ & $0$\\ \hline  
  $w_1$        & $0$     & $0$  & $0$ & $0$ & $0$ \\ \hline
\end{tabular}
  \caption{Fixed points for model E${}_h$, $u,u_1 \to \infty$}
  \label{tab:fpuu1}
\end{table}

There exists a more nontrivial fixed point of model E${}_h$ {when} all charges
{obtain}
nonzero values. The related exact expression was not
calculated in \cite{Danco} because 
the fixed point and its stability 
region depend on the $\epsilon /\delta$ ratio and this makes the $\gamma$ structure
 {very inconvenient for analytical treatment}. 
Our direct numerical calculation in the turbulent regime yields now the unstable point 
with the following location: 
\begin{align*}
  & u_* =0.756, & u_{1*} & =  0.833, & g_{1*} & =  4.808,\\
  &g_{3*} = 1.449, & g_{5*} &  =  -1.021, & g_{4*} & =  32/3,
\end{align*}  
and $\upomega$ {indices}
\begin{align*}
  &\upomega \in \{-12,984, 3.646, -1.889, -2.328 \pm .257 i, 4\}.   
\end{align*}
{Need to say} that
 the corresponding nontrivial fixed point with a physically consistent value ($g_4>0$) is absent in {the} 
equilibrium regime for model E${}_h$.

\medskip

{\subsection {Thermal equilibrium regime, $\epsilon =1$, $\delta =-1$}
\label{Subsec:thermal}}
In this regime the numerical analysis of model F${}_h$ has not exhibited the existence of
the IR stable fixed points of the system (\ref{fps}). Apparently, this is not
 {a} physical result as the one-loop approximation is not sufficient in this case. For example, 
the system (\ref{fps}) reduced to the standard model F leads to the stable fixed
point  $u_*=1.366$,  $b_*=0.655\epsilon $, $g_{1*}=1.199\epsilon $,
$g_{2*}=0.447\epsilon $, $g_{3*}=1.280\epsilon $. Similarly, the four-loop calculations
in the static model C prove the stable scaling regime of the standard model F corresponds
to model E with $b_*=0$, $g_{2*}=0$ \cite {Vasiliev}. Then we can state {that} the 
multi-loop calculations are necessary to make relevant conclusions.

Dynamic Eqs. (\ref{ps1}),(\ref{mm}) {and} 
(\ref{eq1}) demonstrate that the basic fields 
{$\psi$, $\psi^+$, $m$} play a role of passive scalars for the hydrodynamic modes. 
That causes the  exact perturbative statements $g_{4*}=0$, $\gamma _{g_4 *}=0$ at the 
IR stable fixed point for $\delta <0$. Another exact expression $\gamma_{g_4}=-3\gamma_{\nu}$ 
yields the relation $\gamma _{\nu *}=0$ in this regime. The next explicit 
formula $\gamma_{\nu}=\gamma _{\lambda}+\gamma_{u_1}$ leads to the relation
$\gamma _{\lambda *}=-\gamma_{u_1 *}$ for the corresponding fixed point.
Using the fixed point equation $\beta _{u_1}=u_{1*}\gamma_{u_1}=0$ one can observe 
two possibilities. The {former} one is $\gamma _{\lambda 
*}=-\gamma_{u_1 *}=0$ and {the} dynamic index $z$ is 
rigorously equal to $2$ {and in the latter} $u_{1*}=0$. As renormalization 
constants depend on the combination $g_4 u_1^2$, in this case the elements of
the $\Omega$ matrix related to $\partial\beta_k/\partial g_4$
{are} equal to $0$.

Thus, in this case the parameter $u_1$ does not affect the fixed points and its stability.
  {However,} in general the hydrodynamic modes have influence
on the stability analysis as they
lead to the new multiplicatively renormalized charges  $g_5, g_6, g_7$. These new 
charges produce  new columns and rows in the $\Omega$ matrix and then they are 
essential in the analysis of the fixed points stability. Let us remind that this 
analysis is the main problem of model F {at this point.}

Our most interesting achievement in the equilibrium regime is a new way to analyze 
the stability for the standard model E. Indeed, some of the fixed points of model E${}_h$ 
 presented in Tables~\ref{tab:fp_stab}-\ref{tab:fpuu1} correspond to the
standard model E \cite{Vasiliev}.
They must obey $g_4=u_1=0$ and $g_3=g_5$, {in accordance} with Table~\ref{mod}. Besides,
they are stable in the thermal equilibrium regime, i.e. in the region
$\epsilon>0$, $\delta<0$. It was the points FP3 and FP2${}^I$ that obeyed these constraints.
This coincides with a well known two-loop result \cite{Vasiliev}, though it is 
unknown which of these two points is stable for the standard model E. In the 
framework of model F${_h}$ these two points have the following $\upomega$ indices:
\begin{align*}
  {FP3:}\quad  \upomega\in &\{-0.1\epsilon,0, 0.055\epsilon,0.25\epsilon,0.75\epsilon,\epsilon,\\
   &1.5\epsilon,1.92\epsilon,-\delta\},\\
  {FP2}^I:\quad  \upomega\in &\{-0.333\epsilon,-0.01\epsilon, -0.05\epsilon, 0.666\epsilon, \epsilon,\\
  &1.3\epsilon, 2.15\epsilon, -\delta\}.
\end{align*}
That means that the point FP3 seems to be more IR stable with respect to the hydrodynamics effects.
\bigskip

{\section{Summary and Conclusions}  \label{Sec:conclusion}}
%
%
In the previous chapters we have shown that all {values of the}
critical exponents are drastically changed as a result of the turbulent background.
Specifically, charges of the fields $\psi$, $\psi^+$, $m$ that govern the standard critical behavior vanish 
at {the} IR stable critical point in the presence of developed 
turbulence (see (\ref{nontr})).
 In other words, the developed turbulence destroys critical fluctuations. This fact has a simple 
physical ex\-pla\-na\-tion.
It is well known that critical behavior of the system is accompanied by {an}  
unbounded growth of {a} correlation 
radius. On the other hand, in  the background of developed turbulence, the cascade of the eddies 
takes place {\cite{Frisch}}. Due to the decay {of} 
large eddies into smaller ones, the kinetic energy is transferred
from the largest to the smallest scales and dissipates. Thus it is reasonable to expect that precisely these 
eddies (and related cascade mechanism) confine the growth of the correlation radius which crucially 
changes the critical behavior.
Moreover,  for { the} above turbulent re\-gi\-me we have calculated 
{ the} scaling exponent of effective 
viscosity which turned out to be equal {to} $4/3$ (see (\ref{43})) 
and, therefore,  
coincides with {the} well-known fully developed turbulence value. 

We have investigated the regime of equilibrium fluctuations, carried out the analysis and 
classification of the corresponding fixed points and 
made some assumptions related to their stability taking into account  the peculiarities of the
extended model F${}_h$. {{Then, {the} critical dimension of viscosity 
vanishes in the regime of equilibrium fluctuations to all orders in the perturbation theory.}} 
 
Nevertheless, we need to be {careful} in the analysis of the 
obtained results. The {corollary} gained from 
 the  analysis  of the thermal equilibrium regime
suggests that one-loop calculations of models E${}_h$ and F${}_h$ are not sufficient to make
a definite conclusion about the stability of fixed points. Indeed, in the one-loop approximation of 
 model {{F the fixed point mentioned at the beginning of Section \ref{Subsec:thermal} is found as IR stable in the equilibrium regime.}}
 However, the comparison of 
 the {four}-loop and {five}-loop static results \cite{Vasiliev} with
 the two-loop expressions
 in  model E \cite{Dominicis} yields the conclusion $g_2=0$. It was the fixed point of model E
 that is suitable for description of {the} true phase transition point. 
 By analogy, next 
 orders of perturbation corrections can change the sign of the $\upomega$ index and modify 
 the stability analysis in the presence of the turbulent background.

{To make a final conclusion, we assert} that the calculations in the regime with the dominance of
equilibrium fluctuations 
seem to be incomplete. For proper analysis of the theory, high-order calculations with inclusion of turbulent 
fluctuations and consequent resummation procedure are needed.  How\-ever, even the one-loop 
approximation leads to $48$ Feynman graphs. The related calculations {were} possible only due 
to {a} multiple cross-checking of results within authors' group. 
The calculations of multi-loop
contributions, evidently, require {algorithmization} of the work. 

{The} corresponding multi-loop algorithms were elaborated in the  cycle of
articles \cite{Adzhemyan}; how\-ever, their applicability is limited by the
models without non-perturbative charges like $u$ and $u_1$ in model F${}_h$. 
More exactly, the calculations are possible as the fixed points 
including non-perturbation charges are calculated at a lower order of 
perturbation theory. Subsequently the obtained values can be  used in calculations of 
Feynman graphs to the next perturbation order. 

In addition to the above stated results, the list of fixed points having a chance
to become stable in the multi-loop approximation can be considered as a starting point
for high-order computer calculations. This information can be considered as 
 {an} important contribution to the final decision {which}
 model (E or F) is suitable for a
description of {a} phase transition to {the}
superfluid state and whether the turbulent 
background gives {a} contribution to the experimentally observed quantities{.}
%
%

\bigskip

\subsection*{Acknowledgments}
The work was supported by VEGA grant {\textnumero  1/0222/13}   of the Ministry of Education,
 Science, Research and Sport of the Slovak Republic,
by the Russian Foundation for Basic Research within the project 12-02-00874-a 
and by St-Petersburg State University research grant 11.38.185.2014.

We would like to thank  Dr. Martin Vala and the project Slovak Infrastructure for High Performance Computing
(SIVVP) ITMS 26230120002.

 {\section{Appendix}
 \label{Sec:appendix} }
We use the following notation for the vertex factors:
\begin{minipage}[c]{\textwidth}   
\parbox[c]{2cm}{\includegraphics[width = 2cm]{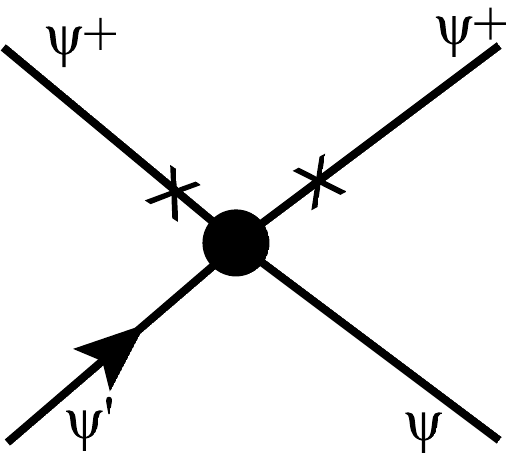}} 
\parbox[c]{\textwidth}{$\displaystyle{=-\lambda \left( \frac{g_1}{3} - i\left(\frac{b g_1}{3} - g_2 g_3\right)\right),}$}
\end{minipage}
\vskip0.25cm
\begin{minipage}[c]{\textwidth}   
\parbox[c]{2cm}{\includegraphics[width = 2cm]{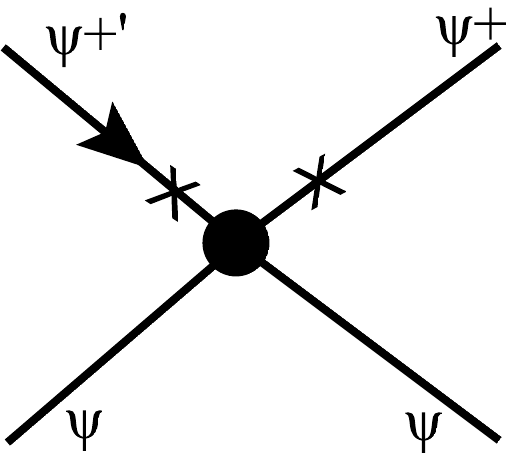}} 
\parbox[c]{\textwidth}{$\displaystyle{= - \lambda \left( \frac{g_1}{3} + i\left(\frac{b g_1}{3} - g_2 g_3\right)\right)},$}
\end{minipage}
\vskip0.25cm
\begin{minipage}[c]{\textwidth}   
\parbox[c]{2cm}{\includegraphics[width = 2cm]{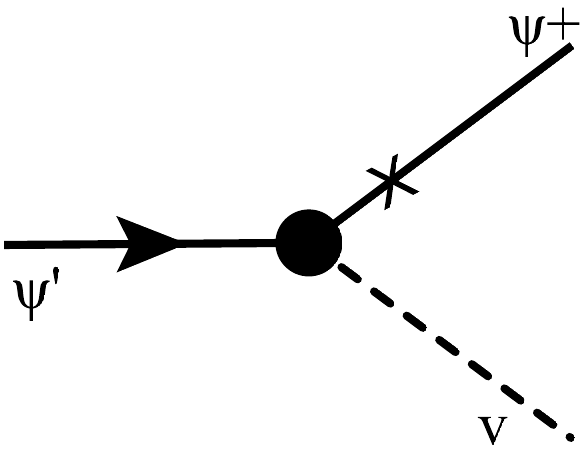}} 
\parbox[c]{0.075\textwidth}{$=1$,}
\parbox[c]{2cm}{\includegraphics[width = 2cm]{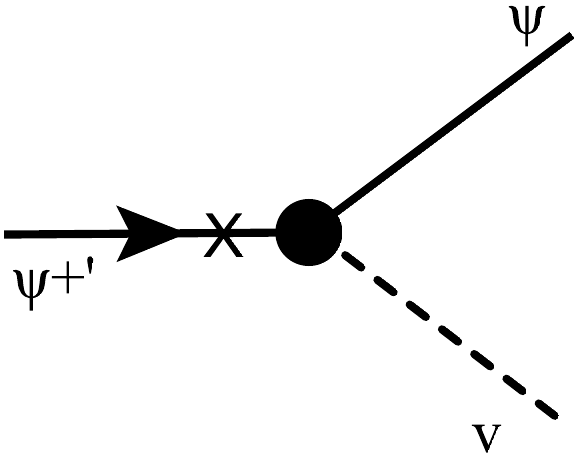}} 
\parbox[c]{0.075\textwidth}{$=1$,}
\end{minipage}
\vskip0.25cm
\begin{minipage}[c]{\textwidth}   
\parbox[c]{2.2cm}{\includegraphics[width = 2.2cm]{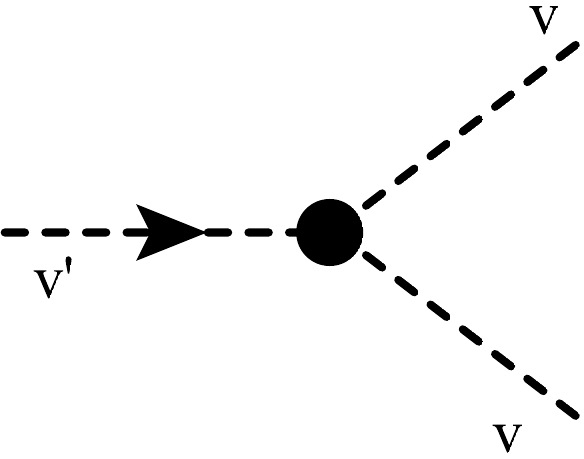}} 
\parbox[c]{0.075\textwidth}{$=1$,}
\parbox[c]{2cm}{\includegraphics[width = 2cm]{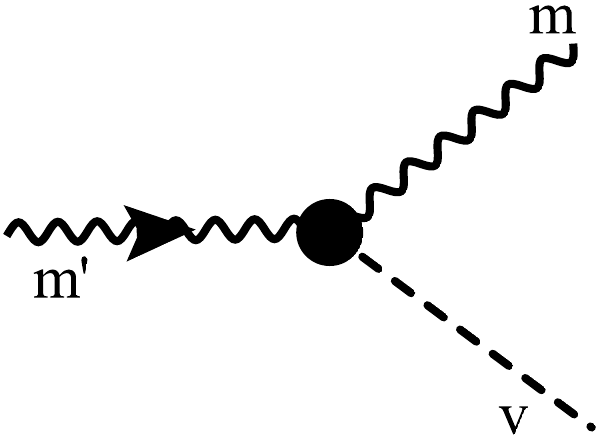}} 
\parbox[c]{0.075\textwidth}{$=1$,}
\end{minipage}
\vskip0.25cm
\begin{minipage}[c]{\textwidth}   
\parbox[c]{2cm}{\includegraphics[width = 2cm]{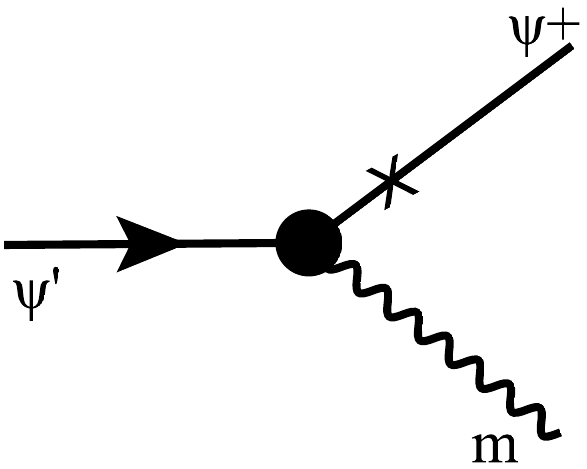}} 
\parbox[c]{\textwidth}{$\displaystyle{=\lambda (g_2 - i ( b g_2 - g_3 ))},$}
\end{minipage}
\vskip0.25cm
\begin{minipage}[c]{\textwidth}   
\parbox[c]{2cm}{\includegraphics[width = 2cm]{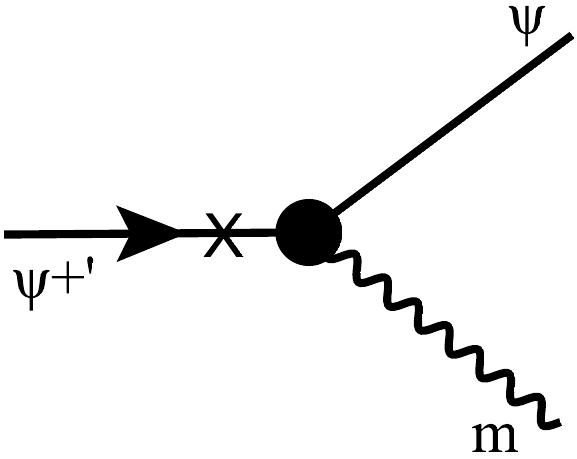}} 
\parbox[c]{\textwidth}{$\displaystyle{=\lambda (g_2 + i ( b g_2 - g_3 ))},$}
\end{minipage}
\vskip0.25cm
\begin{minipage}[c]{\textwidth}   
\parbox[c]{2.5cm}{\includegraphics[width = 2.6cm]{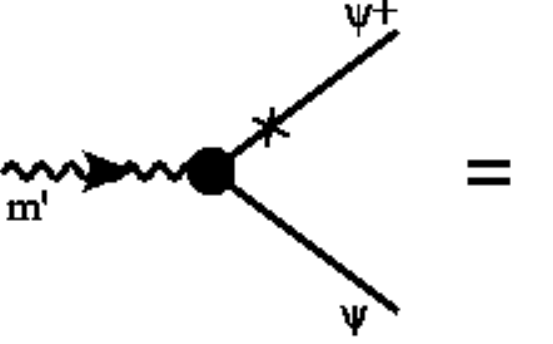}} 
\parbox[c]{\textwidth}{$\displaystyle{i \lambda g_5(\mp^2-\mq^2)- u \lambda g_6(\mp+\mq)^2  }$.}
\end{minipage}
In this graph $\mp$ and $\mq$ are the arguments ascribed to $\psi(\mp)$, $\psi^+(\mq)$, respectively.

The diagrams are numerated as follows:

\noindent
1. \includegraphics[width=3cm]{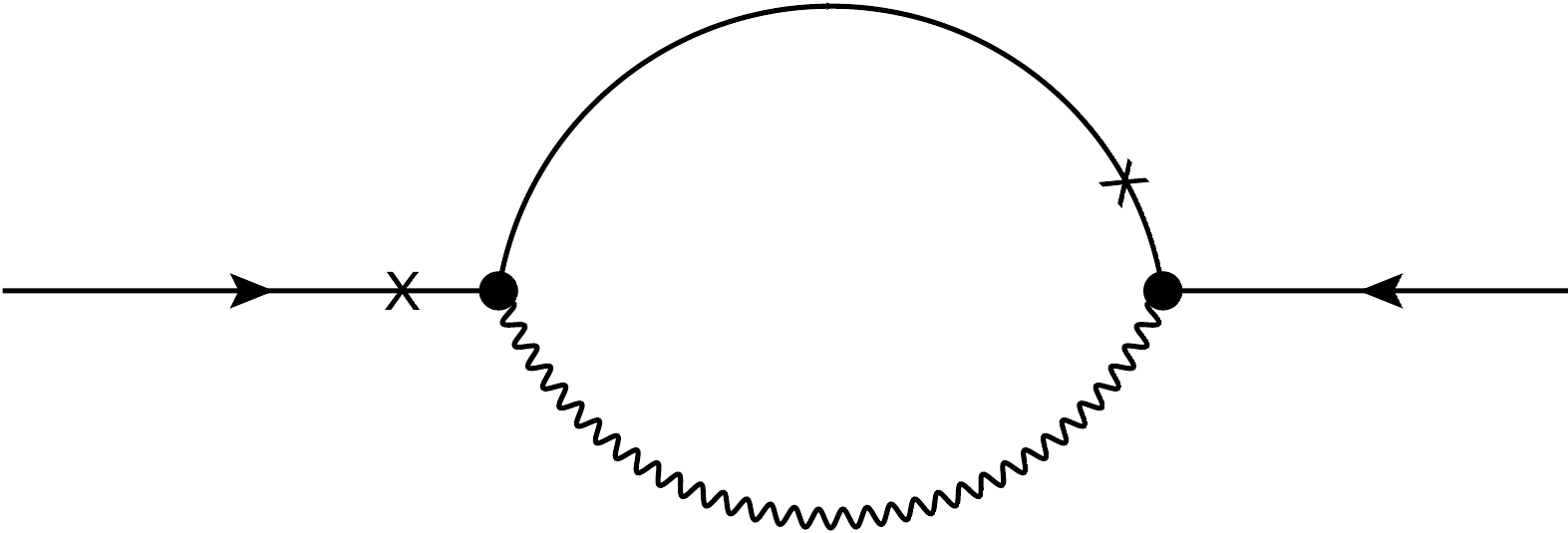}; \quad
2. \includegraphics[width=3cm]{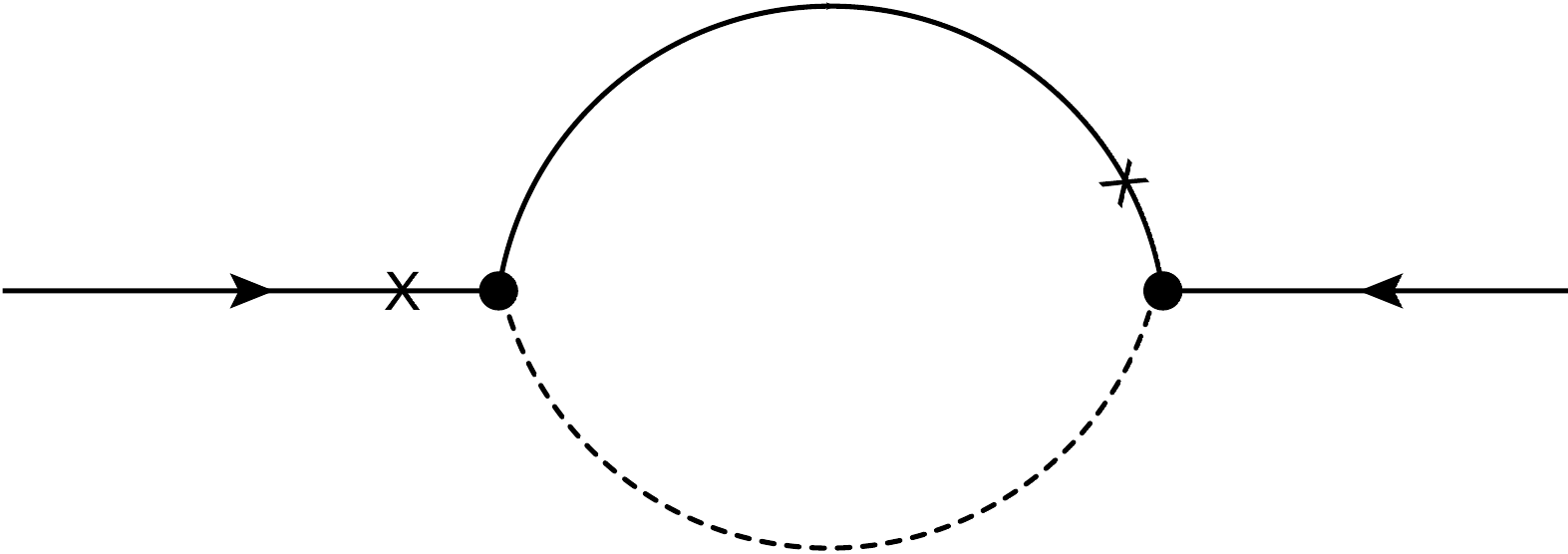}; \quad
\newline
3. \includegraphics[width=3cm]{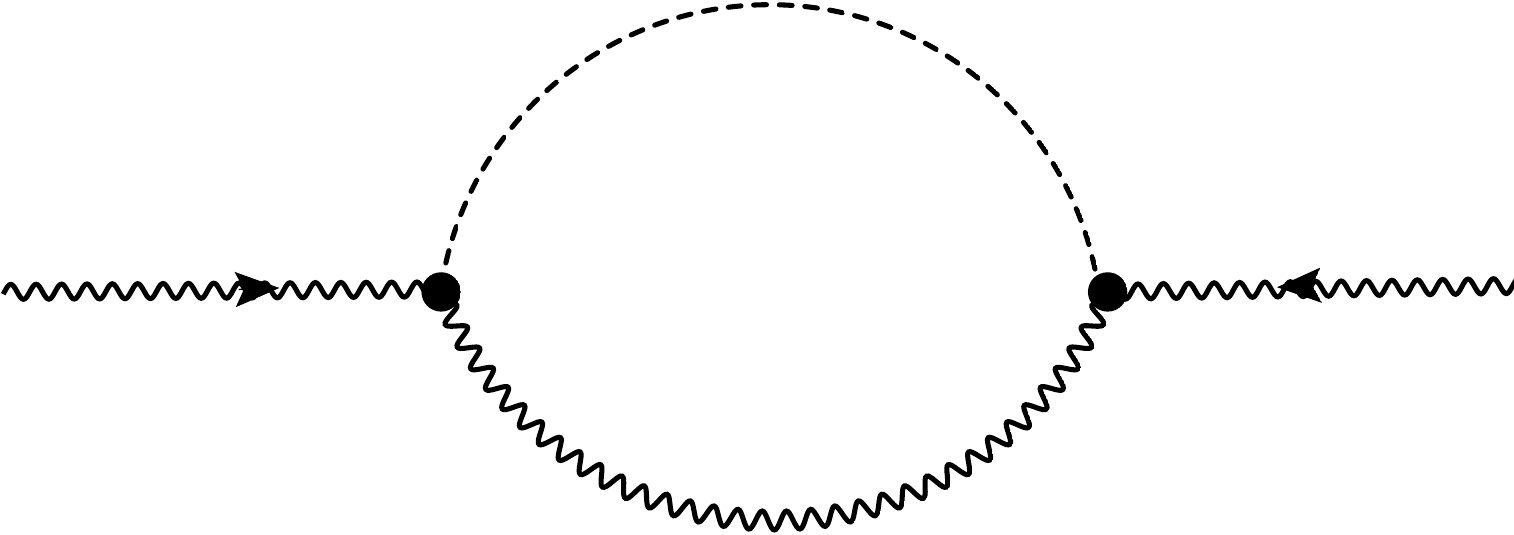}; \quad
4. \includegraphics[width=3cm]{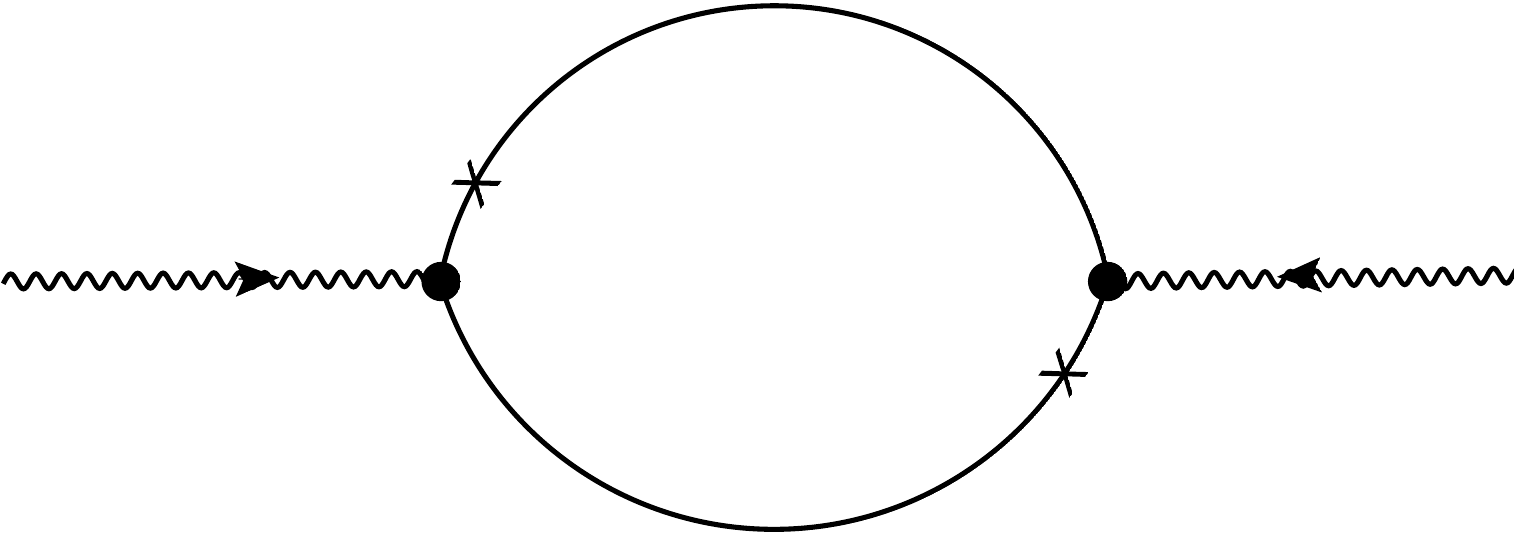}; \quad
\newline
5. \includegraphics[width=3cm]{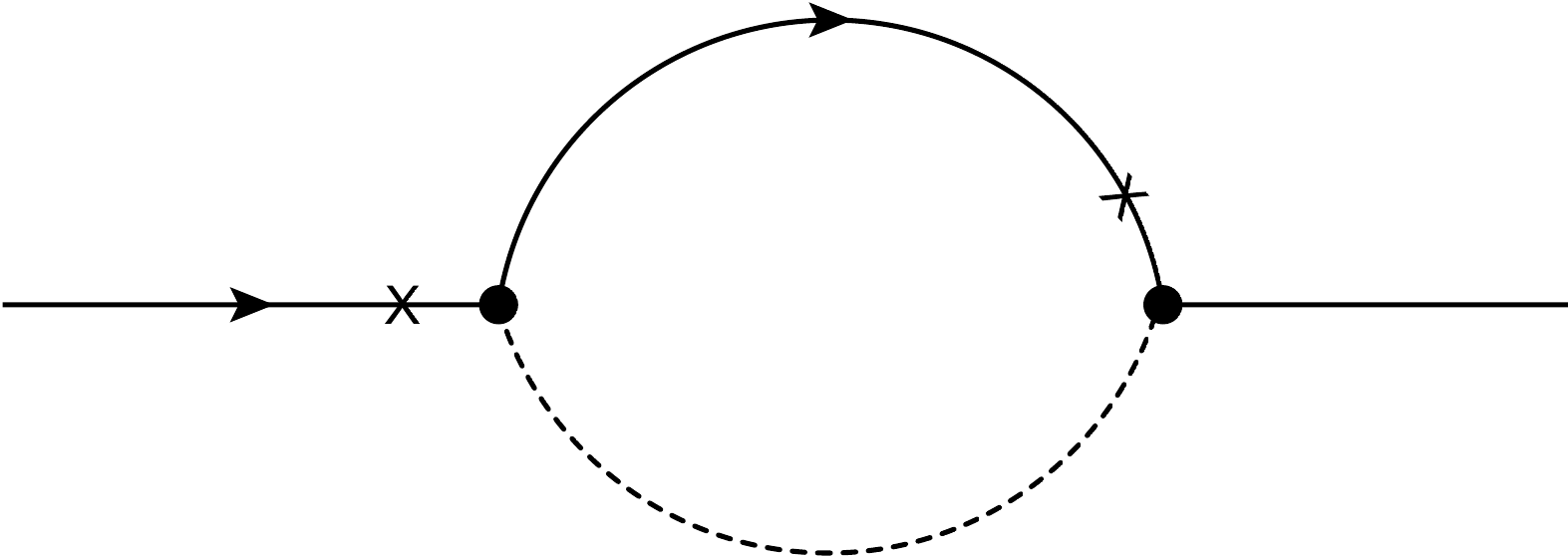}; \quad
6. \includegraphics[width=3cm]{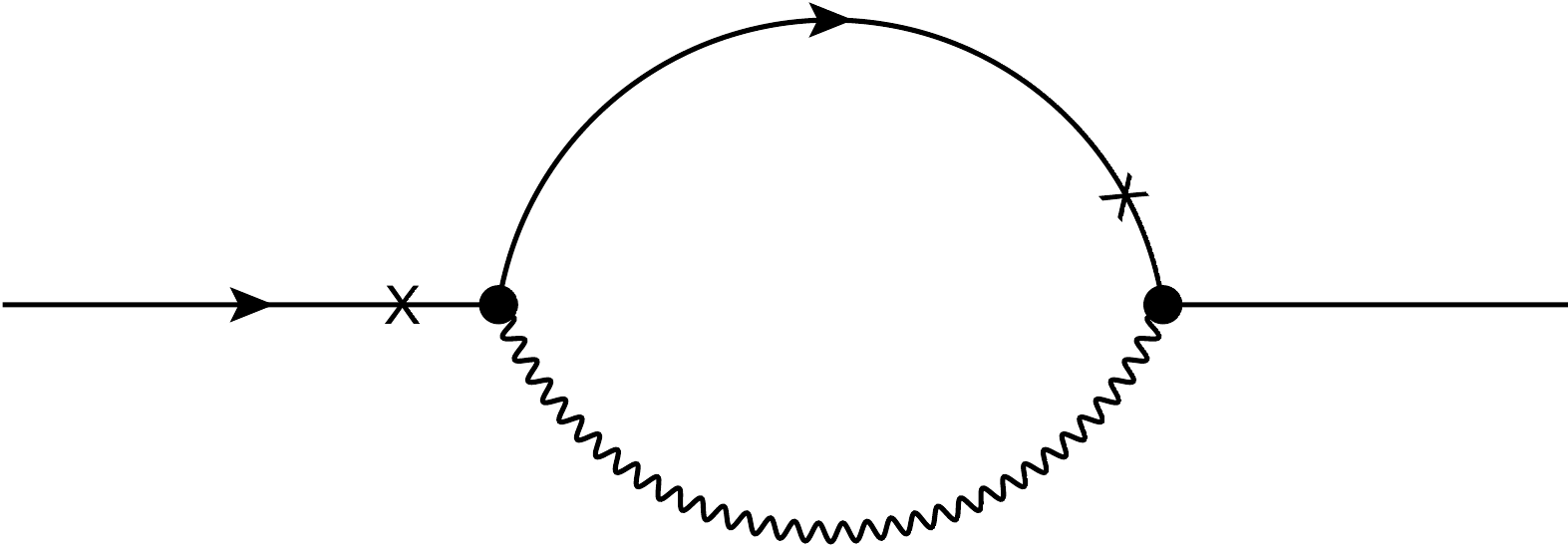}; \quad
\newline
7. \includegraphics[width=3cm, height=1.5cm]{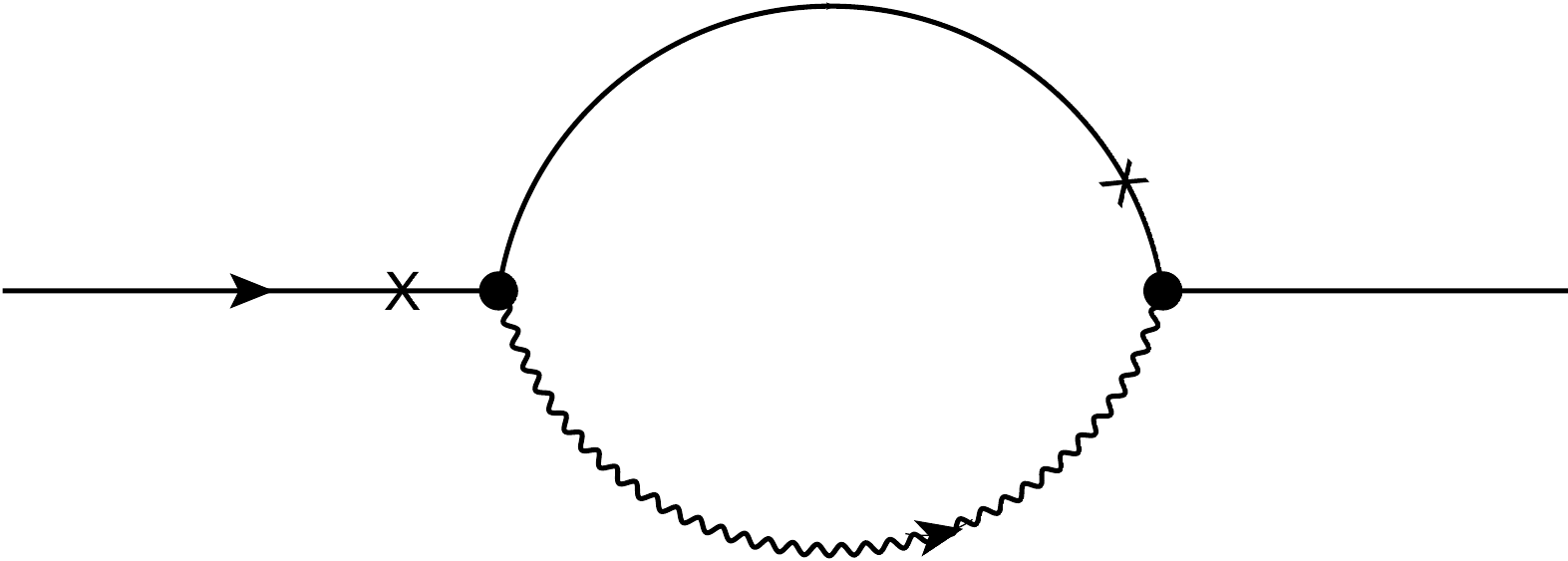}; \quad
8. \includegraphics[width=3cm, height=1.5cm]{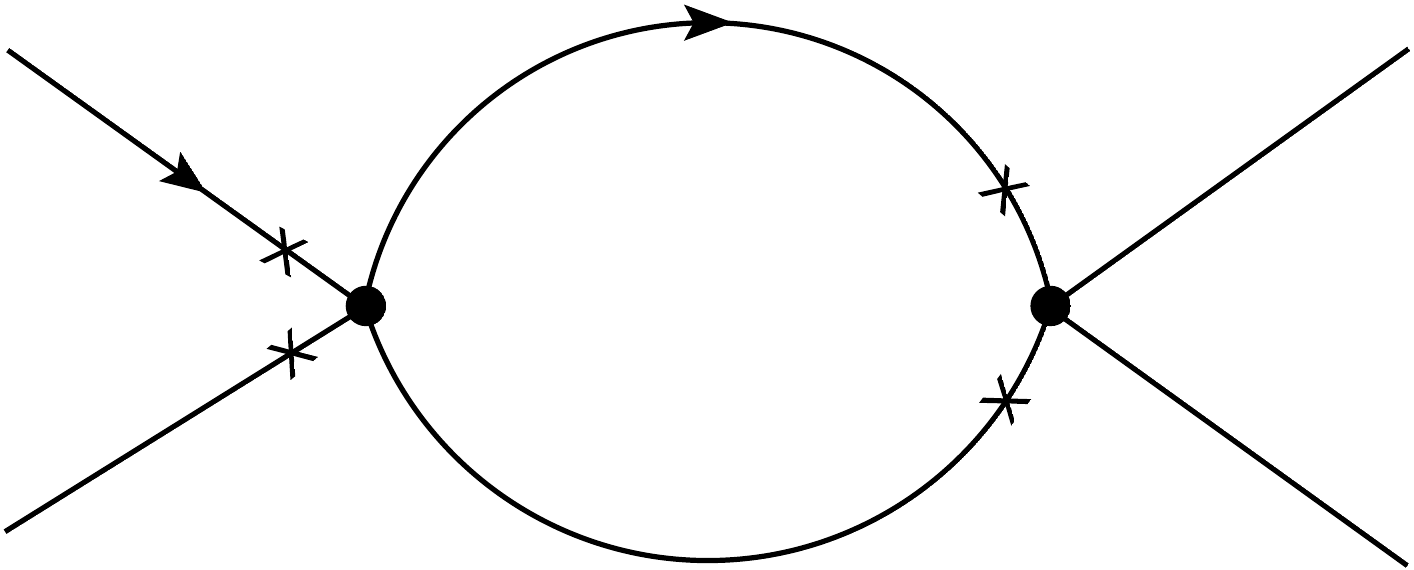}; \quad
\newline
9. \includegraphics[width=3cm, height=1.5cm]{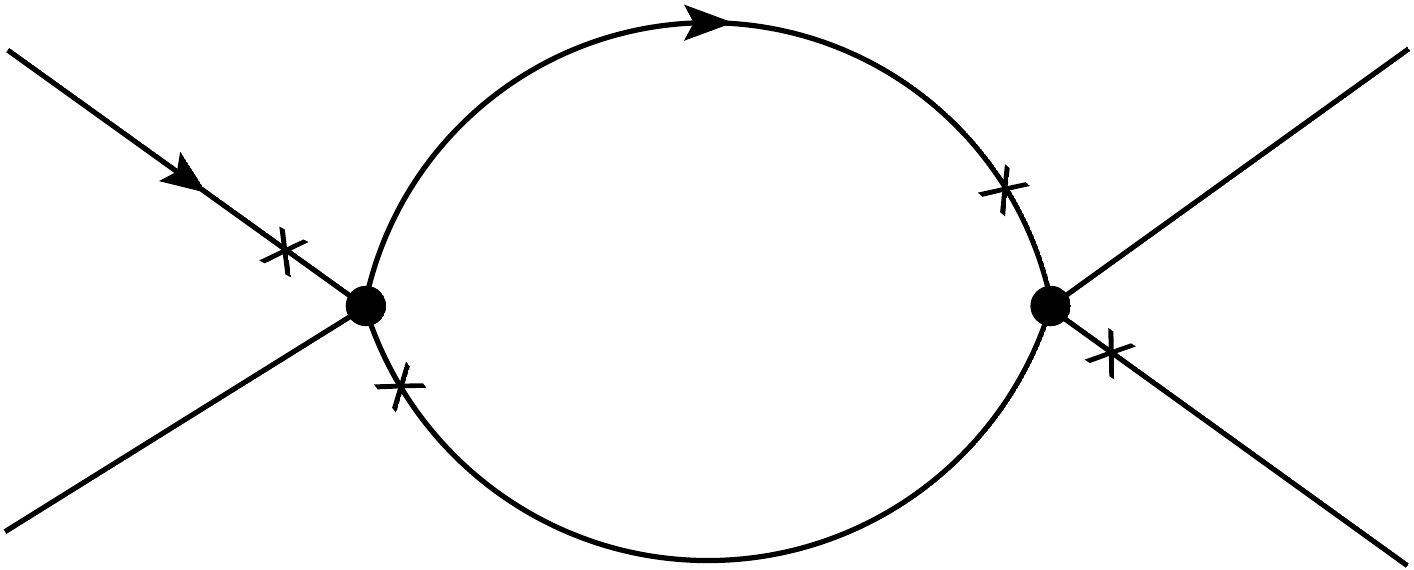}; \quad
10. \includegraphics[width=3cm, height=1.5cm]{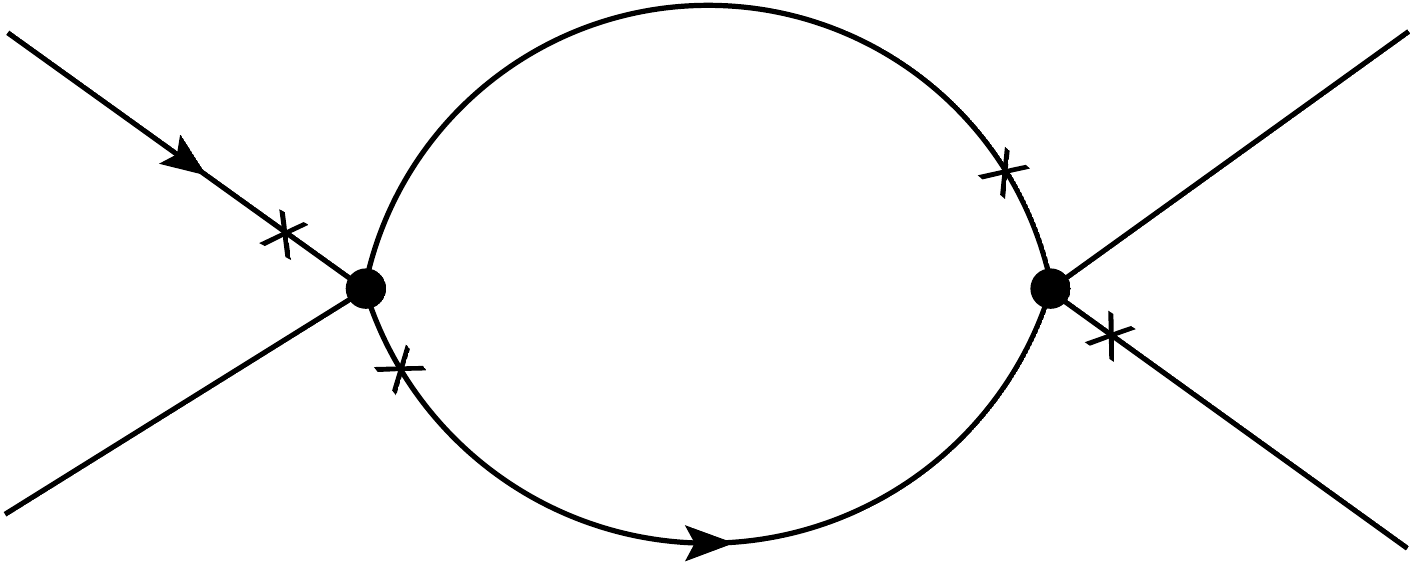}; \quad
\newline
11. \includegraphics[width=3cm, height=1.5cm]{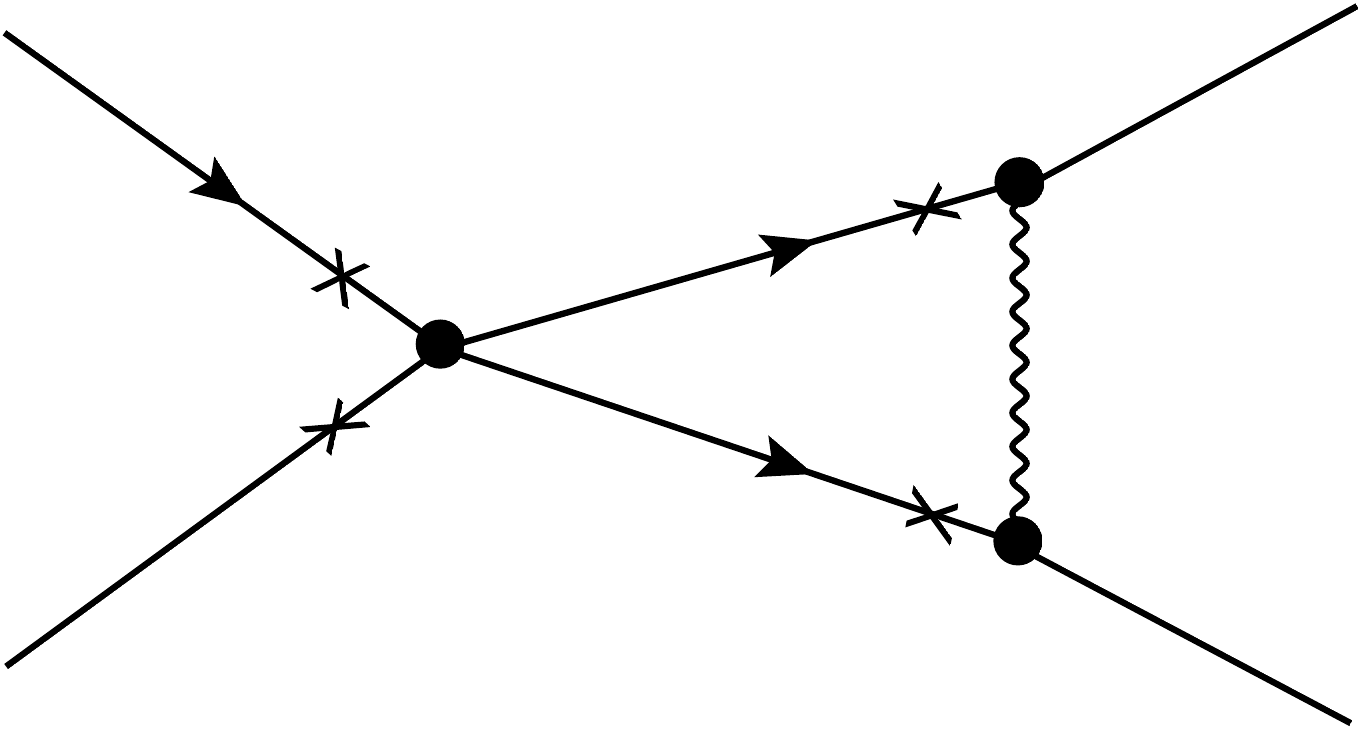}; \quad
12. \includegraphics[width=3cm, height=1.5cm]{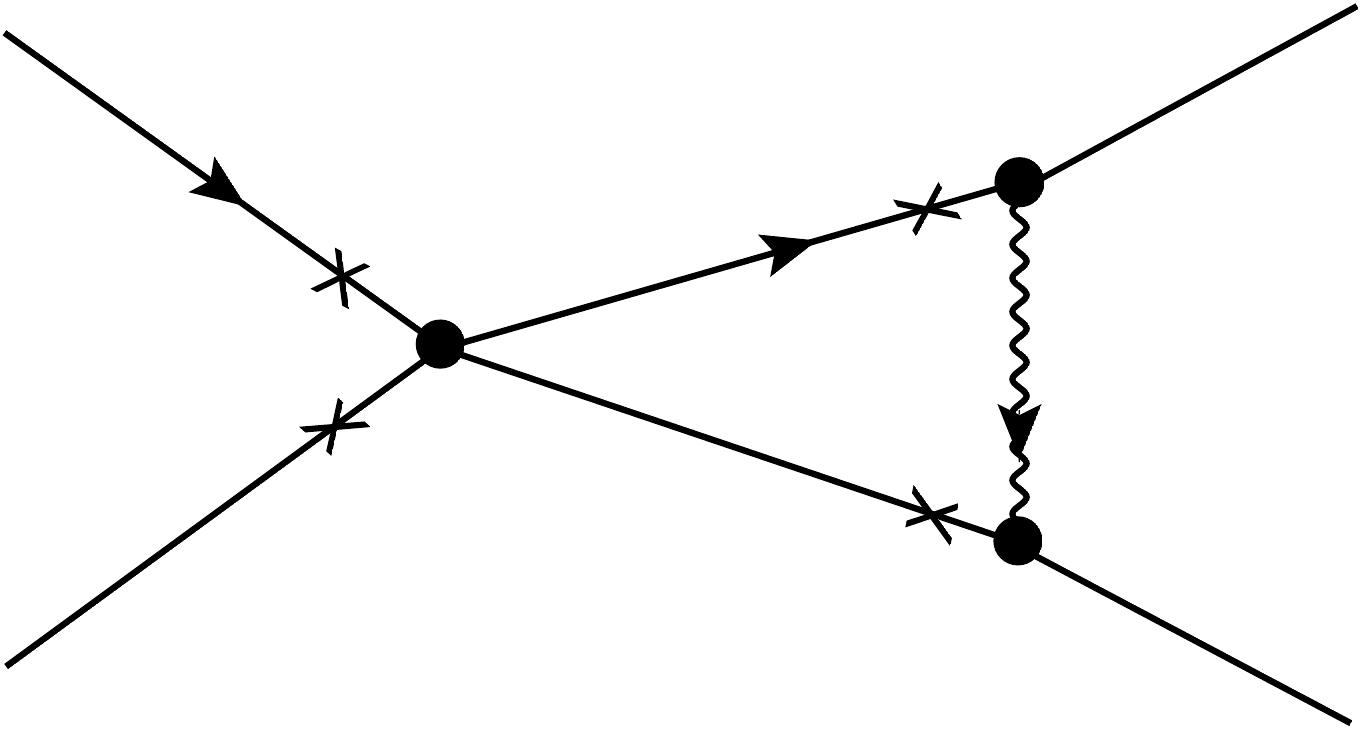}; \quad
\newline
13. \includegraphics[width=3cm, height=1.5cm]{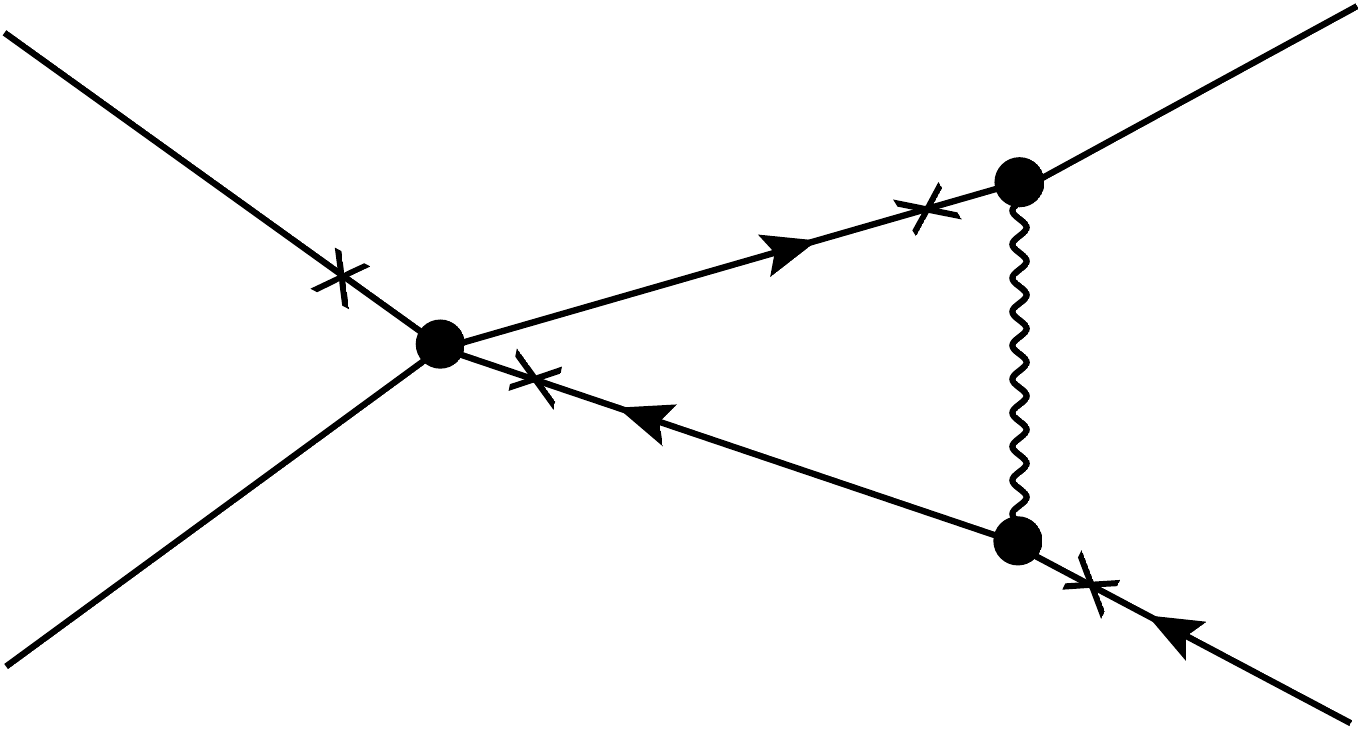}; \quad
14. \includegraphics[width=3cm, height=1.5cm]{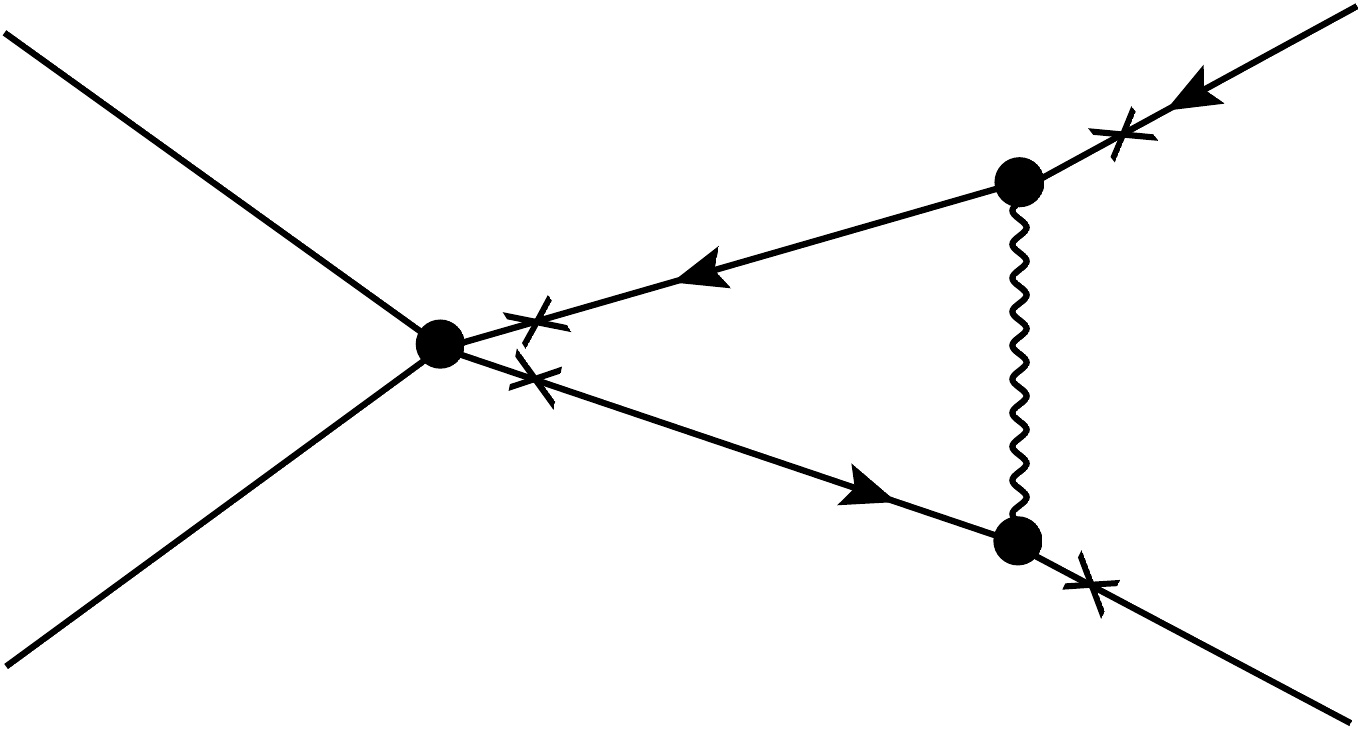}; \quad
\newline
15. \includegraphics[width=3cm, height=1.5cm]{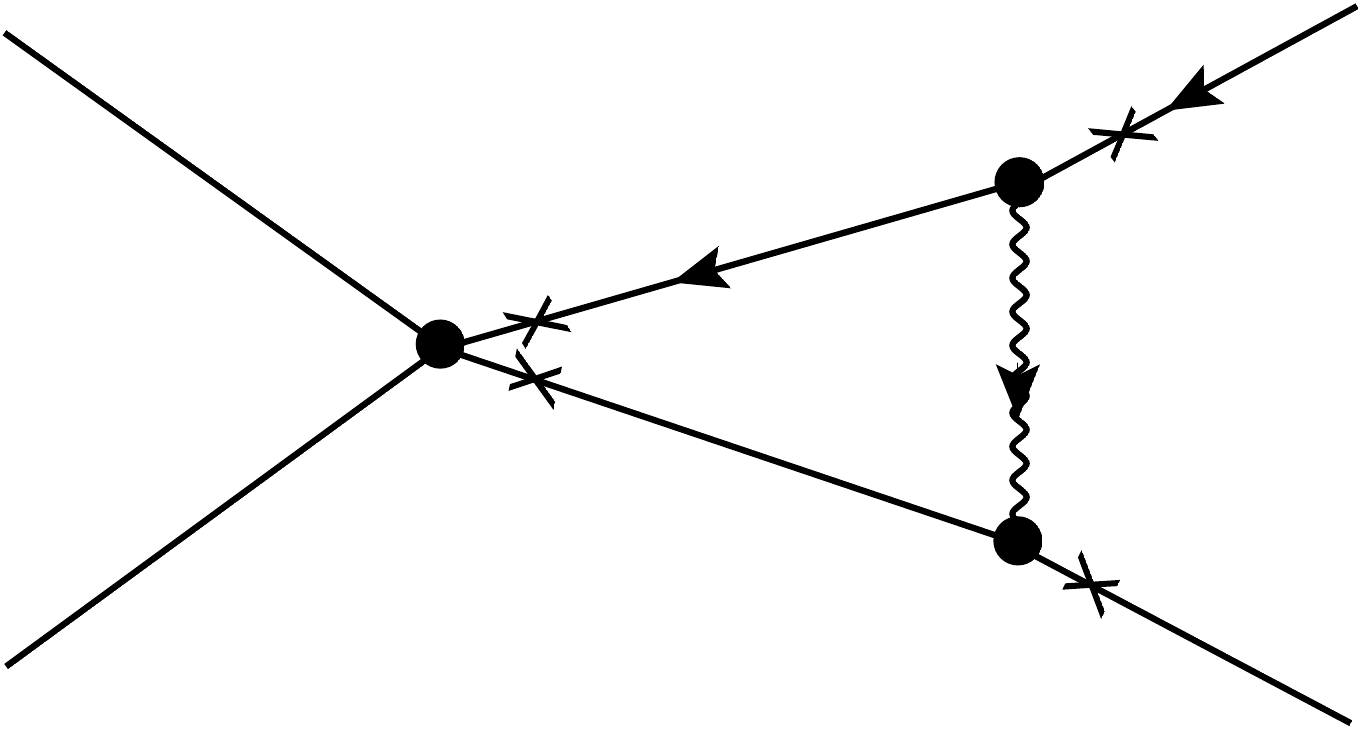}; \quad
16. \includegraphics[width=3cm, height=1.5cm]{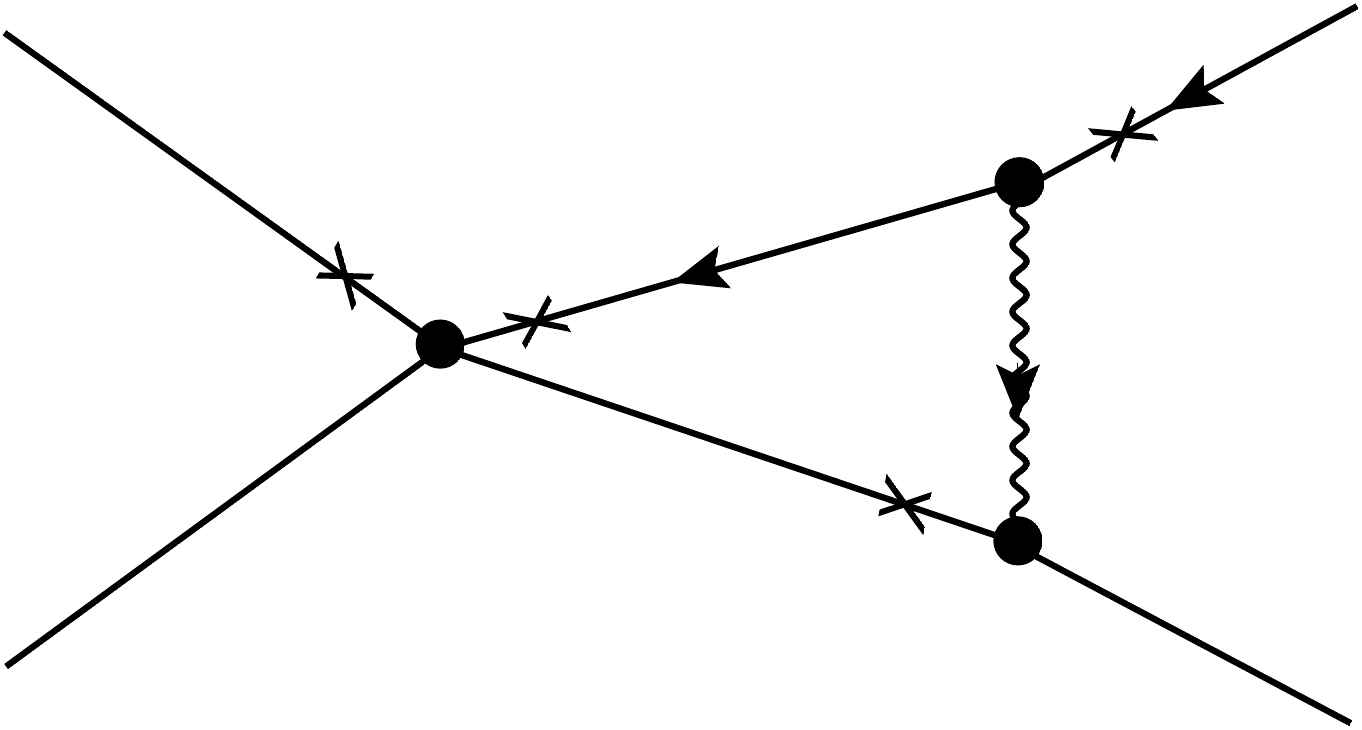}; \quad
\newline
17. \includegraphics[width=3cm, height=1.5cm]{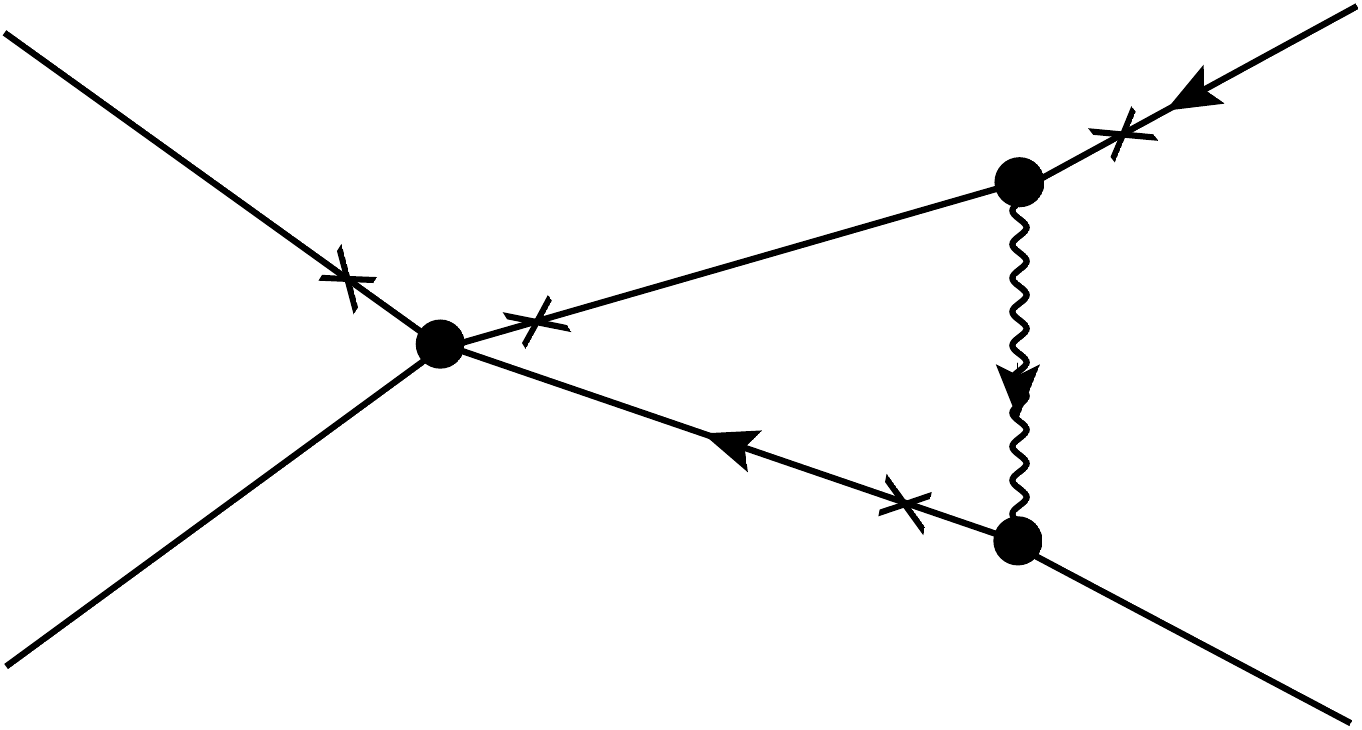}; \quad
18. \includegraphics[width=3cm, height=1.5cm]{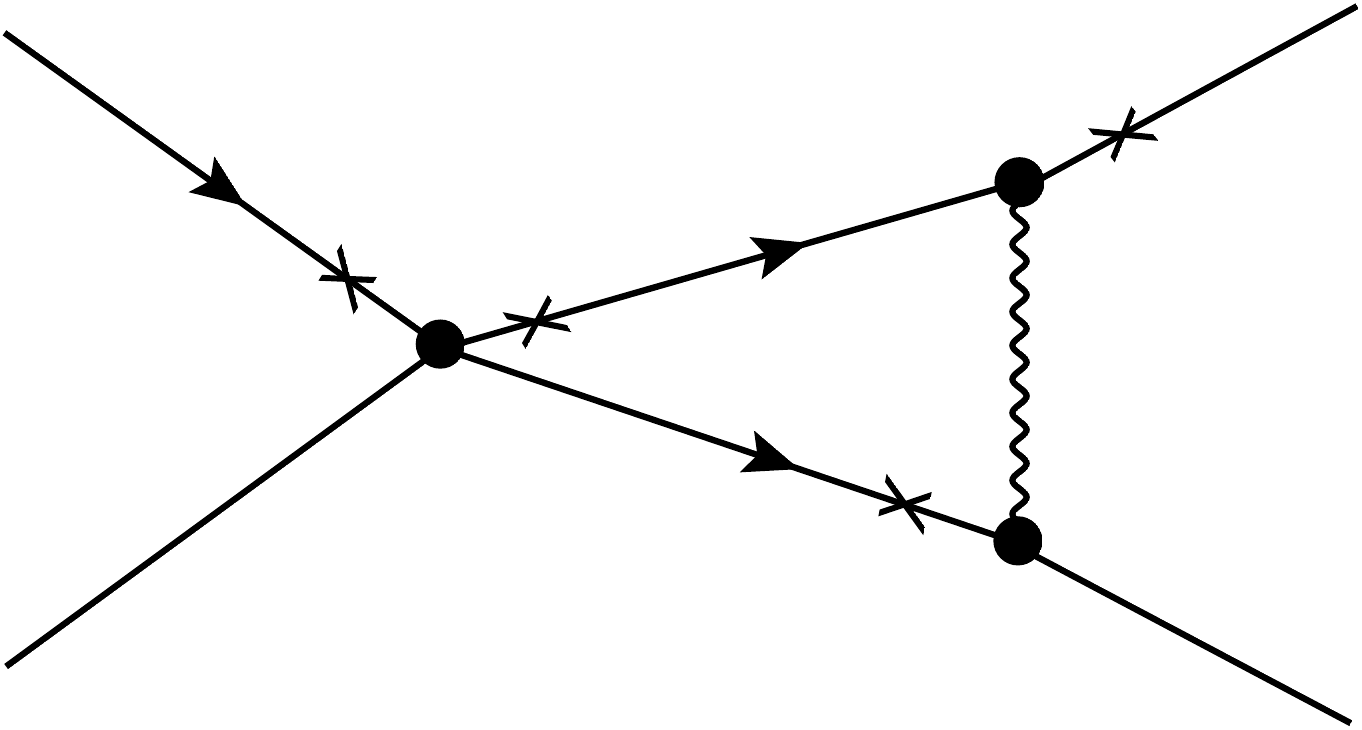}; \quad
\newline
19. \includegraphics[width=3cm, height=1.5cm]{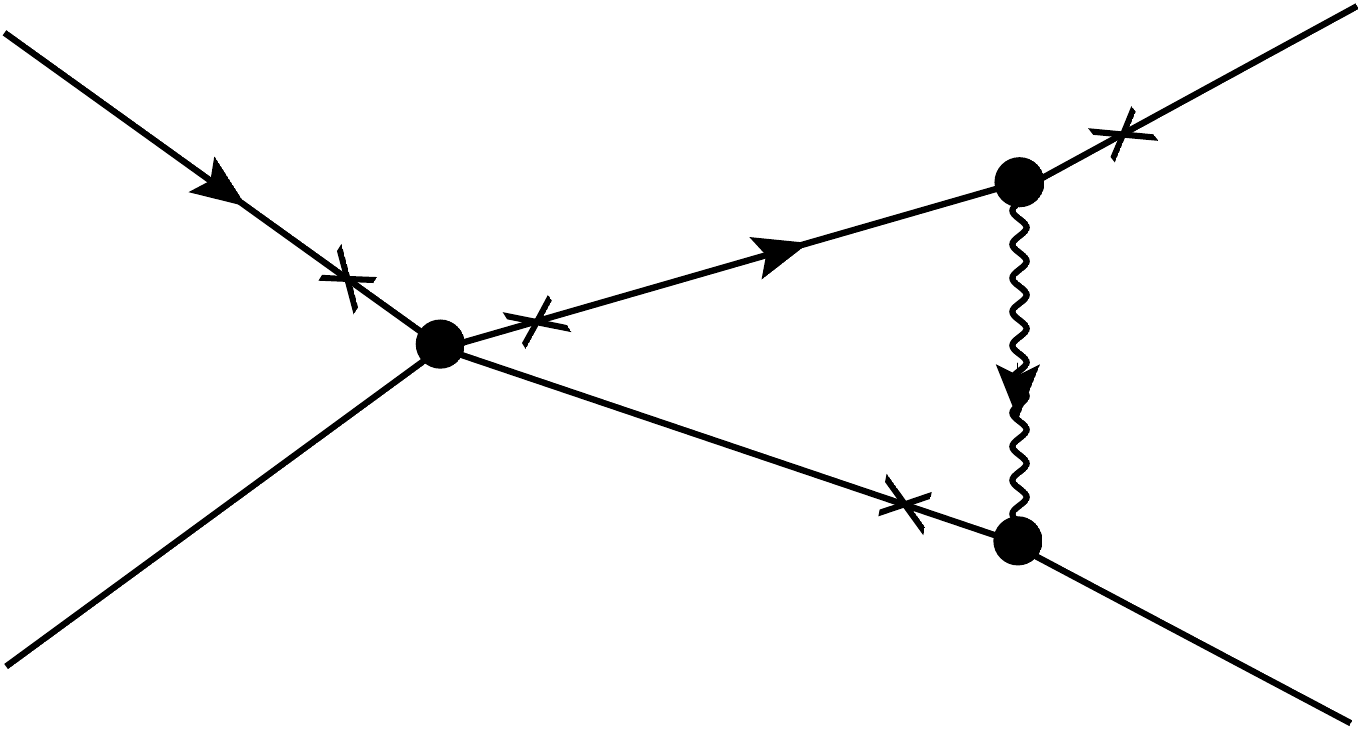}; \quad
20. \includegraphics[width=3cm, height=1.5cm]{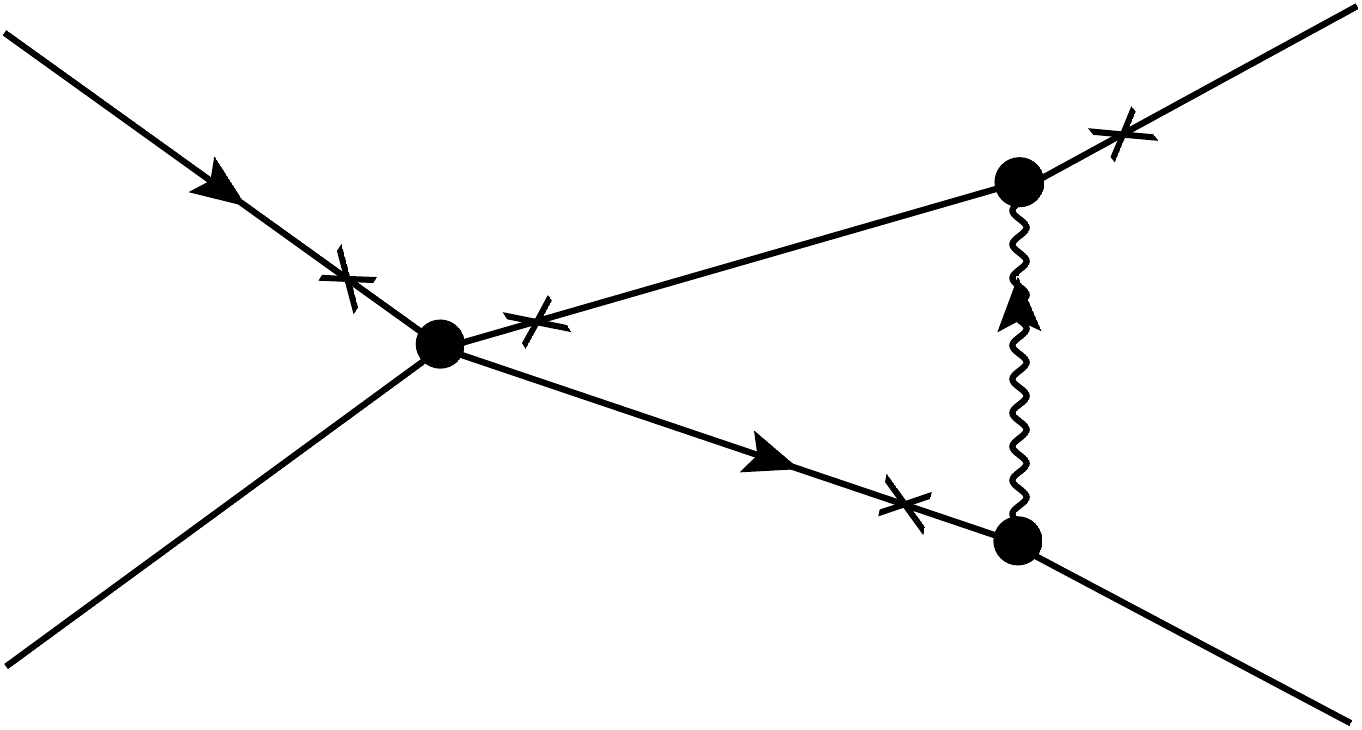}; \quad
\newline
21. \includegraphics[width=3cm, height=1.5cm]{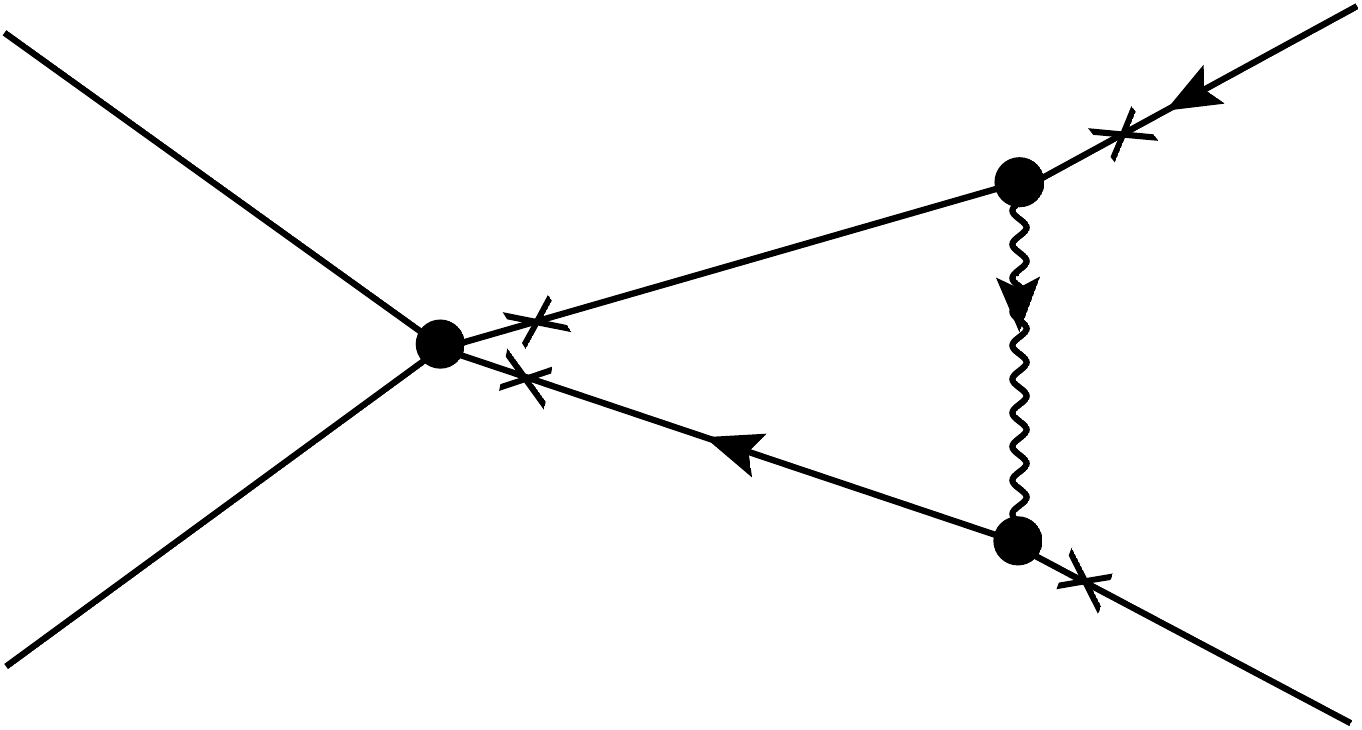}; \quad
22. \includegraphics[width=3cm, height=1.5cm]{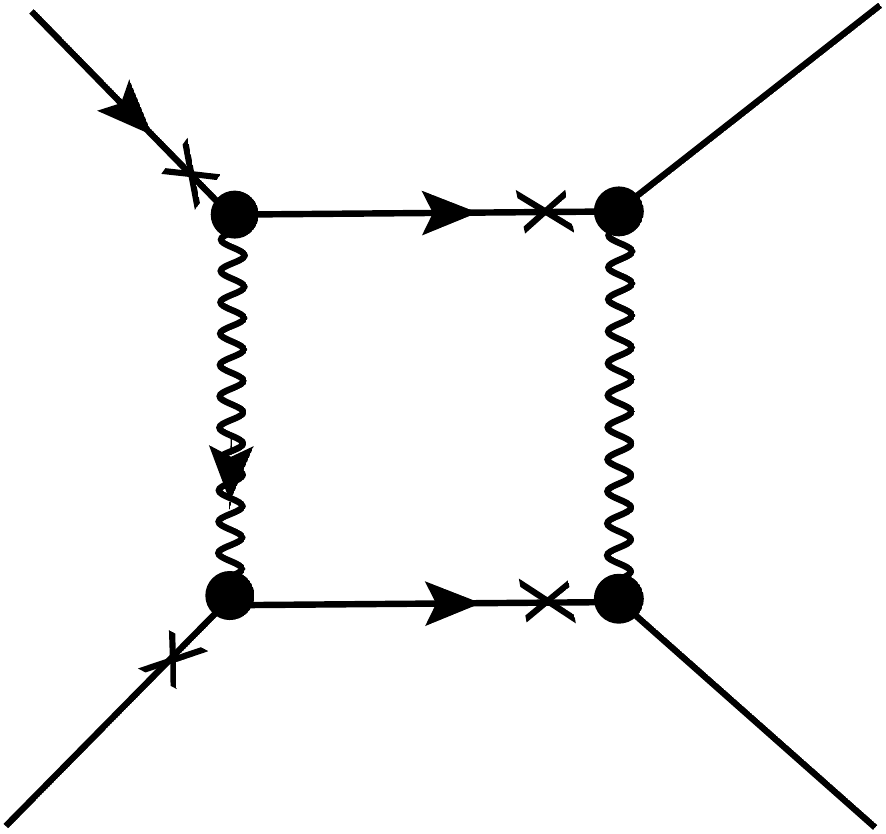}; \quad
\newline
23. \includegraphics[width=3cm, height=1.5cm]{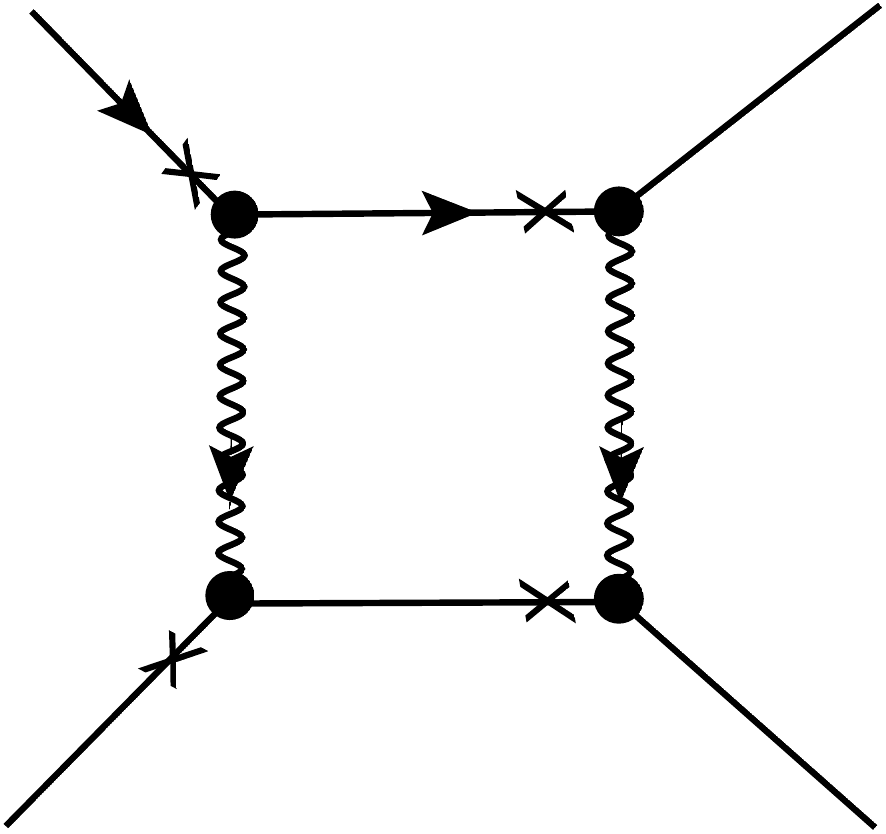}; \quad
24. \includegraphics[width=3cm, height=1.5cm]{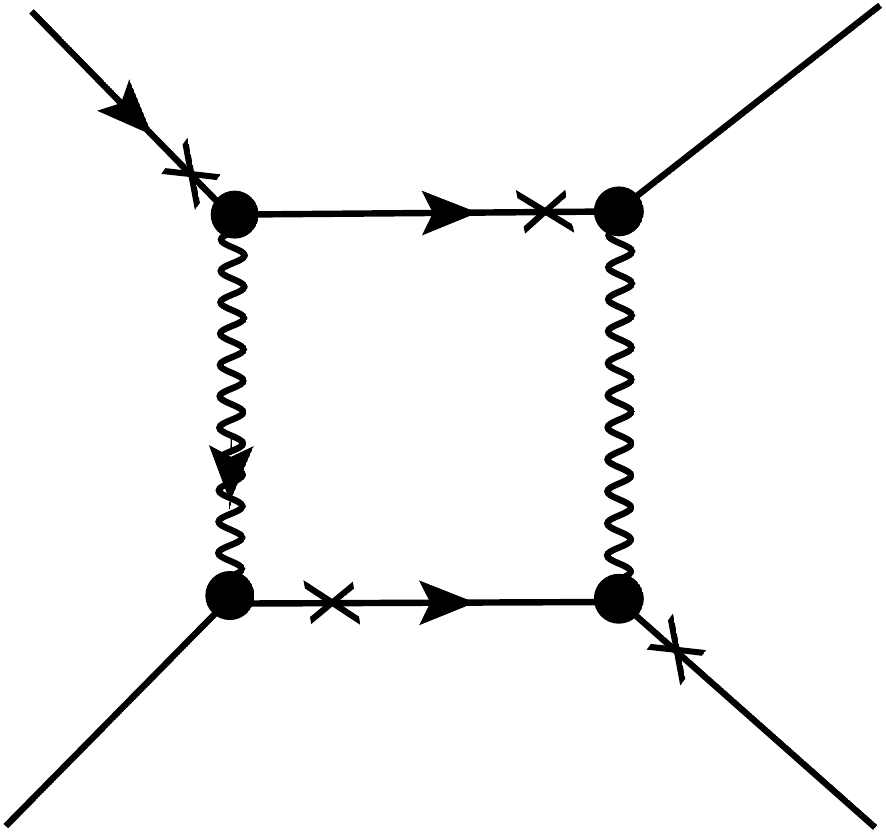}; \quad
\newline
25. \includegraphics[width=3cm, height=1.5cm]{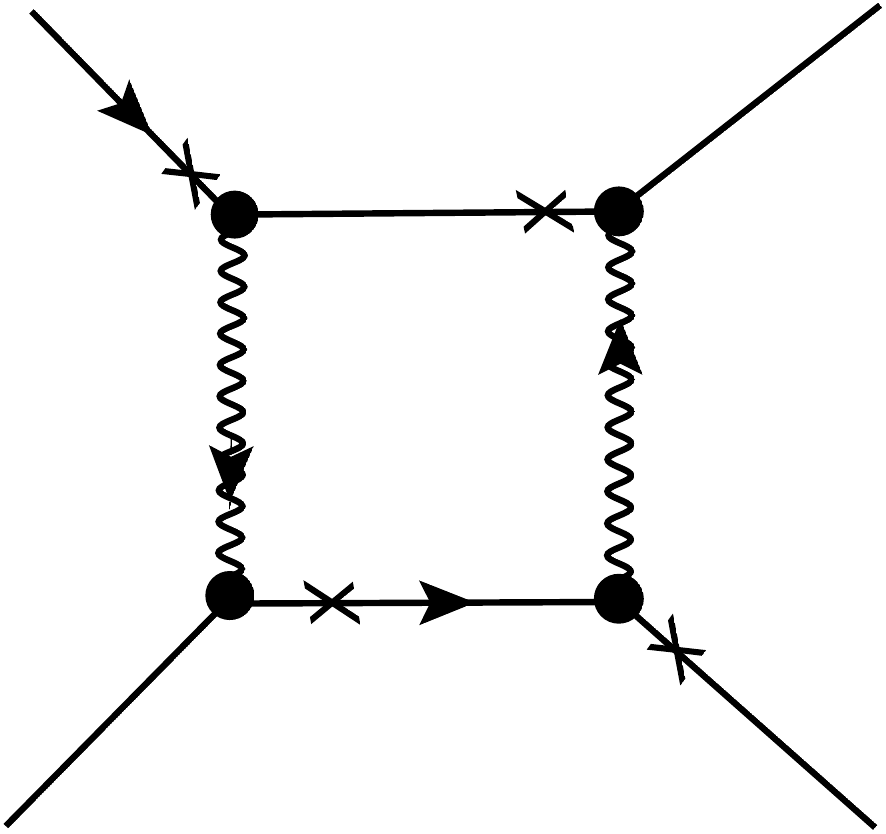}; \quad
26. \includegraphics[width=3cm, height=1.5cm]{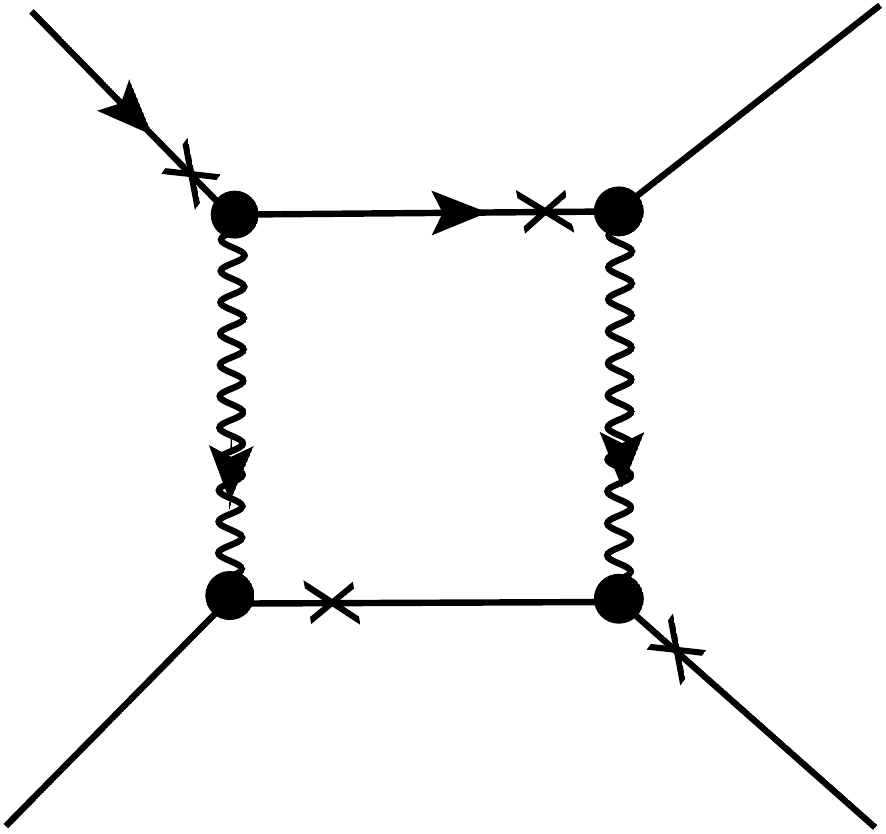}; \quad
\newline
27. \includegraphics[width=3cm, height=1.5cm]{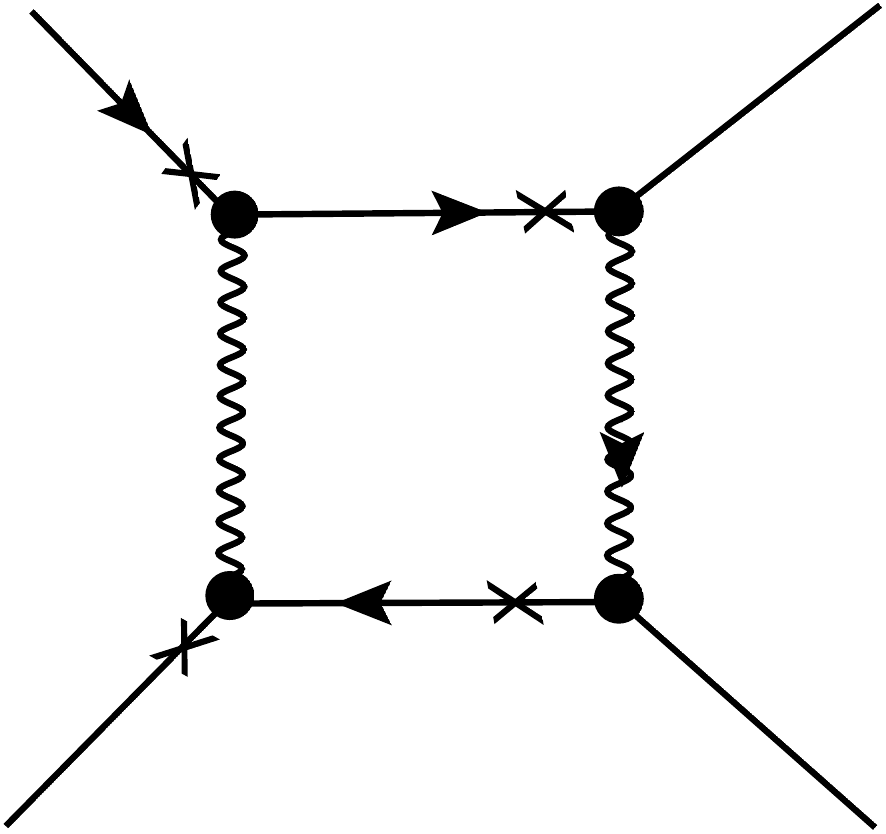}; \quad
28. \includegraphics[width=3cm, height=1.5cm]{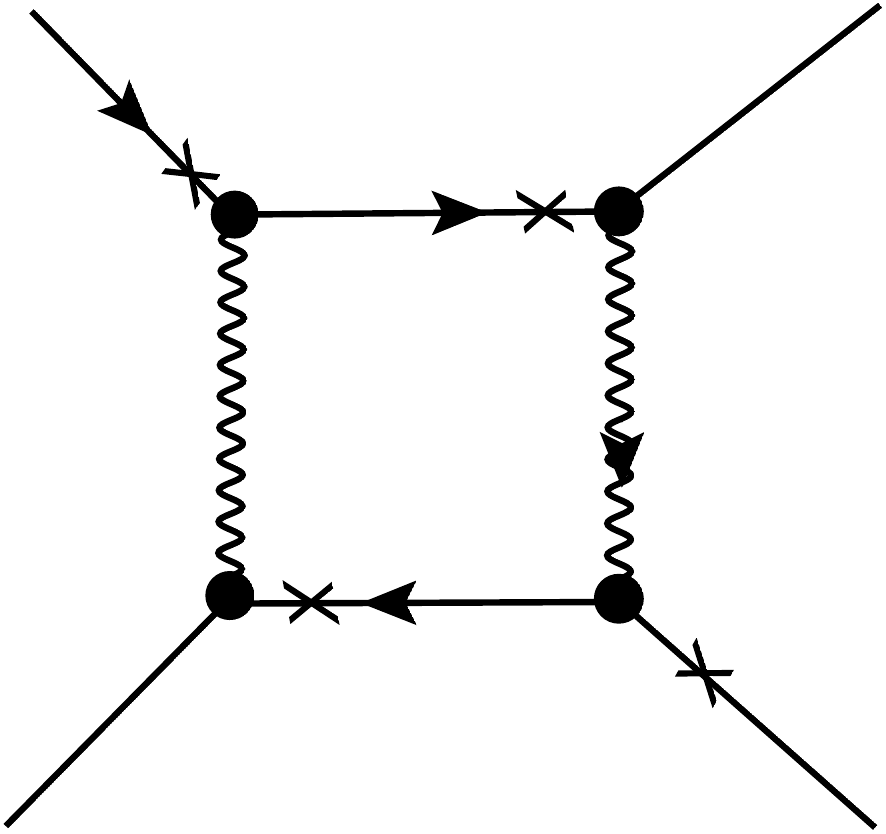}; \quad
\newline
29. \includegraphics[width=3cm, height=1.5cm]{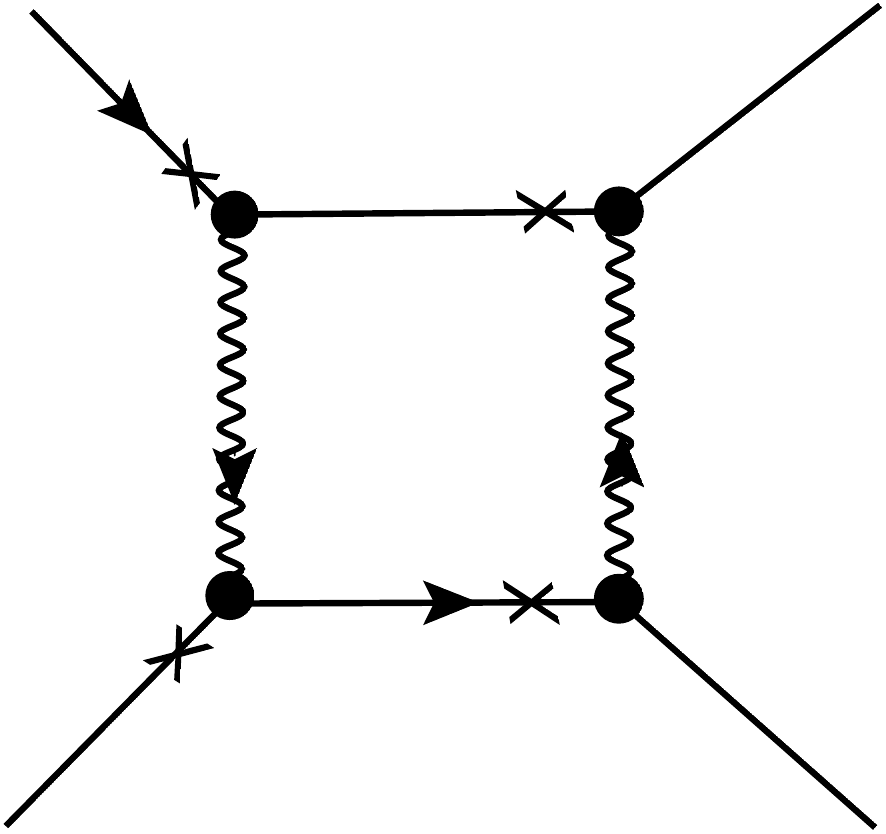}; \quad
30. \includegraphics[width=3cm, height=1.5cm]{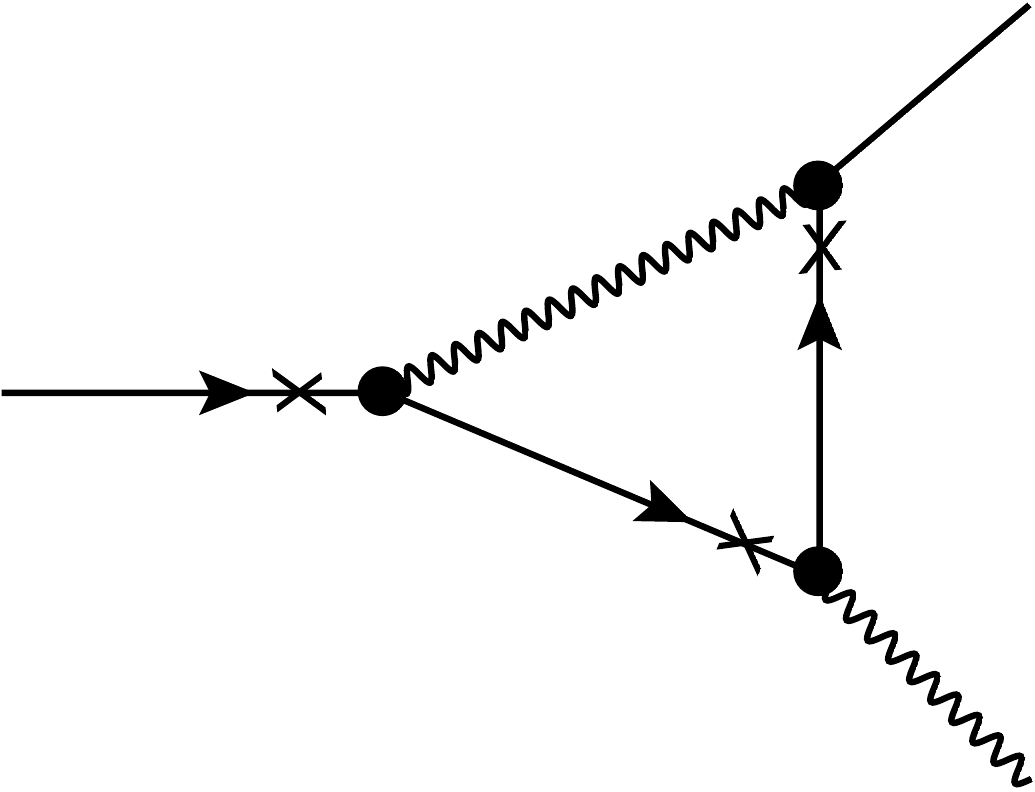}; \quad
\newline
31. \includegraphics[width=3cm, height=1.5cm]{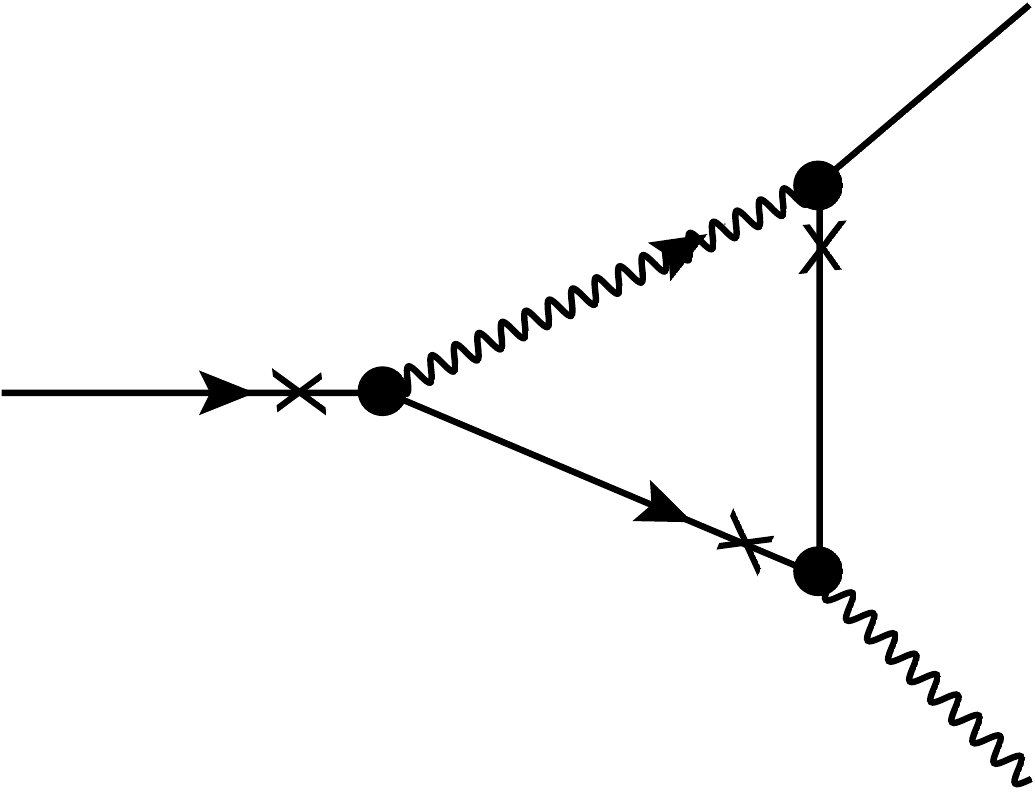}; \quad
32. \includegraphics[width=3cm, height=1.5cm]{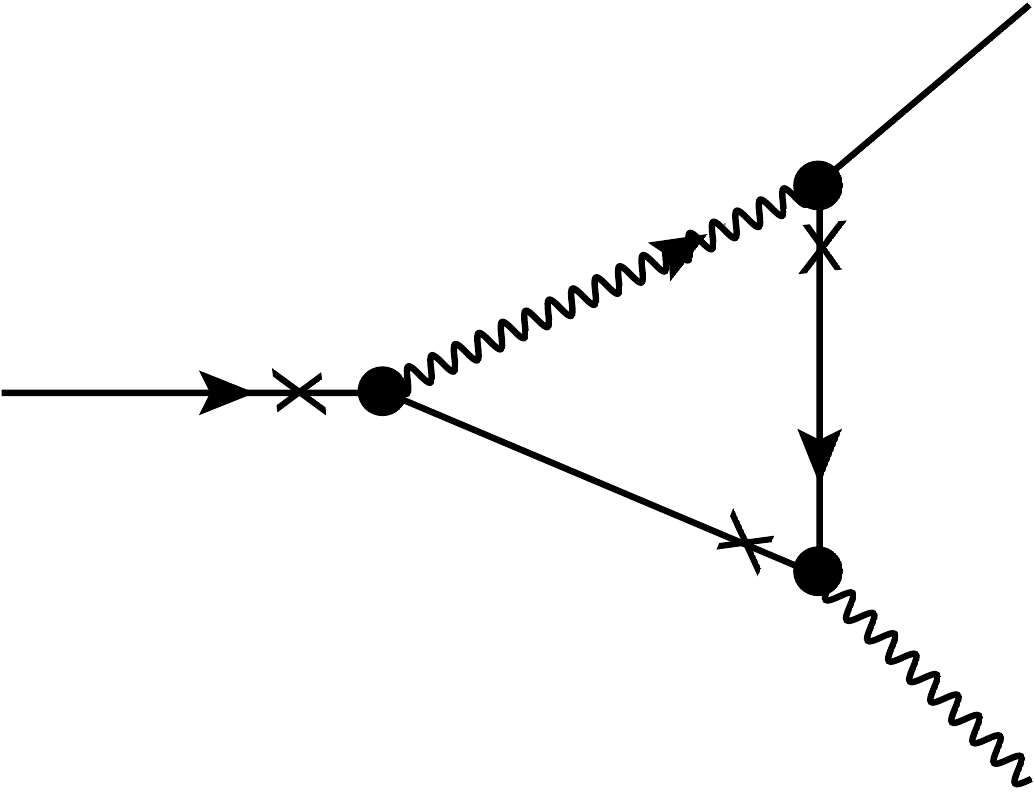}; \quad
\newline
33. \includegraphics[width=3cm, height=1.5cm]{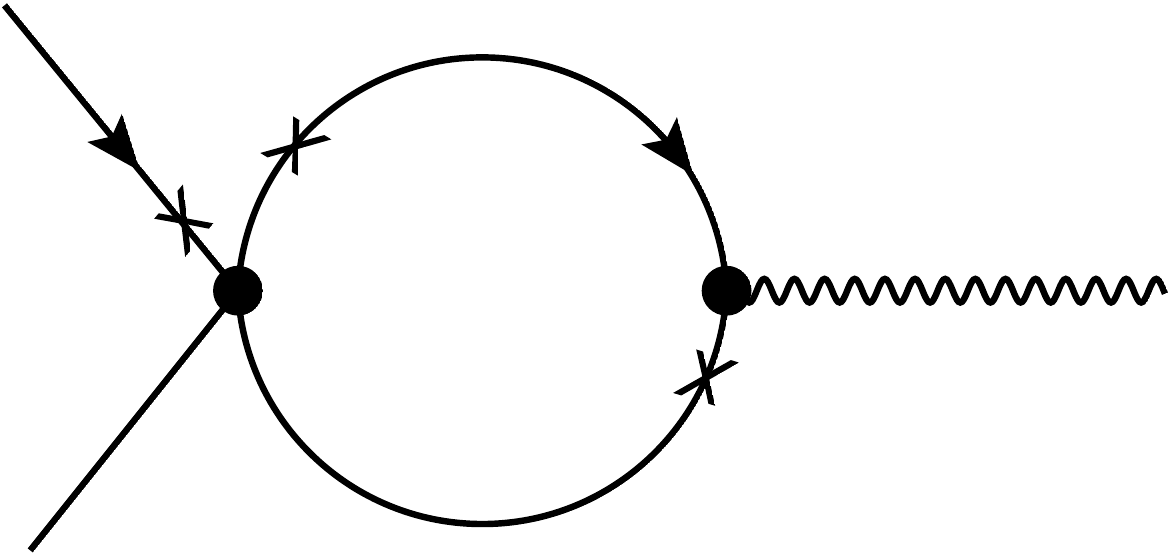}; \quad
34. \includegraphics[width=3cm, height=1.5cm]{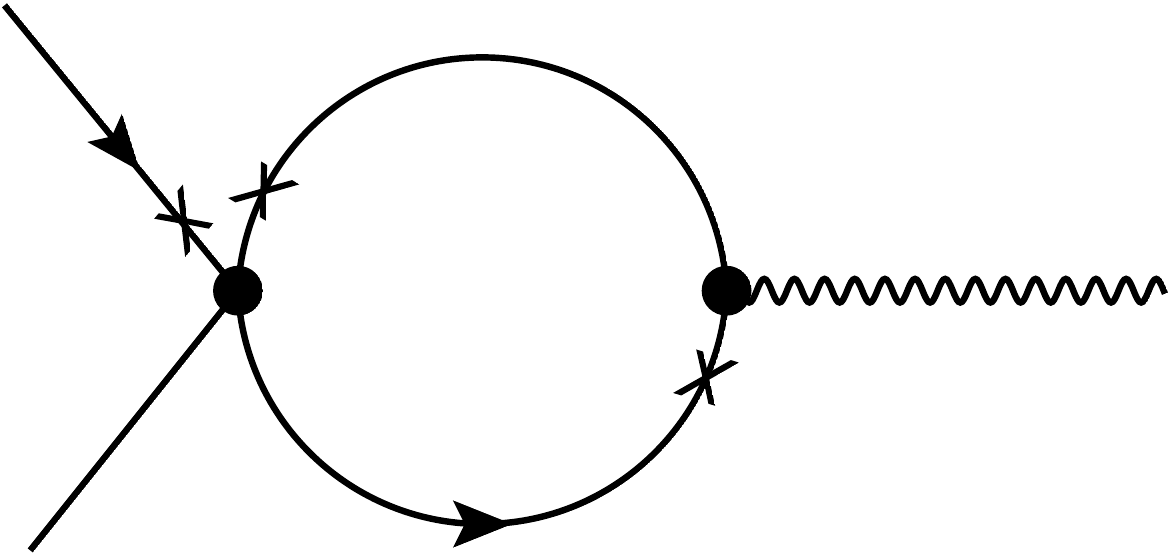}; \quad
\newline
35. \includegraphics[width=3cm, height=1.5cm]{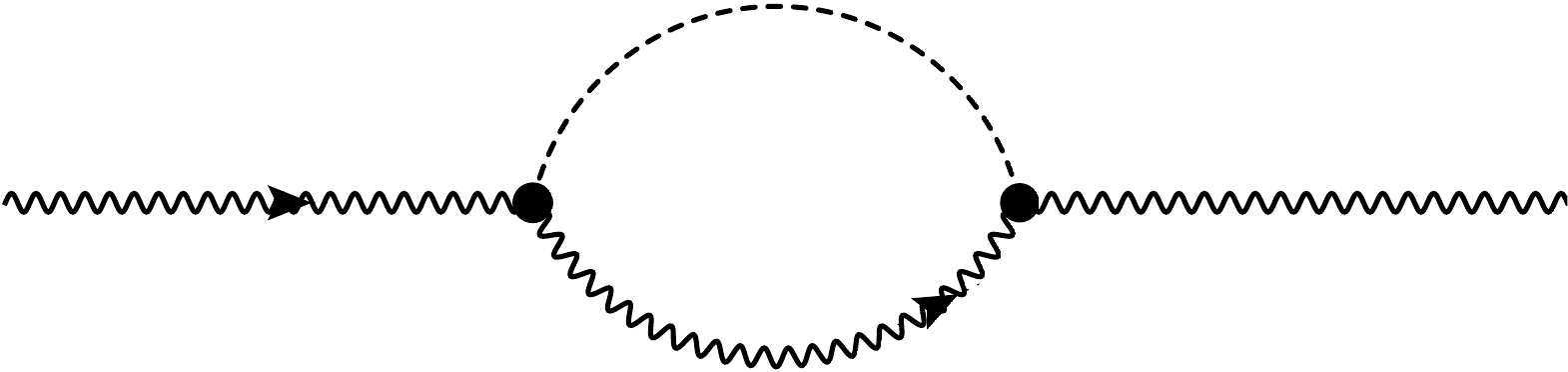}; \quad
36. \includegraphics[width=3cm, height=1.5cm]{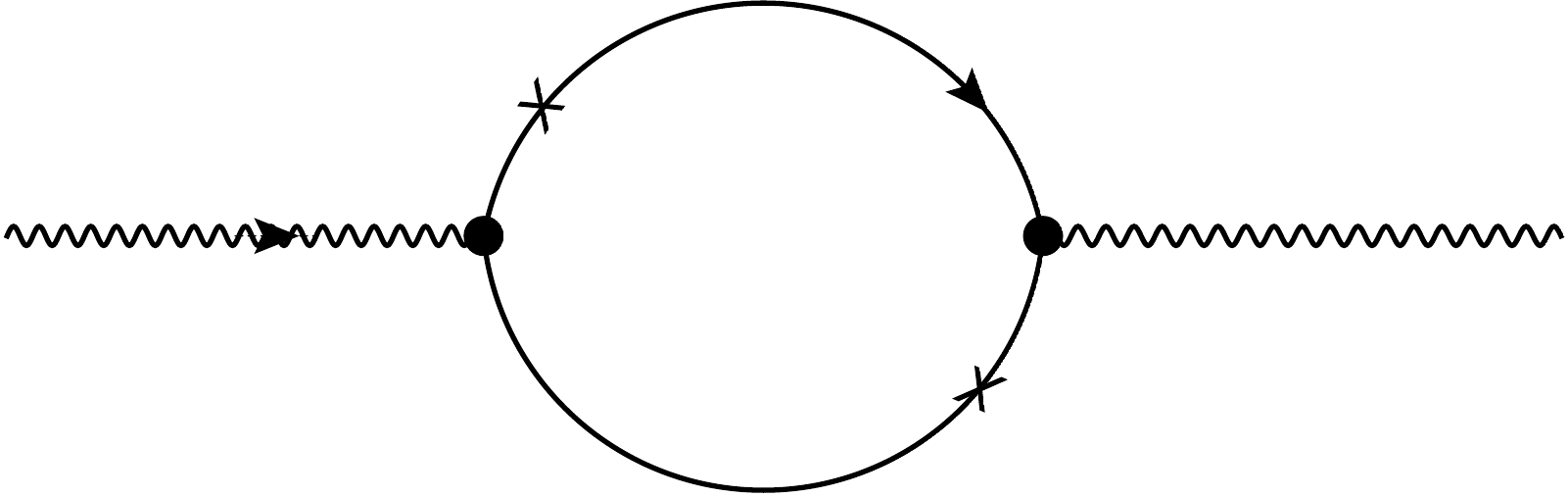}; \quad
\newline
37. \includegraphics[width=3cm, height=1.5cm]{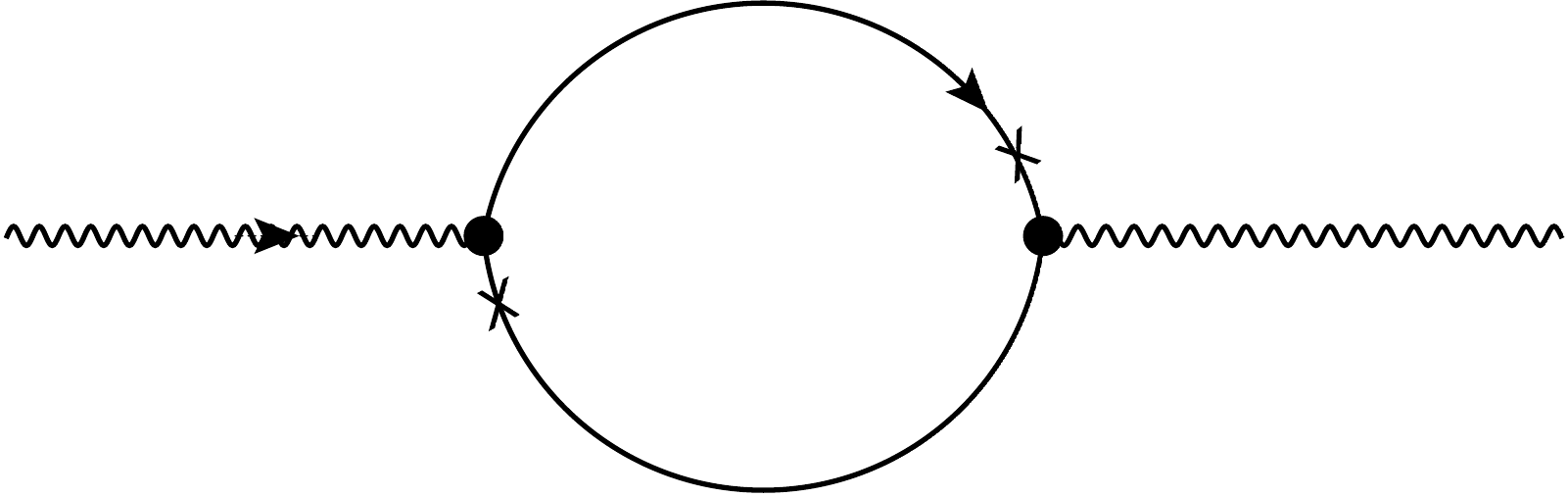}; \quad
38. \includegraphics[width=3cm, height=1.5cm]{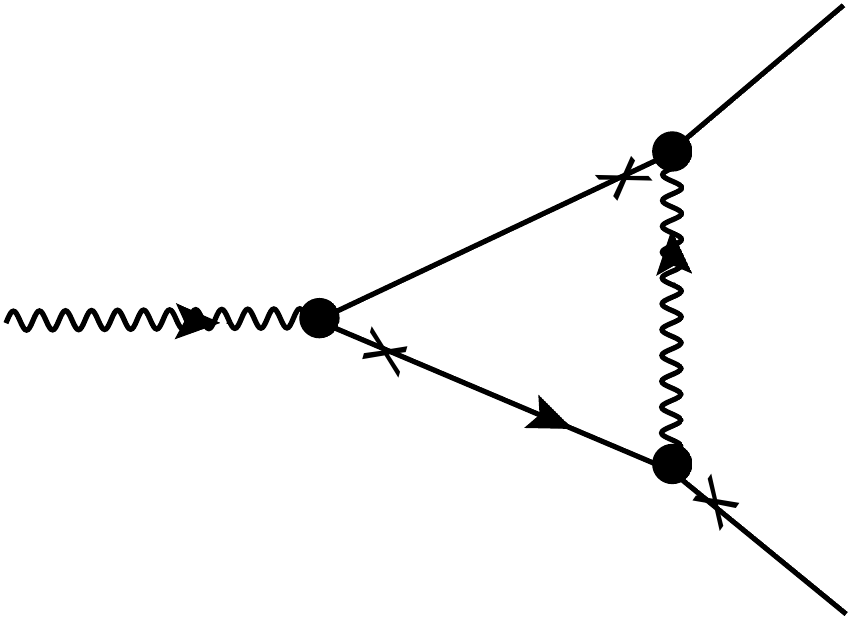}; \quad
\newline
39. \includegraphics[width=3cm, height=1.5cm]{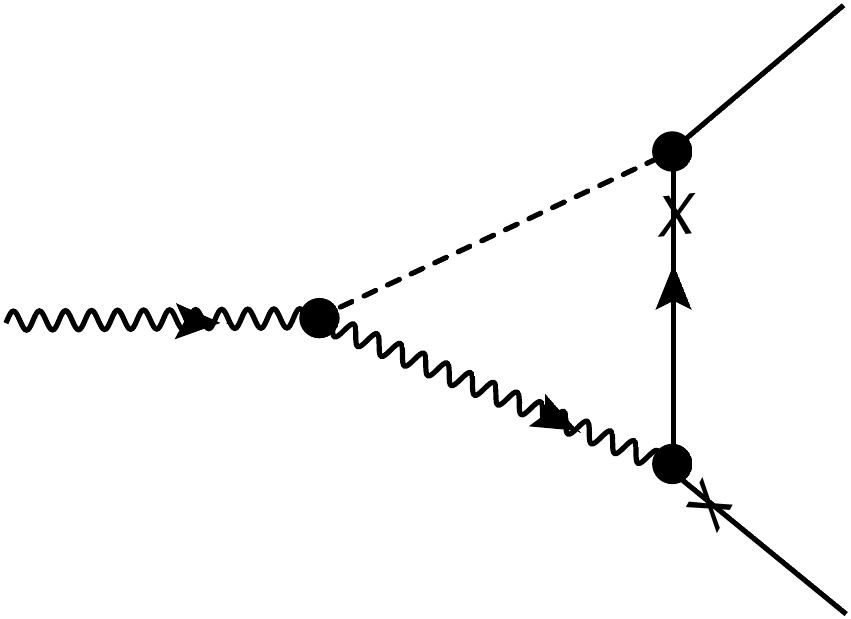}; \quad
40. \includegraphics[width=3cm, height=1.5cm]{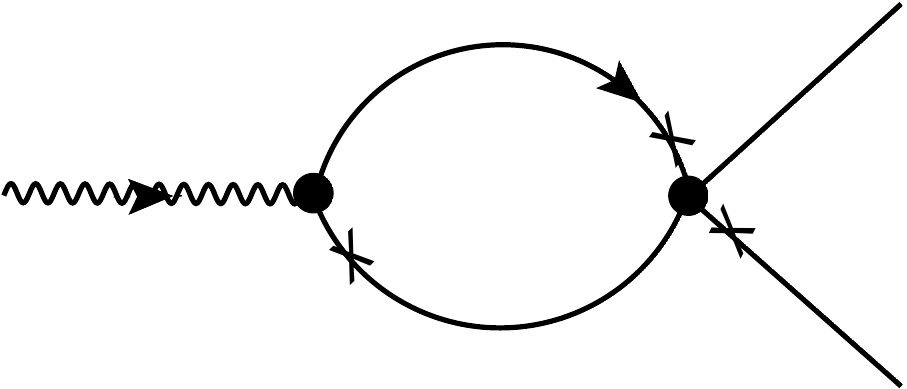}; \quad
\newline
41. \includegraphics[width=3cm, height=1.5cm]{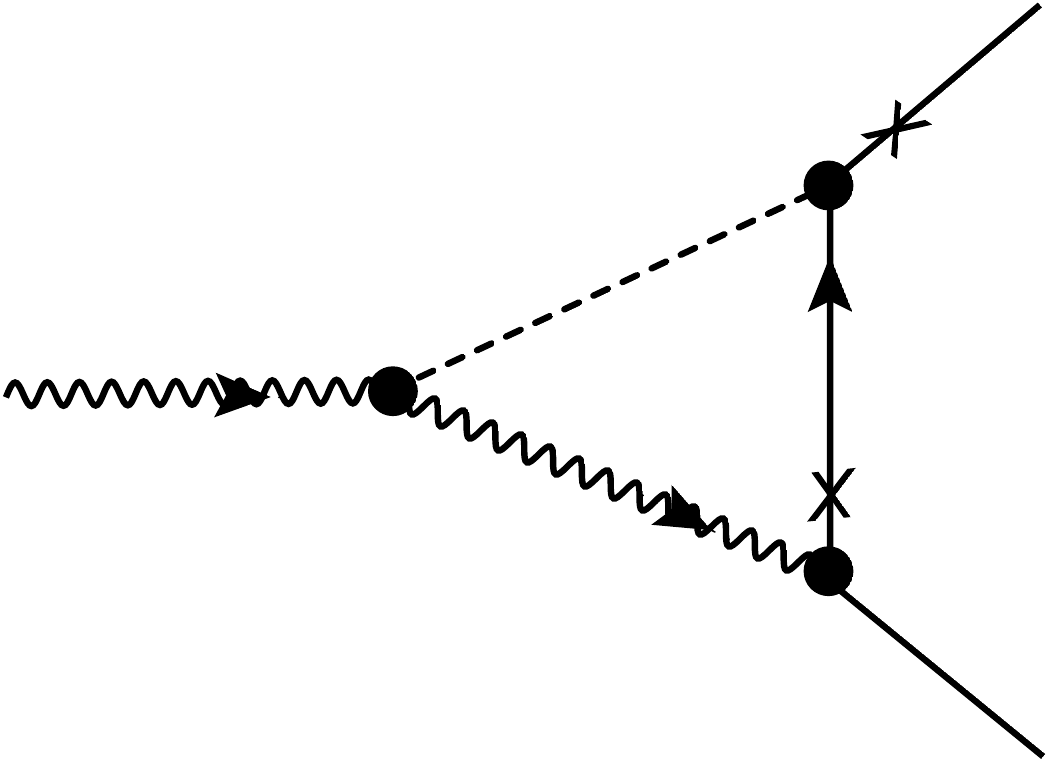}; \quad
42. \includegraphics[width=3cm, height=1.5cm]{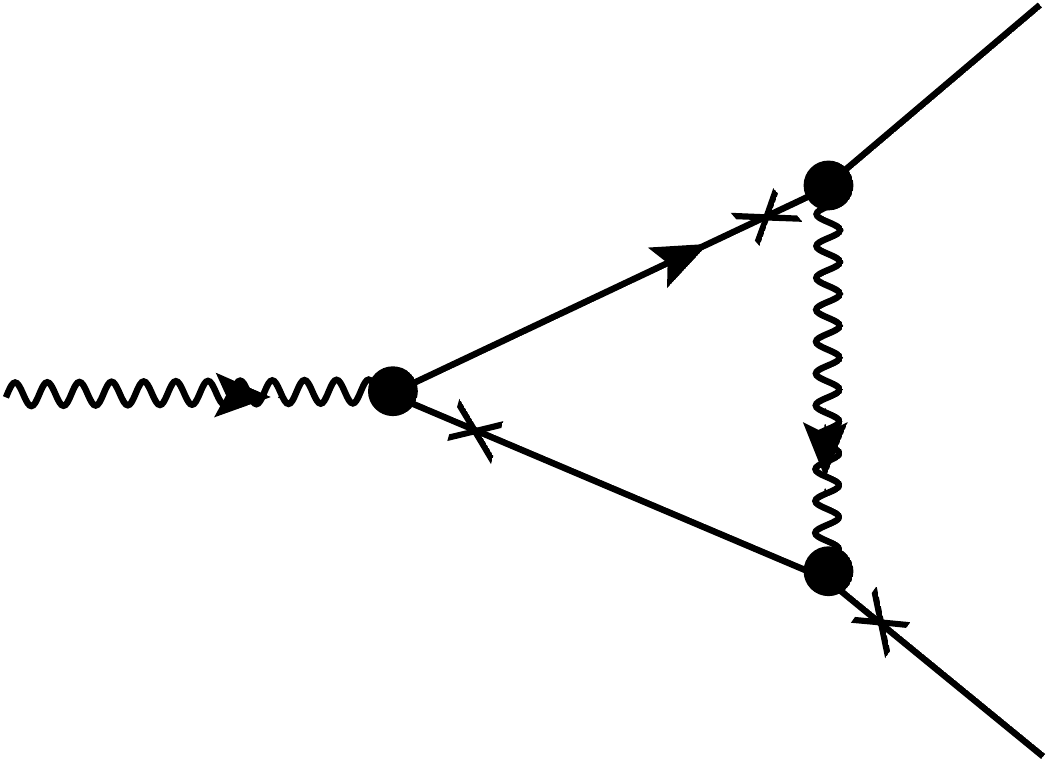}; \quad
\newline
43. \includegraphics[width=3cm, height=1.5cm]{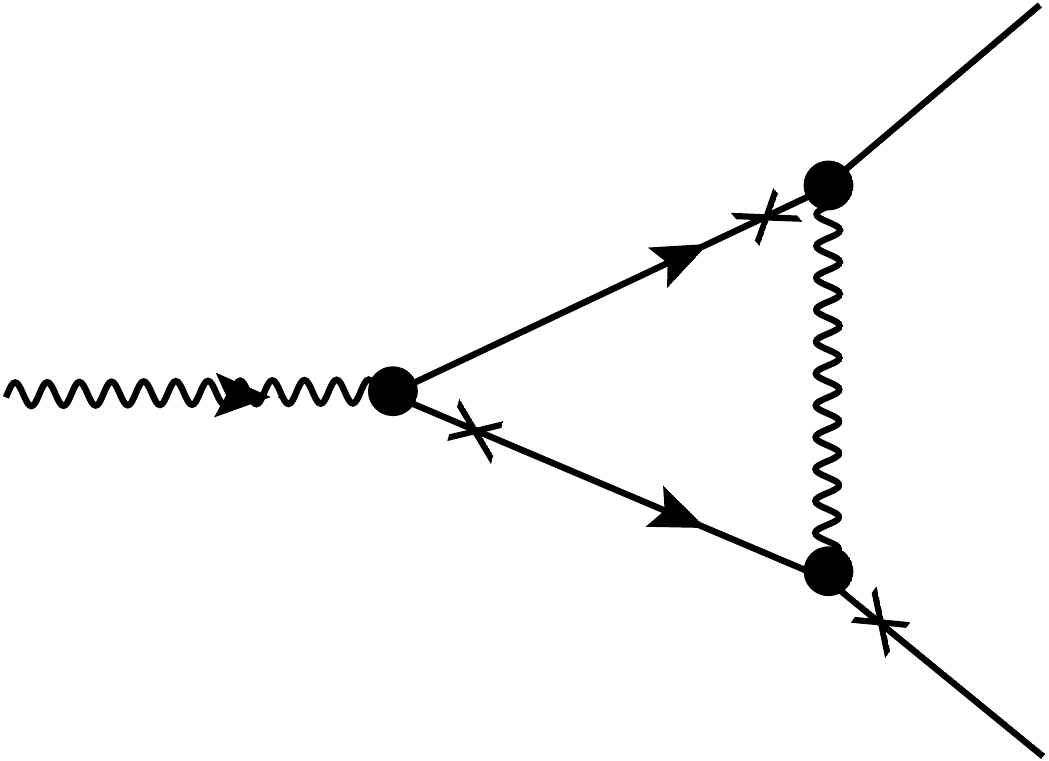}; \quad
44. \includegraphics[width=3cm, height=1.5cm]{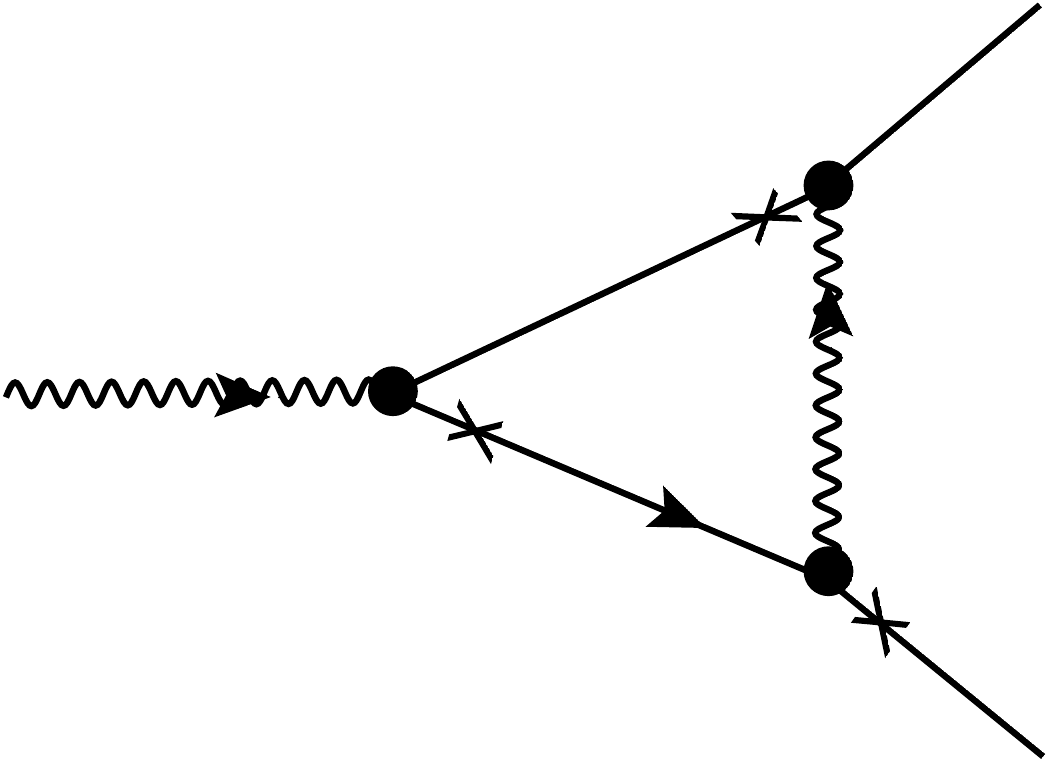}; \quad
\newline
45. \includegraphics[width=3cm, height=1.5cm]{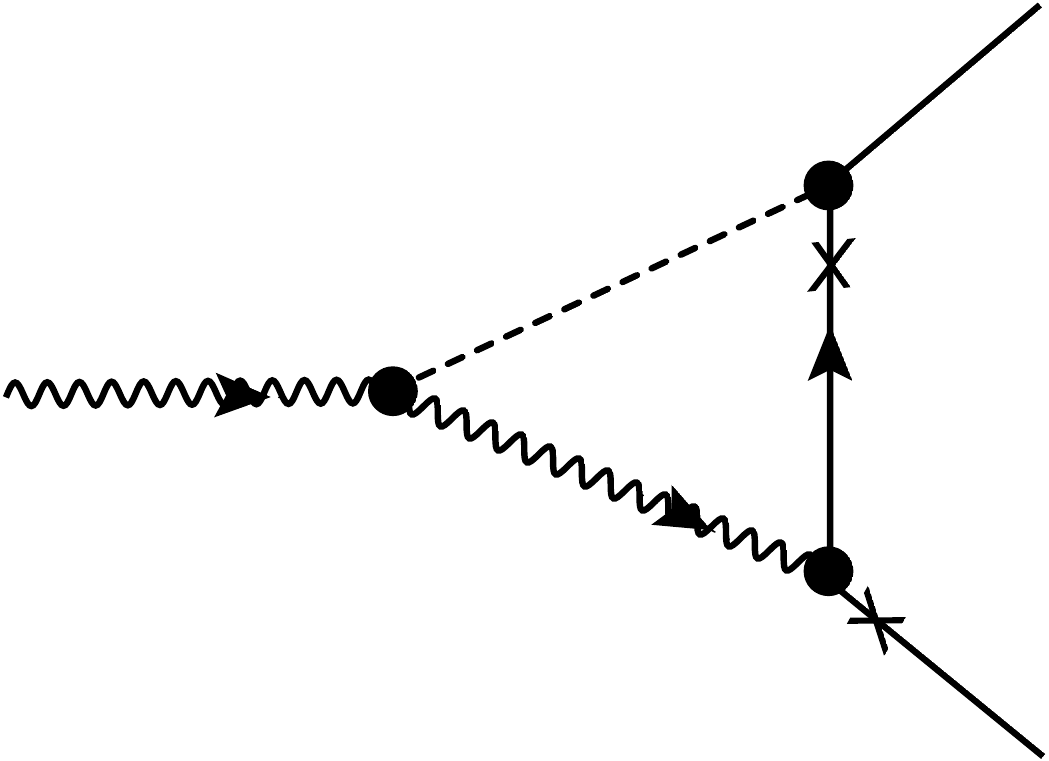}; \quad
46. \includegraphics[width=3cm, height=1.5cm]{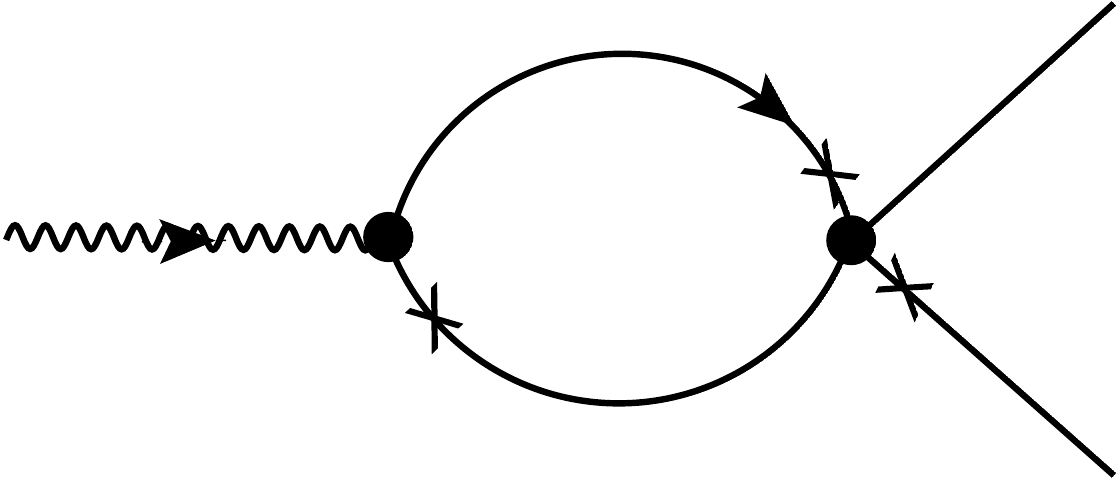}; \quad
\newline
47. \includegraphics[width=3cm, height=1.5cm]{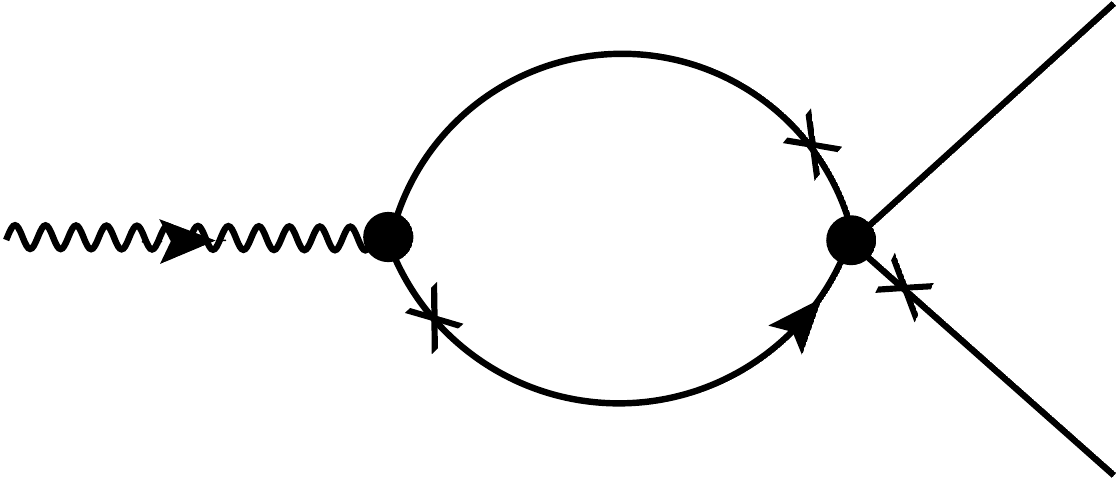}; \quad
48. \includegraphics[width=3cm, height=1.5cm]{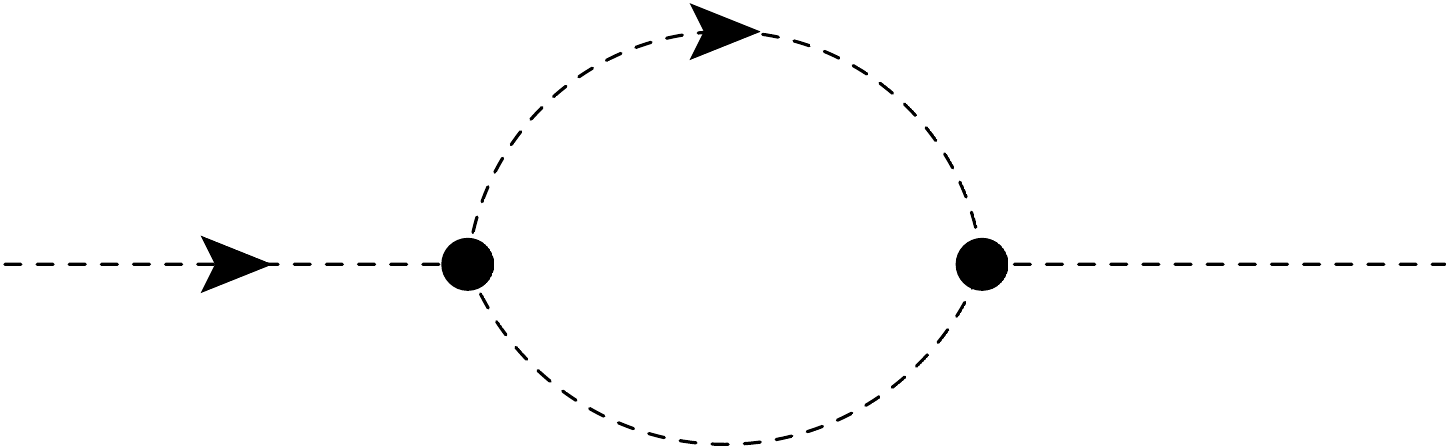}.

{The results for the diagrams} below contain the following {abbreviations}
 \begin{align*}
  G_{1+}&= i g_7+g_1 (1- i b)/3, & G_{1-}& = i g_7-g_1 (1+ i b)/3, \\
  G_{2+} &=g_2+ i (b g_2-g_3),  & G_{2-}& =g_2- i (b g_2-g_3),\\
  G_{5+} &= i g_5+u g_6,  &  G_{5-} & = i g_5-u g_6, \\
  P&=(\mp+\mq)^2,& Q &=(\mp+\mq)\cdot \mq,\\
  b_+&=1+u+ i b, & b_-&=1+u- i b, \\
  b_{1}&=1+ i b, & b_{1+}& =1+u_1+ i  b,\\
  b_{2}&=1- i b, &  b_{2+}& =1+u_1- i  b,\\
 u_+&=u_1+u,& &
\end{align*}
and the coupling constants scaling (\ref{res_char}). 
The results for simple poles of each diagram have the form: \\
\parbox[c]{0.075\textwidth}{\textnumero 1} \parbox{0.45\textwidth}
{$\Rightarrow\quad  - G_{2+} G_{2-}(1+u)/(b_{+}b_{-}\epsilon)$,}\\
\parbox[c]{0.075\textwidth}{\textnumero 3} \parbox{0.45\textwidth}
{$\Rightarrow\quad  -3g_4u_1^2/(8uu_+ \delta)$,}\\
\parbox[c]{0.075\textwidth}{\textnumero 4} \parbox{0.45\textwidth}
{$\Rightarrow\quad  -g_5^2/(2u\epsilon)$,}\\
\parbox[c]{0.075\textwidth}{\textnumero 5$\sim {p^2}$} \parbox{0.45\textwidth}
{$\Rightarrow\quad  -3g_4u_1^2/(8b_{1+}\delta)$,}\\
\parbox[c]{0.075\textwidth}{\textnumero 6$\sim {i\omega}$} \parbox{0.45\textwidth}
{$\Rightarrow\quad  -G_{2+}^2/(b_{+}^2\epsilon)$,}\\
\parbox[c]{0.075\textwidth}{\textnumero 6$\sim {p^2}$} \parbox{0.45\textwidth}
{$\Rightarrow\quad  -G_{2+}^2b_1u/( b_{+}^3\epsilon)$,}\\
\parbox[c]{0.075\textwidth}{\textnumero 7$\sim {i\omega}$} \parbox{0.45\textwidth}
{$\Rightarrow\quad  -G_{5+}G_{2+}/(b_{+}^2\epsilon)$,}\\
\parbox[c]{0.075\textwidth}{\textnumero 7$\sim {p^2}$} \parbox{0.45\textwidth}
{$\Rightarrow\quad  -ig_5G_{2+}[1+3u+u^2+ib$}\\
\parbox[c]{0.075\textwidth}{\phantom{\textnumero }{}\hskip1cm}    \parbox[c]{0.45\textwidth}
{$\times(2+3u+ib)]/(b_{+}^3\epsilon)+ug_6G_{2+} b_1^2/(b_{+}^3\epsilon)$,}\\
\parbox[c]{0.075\textwidth}{\textnumero 8} \parbox{0.45\textwidth}
{$\Rightarrow\quad  G_{1-}^2/(b_1\epsilon)$,}\\
\parbox[c]{0.075\textwidth}{\textnumero 9} \parbox{0.45\textwidth}
{$\Rightarrow\quad  2G_{1-}^2/\epsilon$,}\\
\parbox[c]{0.075\textwidth}{\textnumero 10} \parbox{0.45\textwidth}
{$\Rightarrow\quad  -2G_{1+}G_{1-}/\epsilon$,}\\
\parbox[c]{0.075\textwidth}{\textnumero 11} \parbox{0.45\textwidth}
{$\Rightarrow\quad  G_{2+}^2G_{1-}/(b_{+}b_1\epsilon)$,}\\
\parbox[c]{0.075\textwidth}{\textnumero 12} \parbox{0.45\textwidth}
{$\Rightarrow\quad  G_{5+}G_{2+}G_{1-}/(b_{+}b_1\epsilon)$,}\\
\parbox[c]{0.075\textwidth}{\textnumero 13} \parbox{0.45\textwidth}
{$\Rightarrow\quad  2G_{2+}^2G_{1-}/(b_{+}^2\epsilon)$,}\\
\parbox[c]{0.075\textwidth}{\textnumero 14} \parbox{0.45\textwidth}
{$\Rightarrow\quad  G_{2+}G_{2-}G_{1-}/(b_{+}b_{-}\epsilon)$,}\\
\parbox[c]{0.075\textwidth}{\textnumero 15} \parbox{0.45\textwidth}
{$\Rightarrow\quad  -G_{5-}G_{2+}G_{1-} (b_{+}+2)/(2b_{+}b_{-}b_1\epsilon)$,}\\
\parbox[c]{0.075\textwidth}{\textnumero 16} \parbox{0.45\textwidth}
{$\Rightarrow\quad  G_{5+}G_{2+}G_{1-}( b_{+}+2)/(b_{+}^2\epsilon)$,}\\
\parbox[c]{0.075\textwidth}{\textnumero 17} \parbox{0.45\textwidth}
{$\Rightarrow\quad  -G_{5+}G_{2+}G_{1+}/( b_{+}\epsilon)$,}\\
\parbox[c]{0.075\textwidth}{\textnumero 18} \parbox{0.45\textwidth}
{$\Rightarrow\quad  2G_{1-}G_{2+}G_{2-}(1+u)/(b_{+}b_{-}\epsilon)$,}\\
\parbox[c]{0.075\textwidth}{\textnumero 19} \parbox{0.45\textwidth}
{$\Rightarrow\quad  G_{1-}G_{5+}G_{2-}/(b_{+}\epsilon)$,}\\
\parbox[c]{0.075\textwidth}{\textnumero 20} \parbox{0.45\textwidth}
{$\Rightarrow\quad  -G_{1-}G_{5-}G_{2+}/(b_{-}\epsilon)$,}\\
\parbox[c]{0.075\textwidth}{\textnumero 21} \parbox{0.45\textwidth}
{$\Rightarrow\quad  -G_{1-}G_{5-}G_{2+}/(2b_{+}b_1\epsilon)$,}\\
\parbox[c]{0.075\textwidth}{\textnumero 22} \parbox{0.45\textwidth}
{$\Rightarrow\quad  -G_{5-}G_{2+}^3(u+b_{+})/(2ub_{+}^2b_1\epsilon)$,}\\
\parbox[c]{0.075\textwidth}{\textnumero 23} \parbox{0.45\textwidth}
{$\Rightarrow\quad  -G_{5+}G_{5-}G_{2+}^2(2+u)/(2ub_{+}b_{-}b_1\epsilon)$,}\\
\parbox[c]{0.075\textwidth}{\textnumero 24} \parbox{0.45\textwidth}
{$\Rightarrow\quad  G_{5+}G_{2+}^2G_{2-}(2u^2+2u+b_{+})$}\\
\parbox[c]{0.075\textwidth}{\phantom{\textnumero }{}\hskip1cm}    \parbox[c]{0.45\textwidth}
{$/(2ub_{-}b_{+}^2\epsilon)$,}\\
\parbox[c]{0.075\textwidth}{\textnumero 25} \parbox{0.45\textwidth}
{$\Rightarrow\quad  G_{5+}^2G_{2+}G_{2-}/(2b_{+}^2\epsilon)$,}\\
\parbox[c]{0.075\textwidth}{\textnumero 26} \parbox{0.45\textwidth}
{$\Rightarrow\quad  -G_{5+}G_{5-}G_{2+}^2[(2+u)(1+u)$}\\
\parbox[c]{0.075\textwidth}{\phantom{\textnumero }{}\hskip1cm}    \parbox[c]{0.45\textwidth}
{$+iub]/(2ub_{+}b_{-}b_{+}\epsilon)$,}\\
\parbox[c]{0.075\textwidth}{\textnumero 27} \parbox{0.45\textwidth}
{$\Rightarrow\quad  G_{5+}G_{2+}^2G_{2-}/(2ub_{+}b_{-}\epsilon)$,}\\
\parbox[c]{0.075\textwidth}{\textnumero 28} \parbox{0.45\textwidth}
{$\Rightarrow\quad  -G_{5-}G_{2+}^3/(2ub_{+}^2\epsilon)$,}\\
\parbox[c]{0.075\textwidth}{\textnumero 29} \parbox{0.45\textwidth}
{$\Rightarrow\quad  -G_{5+}G_{5-}G_{2+}^2/(2b_{+}^2b_1\epsilon)$,}\\
\parbox[c]{0.075\textwidth}{\textnumero 30} \parbox{0.45\textwidth}
{$\Rightarrow\quad  -G_{2+}^3/(b_{+}^2\epsilon)$,}\\
\parbox[c]{0.075\textwidth}{\textnumero 31} \parbox{0.45\textwidth}
{$\Rightarrow\quad  -G_{5+}G_{2+}^2(2+b_{+})/(2b_{+}^2\epsilon)$,}\\
\parbox[c]{0.075\textwidth}{\textnumero 32} \parbox{0.45\textwidth}
{$\Rightarrow\quad  -G_{5+}G_{2-}G_{2+}/(2b_{+}\epsilon)$,}\\
\parbox[c]{0.075\textwidth}{\textnumero 33} \parbox{0.45\textwidth}
{$\Rightarrow\quad  -G_{1-}G_{2-}/\epsilon$,}\\
\parbox[c]{0.075\textwidth}{\textnumero 34} \parbox{0.45\textwidth}
{$\Rightarrow\quad  -G_{1-}G_{2+}/\epsilon$,}\\
\parbox[c]{0.075\textwidth}{\textnumero 38} \parbox{0.45\textwidth}
{$\Rightarrow\quad  -(3g_4u_1^2)/(8uu_+\delta)$,}\\
\parbox[c]{0.075\textwidth}{\textnumero 39} \parbox{0.45\textwidth}
{$\Rightarrow\quad  ig_5G_{2-}b_1/(4u\epsilon)+g_6G_{2-}/(2\epsilon)$,}\\
\parbox[c]{0.075\textwidth}{\textnumero 40} \parbox{0.45\textwidth}
{$\Rightarrow\quad  -ig_5G_{2+}b_2/(4u\epsilon)+g_6G_{2+}/(2\epsilon)$,}\\
\parbox[c]{0.075\textwidth}{\textnumero 41} \parbox{0.45\textwidth}
{$\Rightarrow\quad  -3G_{5+}g_4Q u_1^2b_{1+}/ \{8u_+\delta[(u_1+1)^2$}\\
\parbox[c]{0.075\textwidth}{\phantom{\textnumero }{}\hskip1cm}    \parbox[c]{0.45\textwidth}
{$+b^2 ] \}$,}\\
\parbox[c]{0.075\textwidth}{\textnumero 42} \parbox{0.45\textwidth}
{$\Rightarrow\quad  -g_5^2G_{2+}(b_2P/2-uQ/b_{-})/\{2\epsilon$}\\
\parbox[c]{0.075\textwidth}{\phantom{\textnumero }{}\hskip1cm}    \parbox[c]{0.45\textwidth}
{$\times(1+u-ib)\} - i u g_6g_5G_{2+} P/ (2b_{-}\epsilon)   $}\\
\parbox[c]{0.075\textwidth}{\phantom{\textnumero }{}\hskip1cm}    \parbox[c]{0.45\textwidth}
{$+ u^2g_6^2G_{2+}P/(2b_{-}\epsilon )- i u g_6g_5G_{2+}b_2 $}\\
\parbox[c]{0.075\textwidth}{\phantom{\textnumero }{}\hskip1cm}    \parbox[c]{0.45\textwidth}
{$\times(P/2+Q/ b_{-})/(2b_{-}\epsilon ) $,}\\
\parbox[c]{0.075\textwidth}{\textnumero 43} \parbox{0.45\textwidth}
{$\Rightarrow\quad  -ig_5G_{2+}G_{2-} u[(u^2+2u-b^2+1)$}\\
\parbox[c]{0.075\textwidth}{\phantom{\textnumero }{}\hskip1cm}    \parbox[c]{0.45\textwidth}
{$\times(P-2Q)-ib(u^2+4u+b^2+3)P]$}\\
\parbox[c]{0.075\textwidth}{\phantom{\textnumero }{}\hskip1cm}    \parbox[c]{0.45\textwidth}
{$/(2b_{+}^2b_{-}^2\epsilon)+ug_6G_{2+}G_{2-}[(1+u)P]/(b_{+}b_{-}\epsilon)$,}\\
\parbox[c]{0.075\textwidth}{\textnumero 44} \parbox{0.45\textwidth}
{$\Rightarrow\quad  -g_5^2G_{2-}[b_{1}P/2-u(P-Q)/b_{+}
]/(2b_{+}\epsilon)$}\\
\parbox[c]{0.075\textwidth}{\phantom{\textnumero }{}\hskip1cm}    \parbox[c]{0.45\textwidth}
{$+iug_6g_5G_{2-}b_{1}[P/2+(P-Q)/ b_{+}]/(2b_{+}\epsilon)  $}\\
\parbox[c]{0.075\textwidth}{\phantom{\textnumero }{}\hskip1cm}    \parbox[c]{0.45\textwidth}
{$ +iug_6g_5G_{2-}P/(2b_{+}\epsilon) +u^2g_6^2G_{2-}P/(2b_{+}\epsilon) $,}\\
\parbox[c]{0.075\textwidth}{\textnumero 45} \parbox{0.45\textwidth}
{$\Rightarrow\quad  G_{5-}3g_4u_1^2b_{2+}(P-Q)/(8b_{1+}b_{2+}u_+ \delta)$,}\\
\parbox[c]{0.075\textwidth}{\textnumero 46} \parbox{0.45\textwidth}
{$\Rightarrow\quad  -ig_5G_{1-}Pb_2/(2\epsilon) +ug_6G_{1-}P/\epsilon $,}\\
\parbox[c]{0.075\textwidth}{\textnumero 47} \parbox{0.45\textwidth}
{$\Rightarrow\quad  -ig_5G_{1+}b_{1}P/(2\epsilon )-ug_6G_{1+}P/\epsilon$,}\\
\parbox[c]{0.075\textwidth}{\textnumero 48} \parbox{0.45\textwidth}
{$\Rightarrow\quad  -(g_4u_1)/(8\delta)$.}\\
The other diagram contributions to the renormalization constants ({\textnumero 2, 35, 36, 37}) are equal to zero.

 \bibliographystyle{apsrev}
 \bibliography{mybib}

\end{document}